\documentclass[11pt,a4paper]{article}
\pdfoutput=1
\usepackage[english]{babel}
\usepackage{amsfonts,amsbsy,bm,euscript,mathrsfs}
\usepackage{amssymb,stmaryrd,faktor,slashed}
\usepackage[normalem]{ulem}
\usepackage[x11names]{xcolor}
\usepackage[tbtags]{amsmath}
\usepackage[bookmarks=true,colorlinks=true,linkcolor=black,citecolor=black,urlcolor=black,bookmarksnumbered]{hyperref}
\usepackage[nosort]{cite}
\usepackage{tensor,braket,dsfont}
\usepackage{tikz}
\usetikzlibrary{arrows.meta}

\usepackage{comment}

\newcommand{\dd}{\mathrm{d}}

\newcommand{\tr}{\mathrm{tr}}

\numberwithin{equation}{section}

\newcommand{\hstar}{\, \hat{\star}\, }

\newcommand{\X}{\mathbb{X}}
\newcommand{\Y}{\mathbb{Y}}
\newcommand{\U}{\mathbb{U}}
\newcommand{\V}{\mathbb{V}}
\newcommand{\W}{\mathbb{W}}
\newcommand{\Q}{\mathbb{Q}}

\begin{document}

\phantom{.}
\vspace{50pt}

\begin{center}
{\huge{\bf A new perspective on non-commutative deformations of field and gauge theories}} 

\vspace{36pt}

Riccardo Borsato and Tim Meier

\vspace{24pt}

{
\small {\it 
Instituto Galego de F\'isica de Altas Enerx\'ias (IGFAE),\\[2pt]
and Departamento de F\'\i sica de Part\'\i culas,\\[2pt]
Universidade de  Santiago de Compostela,\\[2pt]
15705 Santiago de Compostela,  Spain\\[4pt]}
\vspace{12pt}
\texttt{riccardo.borsato@usc.es}, \qquad \texttt{tim.meier@usc.es}}\\

\vspace{36pt}

{\bf Abstract}
\end{center}
\noindent
We construct non-commutative deformations of field and gauge theories based on star-products implemented by Drinfel'd twists. We are able to encompass a large family of twists, including those built out of conformal symmetries and supersymmetries. The main idea behind our construction is to work with twists constructed from symmetries of the undeformed theory, that are realised as active symmetry transformations. We argue that our construction amounts to a reformulation of known deformations of gauge theories, and that it significantly extends the range of applicable examples. To ensure consistency with gauge invariance, we also identify a unimodularity condition that is weaker than the one that is normally employed in the literature, so that we can apply twists that would otherwise be left out. Finally, we also prove a planar equivalence theorem stating that the Feynman diagrams of the deformed theories retain an undeformed internal structure, with the twist acting only on their external legs. All these  results are important to identify and work with deformations of $\mathcal N=4$ super Yang-Mills that are proposed to be dual to homogeneous Yang-Baxter deformations of the $AdS_5\times S^5$ superstring, but the applicability of our construction and results goes beyond that.

\newpage 


\tableofcontents


\section{Introduction}
Non-commutative deformations of field and gauge theories have been an active area of research for several years, see~\cite{Szabo:2001kg, Douglas:2001ba,Szabo:2025mxr,Hersent:2022gry,Vitale:2023znb,Wallet:2025xbp} for reviews. They originally attracted attention because of the hope that they could provide a way to remove UV divergences of quantum field theories \cite{Snyder:1946qz}. Soon, however, it became clear that divergences could not be completely removed because of the phenomenon of UV/IR mixing~\cite{Minwalla:1999px}. More recently, braided field theories are being studied as a way to avoid UV/IR mixing, thanks to a different type of quantisation, see e.g.~\cite{Giotopoulos:2021ieg}. In this work, we do not consider the issue of UV/IR mixing, as our analysis is restricted to the planar sector of non-commutative gauge theories.

The non-commutative deformations can be encoded into an operation, the star-product $\star$, that replaces the ordinary product of fields by a non-commutative one. A systematic way to construct non-commutative star-products is to reformulate the field theory in the language of Hopf algebras, and to twist  via the well known method of Drinfel'd twists~\cite{drinfeld_YBESolutions_1983}. Their properties ensure that the resulting star-product  is still associative albeit being non-commutative. A consequence of the construction, then, is that the symmetries of the original undeformed field theory are in general broken by the deformation; at the same time, these symmetries are still secretly realised in a twisted way, because the standard Leibniz rule for the action of charges on products of fields is modified by the twisted coproduct of the Hopf algebra.

In the standard point of view, the generators entering the Drinfel'd twists are realised as vectors acting as Lie derivatives on the fields, see e.g.~\cite{Aschieri:2005zs,Dimitrijevic:2011jg}, and in this way we may understand also the implementation of the celebrated Groenewold-Moyal deformation~\cite{Moyal:1949sk,Groenewold:1946kp,Douglas:2001ba,Szabo:2001kg}. As long as an action of the Drinfel'd twists on the fields exists, no further requirement is demanded, and in principle one may construct Drinfel'd twist deformations even out of vectors that do not correspond to symmetries of the original undeformed field theory. 

Issues start to arise when attempting to construct non-commutative twist deformations of \emph{gauge} theories. While it is certainly possible to construct the Groenewold-Moyal deformation of gauge theories~\cite{Szabo:2001kg} (as well as other kinds of deformations, for example the dipole deformation of~\cite{Guica:2017mtd} or the angular dipole of~\cite{Meier:2023kzt}) severe obstructions appear for more general Drinfel'd twists. First,  compatibility with gauge invariance requires that the star product, despite being non-commutative, should nevertheless be  cyclic under integration; this requirement led the authors of ~\cite{Aschieri:2009ky} to a unimodularity condition for the Drinfel'd twist that in this paper we will call the ``$\mathcal{F}$-unimodularity condition''. As we will show here, the $\mathcal{F}$-unimodularity condition of~\cite{Aschieri:2009ky} is in fact too restrictive, because it implements a stronger condition than cyclicity. In fact, it would forbid us to use examples of Drinfel'd twists that are of interest, and  it would select only certain representatives in the equivalence classes under which Drinfel'd twists are organised.  In this paper we will show that the minimal requirement to achieve cyclicity under integration is what we call the ``$\mathcal{R}$-unimodularity condition'', which is naturally compatible with the organisation into equivalence classes of Drinfel'd twists, and that is satisfied by twists of physical interest.

Another  issue  when constructing non-commutative deformations of gauge theories is the fact that in general also the action of partial derivatives (as that of charges) on products of fields fails to follow the standard Leibniz rule, see e.g.~\cite{Aschieri:2005yw,Dimitrijevic:2011jg}. In general, this means that it may be difficult to define covariant derivatives, resulting with an obstruction to promote the gauge invariance to the deformed setup. We refer  to~\cite{Aschieri:2005zs,Wess:2006cm,Dimitrijevic:2011jg ,Dimitrijevic:2014dxa} for examples discussing this issue.
To solve this problem, in this paper we propose to change the perspective in how we realise the Drinfel'd twists. First of all, we require that the Drinfel'd twists are constructed out of elements of the symmetry algebra of the original undeformed theory. Second, we realise these symmetry generators as \emph{active transformations} of the fields. By definition, as opposed to passive ones, active transformations leave the spacetime dependence invariant and only change the fields. This observation therefore automatically implies that, in the presence of a star product built out of active transformations, the standard Leibniz rule still holds for partial derivatives. In our setup, we can therefore construct covariant derivatives in the natural way and promote the gauge theories to the non-commutative setup. Importantly, we show that all non-commutative deformations of gauge theories constructed so far in the literature can be recast in our language, and we considerably extend the family of twist-deformed gauge theories.

This paper is, in fact, motivated by a long-standing issue in the construction of integrable deformations in the context of the AdS/CFT correspondence. The starting point is the fact that the canonical dual pair of $AdS_5/CFT_4$ appears to be integrable in the planar limit~\cite{Beisert:2010jr}. On the one side, the superstring on $AdS_5\times S^5$ admits a Lax connection~\cite{Bena:2003wd}, while on the other side the calculation of anomalous dimensions in $\mathcal N=4$ super Yang-Mills is encoded in an integrable quantum spin-chain~\cite{Minahan:2002ve}. At finite values of the 't Hooft coupling, the two sides of $AdS_5/CFT_4$ admit a description in terms of an integrable model, with the existence of an all-loop S-matrix~\cite{Beisert:2005tm}, and the exact description of the spectrum in terms of a Thermodynamic Bethe Ansatz~\cite{Bombardelli:2009ns,Arutyunov:2009ur,Gromov:2009tv} or a Quantum Spectral Curve~\cite{Gromov:2013pga}.

\begin{figure}
    \centering
    \begin{tikzpicture}[
        every node/.style={
            draw,
            rectangle,
            thick,
            align=center,
            inner sep=4pt
        }
    ]

    \node (sym) at (0,0) {Non-commutative twist-deformed \\ $\mathcal N=4$ super Yang-Mills};

    \node (hyb) at (4,-3)
        {Homogeneous Yang-Baxter \\  deformations of $AdS_5\times S^5$};

    \node (twist) at (-4,-3)
        {Drinfel'd twists of \\  the integrable model};

\draw[{Stealth}-{Stealth}] (sym) -- (hyb);
\draw[{Stealth}-{Stealth}] (sym) -- (twist);
\draw[{Stealth}-{Stealth}] (twist) -- (hyb);

    \end{tikzpicture}
    \caption{Different realisations of twist deformations in $AdS_5/CFT_4$.}
    \label{fig:def-adscft}
\end{figure}
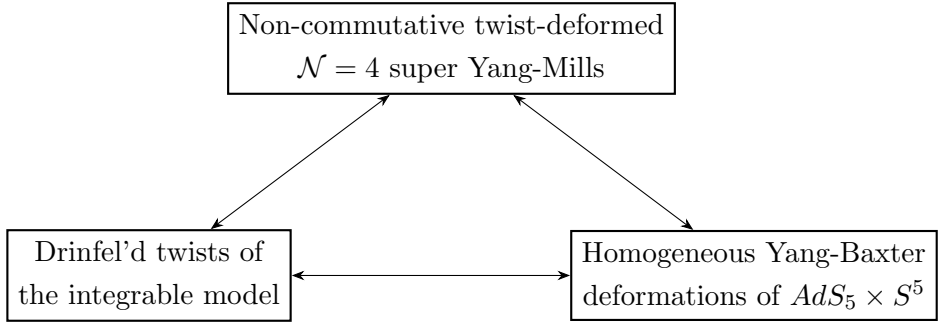

One well-defined way to deform the superstring on $AdS_5\times S^5$ while retaining classical integrability is the known method of homogeneous Yang-Baxter deformations~\cite{Klimcik:2002zj,Klimcik:2008eq,Delduc:2013qra,Kawaguchi:2014qwa,Matsumoto:2014gwa,Matsumoto:2015jja,vanTongeren:2015soa}, characterised by classical $r$-matrices satisfying the classical Yang-Baxter equation in the $\mathfrak{psu}(2,2|4)$ superalgebra. We recall that such classical $r$-matrices identify equivalence classes of Drinfel'd twists;  in turn, it is well known that Drinfel'd twists can be used to deform integrable models, so that it is natural to construct the corresponding twist deformations of the spin-chain that appears in the AdS/CFT correspondence~\cite{Beisert:2005if,vanTongeren:2013gva,Guica:2017mtd,Borsato:2025smn,Driezen:2025dww,Driezen:2025izd,Borsato:2026ypo} or even of the dual sigma model~\cite{vanTongeren:2018vpb,Borsato:2021fuy}.
At the same time, certain deformations that were studied to construct holographic dualities for non-commutative deformations of gauge theories turn out to fall into the class of homogeneous Yang-Baxter deformations. Examples are the Lunin-Maldacena background~\cite{Lunin:2005jy} dual to the Leigh-Strassler deformation\footnote{In this case the twist that gives rise to the star-product is constructed out of internal symmetry ($R$-symmetry) generators, and as such it only introduces some phases. While it is still useful to interpret the deformation in terms of a non-commutative star-product in field space, it is worth remarking that the resulting theory is an ordinary gauge theory with no spacetime non-commutativity.}~\cite{Leigh:1995ep} or, as shown in~\cite{Matsumoto:2014gwa}, the Hashimoto-Itzhaki-Maldacena-Russo background~\cite{Maldacena:1999mh,Hashimoto:1999ut} dual to the Groenewold-Moyal deformation. All these observations led to the conjecture that there should be realisations of the same kind of Drinfel'd twist deformations in the three different corners of Figure~\ref{fig:def-adscft}, and that via a generalisation of the usual holographic duality they should all give rise, for example, to the same spectrum~\cite{vanTongeren:2015uha,vanTongeren:2016eeb}, see also~\cite{Araujo:2017jkb,Araujo:2017jap}.

In one corner of the figure, one expects homogeneous Yang-Baxter deformations of $AdS_5\times S^5$, in another corner Drinfel'd twist deformations of the underlying integrable model, and finally  non-commutative twist-deformations of $\mathcal N=4$ super Yang-Mills.  For a long time, in this last corner several twists could not be considered because of the lack of a consistent construction. In this paper, we go beyond the class of Poincar\'e twists for which a construction was given in~\cite{Meier:2023lku}, and we finally provide the method to construct all the non-commutative deformations of $\mathcal N=4$ super Yang-Mills with twists that are built out of linearly-realised symmetries. 

At the same time, our method is more general and it can be applied to other instances of AdS/CFT where integrability is present, like in $AdS_4/CFT_3$, or to field and gauge theories that are not related to integrability nor holography at all. 

An important result of this paper is the extension of the planar equivalence theorem of~\cite{FILK199653,Meier:2023lku}, which allows us to express, for all the twist-deformations of gauge theories constructed here, planar Feynman diagrams of the deformed theory in terms of their undeformed counterparts.

This paper is organised as follows. In section~\ref{sec:symm} we review some aspects of symmetry transformations in field theories that are crucial for our construction. In section~\ref{sec:starProducts},  we construct the star-product in the picture of active transformations and compare it with the usual construction. In section~\ref{sec:star-theories} we explain the resulting properties of twist-deformed field and gauge theories, and we relate the property of cyclicity under integration to the $\mathcal R$-unimodularity condition. In section~\ref{sec:quant-plan} we move to the quantum setup and prove a planar equivalence theorem for the Feynman diagrams of the deformed gauge theories. In section~\ref{sec:class} we provide a rough classification of Drinfel'd twists that may be applied to gauge theories, having in mind in particular the symmetries of $\mathcal N=4$ super Yang-Mills. Finally, in section~\ref{sec:concl} we collect our conclusions and outlook. We also have several appendices providing extra details to the calculations of the main text.

\section{On symmetry transformations of fields}\label{sec:symm}
As explained in the introduction, the philosophy behind the construction is to use \emph{symmetries} of a seed field theory to construct its twist-deformations. To set the notation and to make some remarks that will be crucial in the following construction, we will therefore start by recapping some facts about the symmetry transformations that we will use.

Importantly, the symmetries that we will exploit to construct the twists are only the \emph{global} ones. In principle, any set of global symmetries may be used to generate twist deformations of the field theory, as long as the corresponding Lie algebra admits solutions to the classical Yang-Baxter equation, see appendix~\ref{app:Drinf}. The spectrum of possibilities, then,  is richer when the Lie algebra of symmetries is large, and to have a discussion that is as general as possible within reasonable limits, we will allow the seed gauge theory to be invariant under a superconformal symmetry. This means that, for example, when discussing spacetime symmetry transformations we do not need to limit ourselves to invariance under Poincar\'e only, and to construct twists we may include also the rest of the conformal symmetry generators. Similarly, we may allow the seed field theory to be supersymmetric, therefore allowing access to an even richer set of twists. This last option is in fact important for the correct implementation of certain twists of \emph{gauge} theories, see section~\ref{sec:class}. In general, we may have also internal symmetries that contribute to enlarging the family of twists that we can consider. Obviously, although our presentation is made by having in mind a rich setting with an extended (super)algebra of symmetries,  our discussion will of course be valid also when restricting the scope and considering, for example, theories that are  invariant under Poincar\'e but are neither conformal nor supersymmetric.

In general, we will denote by $\mathfrak g$ the Lie (super)algebra of symmetries of the undeformed seed field theory. The generators of $\mathfrak g$ will be denoted by $\X_A$, so that $[\X_A,\X_B]=f_{AB}{}^C\, \X_C$ with $f_{AB}{}^C$ the structure constants. To have a lighter notation, in the following we will often write just $\X$ with no index to denote a generic element of $\mathfrak g$.

\subsection{Spacetime symmetry transformations}
To each element $\X$ in the conformal algebra, one may associate a spacetime vector field $X^\mu(x)$, so that the infinitesimal transformation of spacetime points is
\begin{equation}
    x'{}^\mu\approx x^\mu+X^\mu(x).
\end{equation}
In particular, one has
\begin{equation}
    \begin{aligned}
    &\text{Translations:} \ && X^\mu(x)=a^\mu,\\
    &\text{Lorentz:} \ && X^\mu(x)=\omega^{\mu\nu}x_\nu,\\
    &\text{Dilatation:} \ && X^\mu(x)=\lambda x^\mu,\\
    &\text{Special conformal:} \ && X^\mu(x)=x^2\lambda^\mu-2x^\mu x^\nu\lambda_\nu,\\
    \end{aligned}
\end{equation}
where $a^\mu, \omega^{\mu\nu},\lambda,\lambda^\mu$ are the constant parameters of the transformations.
To simplify the notation, in the following we will often omit the $x$-dependence of the vector field and we will  simply write $X^\mu$.

In general, in field theory there are different notions of symmetry transformations because the fields themselves, and not just the spacetime points, may transform. Given a generic field\footnote{We will always use $\Phi$ to denote a generic field, so that we include scalar, vector, tensor and spinor fields.} $\Phi(x)$, the infinitesimal \emph{total transformation} of the field is
\begin{equation}
    \text{Total transformation:}\quad \X_{tot}(\Phi(x))\equiv \Phi'(x')-\Phi(x),
\end{equation}
where in the first term we put a prime both on $\Phi$ and on $x$ because we are considering the \emph{transformed field} evaluated at the \emph{transformed spacetime point}. In this paper, even though we do not use the symbol $\approx$, we always assume that the symmetry transformations are considered just at the infinitesimal level.

In field theory, one has also the concepts of \emph{passive and active transformations}
\begin{equation}
    \begin{aligned}
        &\text{Passive transformation:}\quad &&\check\X(\Phi(x))\equiv \Phi(x')-\Phi(x),\\
        &\text{Active transformation:}\quad &&\hat\X(\Phi(x))\equiv \Phi'(x)-\Phi(x).
    \end{aligned}
\end{equation}
In the first case only the spacetime point is transformed, while in the second case only the field. 
The  three concepts above are related by noticing that the total transformation is the sum of the passive and active ones
\begin{equation}\label{eq:tot-pas-act}
    \begin{aligned}
        \X_{tot}(\Phi(x))&=\Phi'(x')-\Phi'(x)+\Phi'(x)-\Phi(x)\\
        &=\check\X(\Phi'(x))+\hat\X(\Phi(x))\\
        &\approx\check\X(\Phi(x))+\hat\X(\Phi(x)),
    \end{aligned}
\end{equation}
because  the difference between $\check\X(\Phi'(x))$ and $\check\X(\Phi(x))$ is subleading when considering the leading order of the infinitesimal transformations.

From the Taylor expansion, the passive transformation of a generic field is always
\begin{equation}
    \check\X(\Phi(x))=X^\mu\partial_\mu\Phi(x).
\end{equation}
The expressions for total and active transformations of fields, instead, are not universal, and take different forms depending on whether the field is a scalar, vector, tensor or spinor. See appendix~\ref{app:symm} for our conventions.

An important property that we will use later---and that is universally true for a generic field---is that active transformations commute with partial derivatives
\begin{equation}
    [\partial_\mu,\hat\X](\Phi(x))=0.
\end{equation}
This follows directly from the definition of active transformations, since the spacetime coordinates are not changed
\begin{equation}
    \hat\X(\partial_\mu\Phi(x))=\partial_\mu\Phi'(x)-\partial_\mu\Phi(x)=\partial_\mu(\hat\X(\Phi(x))).
\end{equation}
We refer to appendix~\ref{sec:descendantFields} for an alternative proof of this property. In fact, this property is what justifies the choice of active symmetry transformations (rather than the passive ones) when proving Noether's theorem. 

The above observation should be compared to what happens in the case of passive transformations, that instead do not commute with partial derivatives. In fact,

\begin{equation}
    \begin{aligned}
        \check\X(\partial_\mu\Phi(x))&=\partial'_\mu\Phi(x')-\partial_\mu\Phi(x)\\&=\partial_\mu\Phi(x')-\partial_\mu\Phi(x)-\partial_\mu X^\nu\partial_\nu\Phi(x)\\
        &=\partial_\mu(\Phi(x')-\Phi(x))-\partial_\mu X^\nu\partial_\nu\Phi(x)\\
        &=\partial_\mu(\check\X(\Phi(x)))-\partial_\mu X^\nu\partial_\nu\Phi(x)
    \end{aligned}
\end{equation}
In general, the last term is not zero, but it vanishes for the case of translations. 

In the rest of the paper, we will implement the active symmetry transformations  via an object that we call  the ``Weyl-Lie derivative'' $\mathcal L^W$.
This can be understood as a modification of the standard Lie derivative $\mathcal L$ that takes into account also a weight $w$ of the fields
\begin{equation}
    \mathcal L^W_\X\equiv \mathcal L_\X+\frac{w}{d}\partial_\mu X^\mu,
\end{equation}
where $d$ is the number of spacetime dimensions.
The weight $w$ is in general related to the scaling dimension $\Delta$ of the field on which $\mathcal L^W_\X$ is acting.
We refer to appendix~\ref{app:symm} for the explicit formulas in the cases of scalar, vector, tensor and spinor fields.

Apart from the details of the explicit expressions that differ for each type of field, here we want to remark that the  active transformation $\hat\X$ of a generic field $\Phi$ will be implemented by the Weyl-Lie derivative as
\begin{equation}\label{eq:hatX-LW}
    \hat\X(\Phi(x))=\mathcal L^W_{-\X}(\Phi(x)).
\end{equation}
Importantly,  the active transformation is implemented as the Weyl-Lie derivative of \emph{minus} $\X$. In section~\ref{sec:comp} we will come back to this point.

\subsection{Internal symmetries}
In general, we may also have internal symmetries that only act on the space of fields while leaving the spacetime coordinates invariant. It is useful to consider also this kind of symmetry transformations because they can mix with spacetime symmetries to give rise to interesting twists. 

Given a collection of fields $\Phi^i(x)$ with $i=1,\ldots,N$, we will consider linearly-realised symmetries acting as $\Phi'{}^i(x)=U^i{}_j\Phi^j(x)$ for a constant matrix $U^i{}_j$. For symmetry transformations continuously connected to the identity $U^i{}_j=\delta^i{}_j+u^i{}_j+\ldots$,  we may write that the symmetry generator $\X$ of an internal symmetry acts infinitesimally as
\begin{equation}
    \X\Phi^i(x)=u^i{}_j\Phi^j(x).
\end{equation}
Given that internal symmetries do not change spacetime coordinates, it is natural to consider them together with \emph{active} transformations of spacetime symmetries, rather than their total or passive versions.

To have a unifying notation, one may even extend the notion of $\mathcal L^W$ and define it also for the case of internal symmetries to be
\begin{equation}
    \mathcal L^W_\X\Phi^i(x)=-u^i{}_j\Phi^j(x),
\end{equation}
where the minus sign is added by hand so that, like in the case of active transformations of spacetime symmetries, we may write 
\begin{equation}
    \hat\X\Phi^i(x)=\mathcal L^W_{-\X}\Phi^i(x).
\end{equation}

\subsection{Supersymmetry transformations}\label{sec:susy-symm}
The goal of this subsection is to discuss to which extent the action of supersymmetry transformations shares the same qualitative features as that of the spacetime and internal symmetries previously discussed, in order to have a unifying language that allows us to consider Drinfel'd twists that include all these symmetry transformations at once. Some of the comments that we will make here anticipate properties that we need for the following construction.

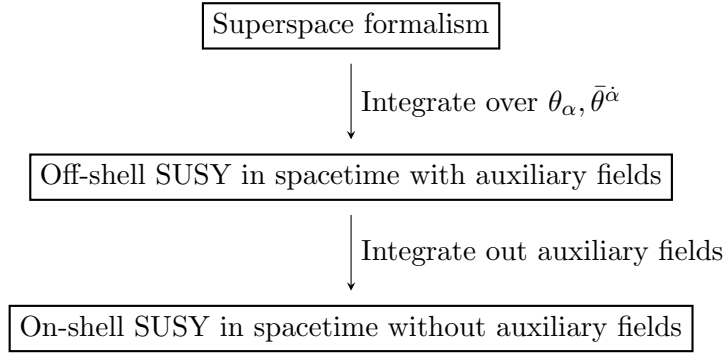
\begin{figure}
    \centering
  \begin{tikzpicture}
   \draw (0,0) node[shape=rectangle,draw,thick] {Superspace formalism};
   \draw [-stealth](0,-0.5) -- (0,-1.5);
   \draw (0,-1) node[right]{Integrate over $\theta_{\alpha},\bar\theta^{\dot\alpha}$};
    \draw (0,-2) node[shape=rectangle,draw,thick] {Off-shell SUSY in spacetime with auxiliary  fields};
   \draw [-stealth](0,-2.5) -- (0,-3.5);
   \draw (0,-3) node[right]{Integrate out auxiliary fields};
    \draw (0,-4) node[shape=rectangle,draw,thick] {On-shell SUSY in spacetime without auxiliary fields};
\end{tikzpicture}
    \caption{Possible realisations of $\mathcal N=1$ supersymmetric field theories.}
    \label{fig:susy}
\end{figure}

A possible path to construct $\mathcal N=1$ supersymmetric field theories may be summarised schematically as in Figure~\ref{fig:susy}.
The most elegant construction of field theories invariant under super-Poincar\'e is via the superspace and superfield formalism. In $d=4$, to realise $\mathcal N=1$ supersymmetry with supercharges $\Q_\alpha,\bar\Q^{\dot\alpha}$, one introduces Grassmann coordinates $\theta_{\alpha},\bar\theta^{\dot\alpha}$, and constructs superfields whose integration in superspace  provides supersymmetric action terms. The explicit integration over the Grassmann coordinates yields standard actions of the form $S=\int d^4x \mathcal L$, where $\mathcal L$ is a Lagrangian density that is invariant under off-shell supersymmetry, i.e.~without the need to imposing the equations of motion. 

This off-shell invariance  is possible thanks to the presence of auxiliary fields. 
For example, in the case of the $\mathcal N=1$ chiral multiplet of super Poincar\'e  (with complex scalar $\phi$ and Weyl spinor $\psi_\alpha$ complemented by the auxiliary field $F$) the off-shell supersymmetry transformations are\footnote{We refer to appendix~\ref{app:symm} for our conventions on the matrices $\sigma^\mu,\bar\sigma^\mu$.}
\begin{equation}
\begin{aligned}
        &\Q_\alpha\phi=-i\sqrt{2}\psi_\alpha,\quad
    &&\Q_\alpha\psi_\beta=i\sqrt{2}\epsilon_{\alpha\beta}F,\\
    &\bar\Q^{\dot\alpha}\psi_\alpha=\sqrt{2}(\sigma^\mu)_\alpha{}^{\dot\alpha}\partial_\mu\phi,\quad
    &&\bar\Q^{\dot\alpha}F=-\sqrt{2}\bar\sigma^{\mu\dot\alpha\alpha}\partial_\mu\psi_\alpha.
\end{aligned}
\end{equation}
In general, we may repackage all physical and auxiliary fields into just one set $\Phi^A$, where $A$ is the index keeping track of each of them. We  have then that the supersymmetry transformations are of the form 
\begin{equation}\label{eq:schematic-action-susy}
    \Q \Phi^A=Q^A{}_B\Phi^B,
\end{equation}
where  $Q^A{}_B$ is at the same time a matrix rotating different fields into each other (as in the case of internal symmetries) and an operator that may  introduce also derivatives acting on the fields (as in the case of spacetime symmetries).  

The auxiliary fields are not part of the supersymmetry representations, and integrating them out one obtains an action only for the dynamical fields. The cost to pay for this, however, is that now the action is invariant under supersymmetry only on-shell, i.e.~when imposing the equations of motion. In fact, when integrating out auxiliary fields, in general the supersymmetry transformations fail to be of the form~\eqref{eq:schematic-action-susy} because extra non-linearities may be introduced by replacing the expressions that are given by the equations of motion.
For our purposes, it would be difficult to construct the twisted gauge theories by dealing with an on-shell formulation of supersymmetry. In fact, several properties of the twisted gauge theories, notably their gauge invariance, will rely on the fact that the undeformed seed theory is symmetric. We expect that constructing twists involving supersymmetric generators in an on-shell formulation of supersymmetry may be possible but cumbersome, because of the related discussion of the gauge invariance of the twisted gauge theories. It is for this reason that our discussion assumes off-shell supersymmetry.

There is in fact another subtlety to which one should pay attention.\footnote{We thank Stijn van Tongeren and Julio Cabello Gil for important discussions on this point.} In the vector superfield that is used to construct supersymmetric \emph{gauge} theories, there are gauge degrees of freedom that in general one may want to gauge fix by using suitable gauge choices like the Wess-Zumino gauge, for example. In general, however, this gauge fixing  may imply that certain supersymmetry transformations are not manifest, since the supersymmetry invariance of the action may be true only up to gauge transformations that move us away from the gauge choice. Notice that also when fixing a gauge, in general, the form of the action of the supercharges on the fields may not be as in~\eqref{eq:schematic-action-susy}, because non-linear terms may appear on the right-hand-side and because extra compensating gauge transformations may be needed.
We refer to~\cite{StijnJulio} for more detailed discussions on the role of supersymmetry transformations in the context of Drinfel'd twisted gauge theories.
Overall, we will assume that the extra gauge degrees of freedom have not been gauge fixed, so that one can still work with~\eqref{eq:schematic-action-susy}.

When moving to $\mathcal N=2$ supersymmetric gauge theories, in the case of $\mathcal N=2$ vector multiplets the generalisation from $\mathcal N=1$ does not have significantly different qualitative features; while having in mind the previous remarks, one can construct an $\mathcal N=2$ superspace with $\theta^I_{\alpha},\bar\theta^{I\dot\alpha}, I=1,2$ Grassmann coordinates, and an action principle from a holomorphic prepotential. The case of the $\mathcal N=2$ hypermultiplets is different because an off-shell formulation of supersymmetry requires an infinite number of auxiliary fields, or equivalently the employment of harmonic superspace~\cite{Galperin:2001seg}. In either case, we can still argue that, if (a possibly infinite number of) auxiliary fields are not integrated out and the gauge is not fixed, all supersymmetry transformations are still realised linearly and off-shell as in~\eqref{eq:schematic-action-susy}.

The case of $\mathcal N=4$ supersymmetry is special, as there is no off-shell $\mathcal N=4$ superspace formulation. Although $\mathcal N=4$ super Yang-Mills may be formulated as an $\mathcal N=1$ or an $\mathcal N=2$ theory, the supersymmetry transformations that escape the relevant $\mathcal N=1$ or $\mathcal N=2$ superspace description will be realised only on-shell. For them, therefore, we cannot assume the form~\eqref{eq:schematic-action-susy}.

Theories that are invariant under conformal symmetry and are also supersymmetric enjoy a $\mathfrak{su}(2,2|\mathcal N)$ superconformal symmetry.\footnote{In the case of $\mathcal N=4$ there is a central bosonic generator that can be modded out so that in that case the superconformal algebra is just $\mathfrak{psu}(2,2|4)$.} In the presence of superconformal symmetry, we have additional supercharges normally denoted by $\mathbb S^{I\alpha},\bar{\mathbb S}^I_{\dot\alpha}$. These may be obtained by the sequence of transformations inversion-supercharge-inversion, where ``inversion'' acts as $x^\mu\to x^\mu/|x|^2$ and ``supercharge'' is one of the Poincar\'e supercharges. From this point of view, the discussion for superconformal charges $\mathbb S^{I\alpha},\bar{\mathbb S}^I_{\dot\alpha}$ follows that of Poincar\'e supercharges.

To conclude, through out this paper we will assume that the supersymmetry and superconformal transformations that are used to construct the twists are realised linearly and off-shell, schematically as in~\eqref{eq:schematic-action-susy}. In general, this  means that auxiliary fields should not be integrated out and that gauge degrees of freedom should be retained and  not gauge fixed. In this setup, supersymmetry invariance of the action is ensured without the  requirement of putting the theory on-shell or of implementing compensating  gauge transformations. Validity off-shell and linearity are in fact accompanied also by the property that the supersymmetry transformations act only on the fields while leaving spacetime points invariant;  they therefore match the key properties  of active transformations of spacetime symmetries that we will need to implement the Drinfel'd twists. To use the same notation also in this case, we may define
\begin{equation}
    \mathcal L^W_{\Q}\Phi^A(x)=-Q^A{}_B\Phi^B(x),
\end{equation}
so that 
\begin{equation}
    \hat\Q\Phi^A=\mathcal L^W_{-\Q}\Phi^A(x),
\end{equation}
 and similarly for $\mathbb S$.

\subsection{Universal enveloping algebra and Hopf algebra}\label{sec:univ-comp-Hopf}
In order to construct the twists that deform the field theories, it is not enough to work with $\mathfrak g$, we  actually need its universal enveloping algebra $U(\mathfrak g)$. That means that we introduce the identity element $1$, and we  allow ourselves to consider arbitrary multiplications (i.e.~compositions) 
\begin{equation}
    \Y\, \X\equiv\Y\circ\X
\end{equation}
of symmetry transformations $\X,\Y$, obviously identifying certain combinations by means of the graded Lie bracket $[\cdot,\cdot]$ of $\mathfrak g$, i.e.
\begin{equation}
    \X\Y-(-1)^{F(\X)F(\Y)}\Y\X\equiv[\X,\Y].
\end{equation}
Here and in the rest of the paper we implement the $\mathbb Z_2$ grading of Lie superalgebras with
\begin{equation}
    F(\X)=\left\{\begin{array}{ll}
        0, & \X \text{ is even/bosonic} \\
         1, & \X \text{ is odd/fermionic}
    \end{array}\right.
\end{equation}
The notation we are using allows us to  straightforwardly translate formulas of the abstract universal enveloping algebra to the representations as active or passive transformations. For example, given the abstract element of $U(\mathfrak g)$
\begin{equation}
    \exp(\X)=1+ \X+\frac12 \X^2+\ldots,
\end{equation}
for active and passive transformations we may simply write respectively
\begin{equation}
    \exp(\hat\X)=1+ \hat\X+\frac12 \hat\X^2+\ldots,\qquad
    \exp(\check\X)=1+ \check\X+\frac12 \check\X^2+\ldots.
\end{equation}
In the rest of the paper, we will continue to use $\X,\Y,\ldots$ to denote elements of $\mathfrak g$, while we will tend to use $\U,\V,\ldots$ for elements of $U(\mathfrak g)$.

It is well known that the universal enveloping algebra admits an interpretation as a Hopf algebra.\footnote{We refer to appendix~\ref{app:Hopf} for a short recap on Hopf algebras and for our notational conventions.} In particular, it admits a coproduct $\Delta:U(\mathfrak g)\to U(\mathfrak g)\otimes U(\mathfrak g)$ and an antipode $S:U(\mathfrak g)\to U(\mathfrak g)$ that on $\X\in\mathfrak g$ are
\begin{equation}
    \Delta(\X)=\X\otimes 1+1\otimes \X,\qquad\qquad
    S(\X)=-\X.
\end{equation}
Formulas for generic elements of $U(\mathfrak g)$ may be obtained by knowing $\Delta(1)=1\otimes 1,\ S(1)=1$ and that $\Delta$ is an algebra homomorphism, while $S$ is an algebra antihomomorphism (i.e. it swaps the order)
\begin{equation}
    \Delta(\U\V)=\Delta(\U)\Delta(\V),\qquad\qquad
    S(\U\V)=S(\V)S(\U).
\end{equation}
The language of Hopf algebra turns out to be very useful for the construction, and an example of this is the interpretation of the composition of symmetry transformations, as we are about to see.

\subsubsection{Compositions of symmetry transformations}\label{sec:comp}
In the case of active symmetry transformations, their composition is implemented on fields by swapping their order, i.e.\footnote{Here we are adopting the usual convention used in physics that the rightmost generator is the one acting first on the fields.}
\begin{equation}
    \hat \Y\, \hat\X (\Phi(x))=\mathcal L^W_{-\X}\mathcal L^W_{-\Y}\Phi(x).
\end{equation}
The above property is valid for active transformations of spacetime symmetries, supersymmetries and internal symmetries. It is in fact just a consequence of the fact that these transformations act on the fields only, so that after implementing the transformation $\hat\X$, the other transformation $\hat\Y$ passes through to act just on the field. To see this explicitly, let us first consider the composition of spacetime symmetry transformations. We will consider the sequence of transformations given by
\begin{equation}
    x^\mu\quad\to\quad x'{}^\mu=x^\mu+X^\mu(x),
    \quad\to\quad
    x''{}^\mu=x'^\mu+Y^\mu(x').
\end{equation}
Notice that the second vector field $Y^\mu(x')$ naturally depends on the coordinate system $x'$ obtained after the first symmetry transformation. 
We have
\begin{equation}
   \hat \Y\, \hat\X (\Phi(x))=\hat \Y(\mathcal L^W_{-\X}\Phi(x))=\mathcal L^W_{-\X}(\Phi''(x)-\Phi'(x))=\mathcal L^W_{-\X}\mathcal L^W_{-\Y}\Phi(x),
\end{equation}
and we refer to appendix~\ref{app:CompTF} for an alternative proof of the above property. Similarly, for the case of internal symmetries we have
\begin{equation}
   \hat \Y\, \hat\X (\Phi^i(x))=\hat \Y((u_{\X})^i{}_j\Phi^j(x))=(u_{\X})^i{}_j(\hat \Y\Phi^j(x))=(u_{\X})^i{}_j(u_{\Y})^j{}_k\Phi^k(x))=\mathcal L^W_{-\X}\mathcal L^W_{-\Y}\Phi^i(x).
\end{equation}
We can interpret the above observation by means of the language of Hopf algebras. In fact, if we declare the Weyl-Lie derivative to follow the obvious composition rule
\begin{equation}
    \mathcal L^W_{\Y\X}\equiv \mathcal L^W_\Y\circ \mathcal L^W_\X,
\end{equation}
then it is natural to identify the active transformation  of a generic element $\U$ of the universal enveloping algebra as
\begin{equation}\label{eq:hatU-LWSU}
    \hat \U=\mathcal L^W_{S(\U)},
\end{equation}
where $S$ is the antipode introduced above. In fact, the presence of the antipode explains both the minus sign when looking at the active transformation of $\X\in\mathfrak g$
\begin{equation}
    \hat \X\Phi(x)=\mathcal L^W_{S(\X)}\Phi(x)=\mathcal L^W_{-\X}\Phi(x),
\end{equation}
as well as the composition rule that swaps the order of the symmetry transformations
\begin{equation}
    \hat \Y\, \hat\X (\Phi(x))=\mathcal L^W_{S(\Y\X)}\Phi(x)=\mathcal L^W_{S(\X)S(\Y)}\Phi(x)=\mathcal L^W_{S(\X)}\mathcal L^W_{S(\Y)}\Phi(x)=\mathcal L^W_{-\X}\mathcal L^W_{-\Y}\Phi(x).
\end{equation}
With equation~\eqref{eq:hatU-LWSU} we are then extending the notion of the action of $\hat\U$ to the whole universal enveloping algebra.

Because of the above observations, the map relating an abstract element $\U$ of the universal enveloping algebra to its version $\hat\U$ as an active symmetry transformation is a Hopf algebra homomorphism. For example,  if $\U=\X\Y$ then $\hat\U=\hat\X\hat\Y$, and then going from the abstract to the active picture really just amounts to put hats everywhere, as claimed in the previous section.

\section{Star products}\label{sec:starProducts}
In this section we will implement the construction of non-commutative star products via Drinfel'd twists. While the construction has been known for a long time, the new perspective will be to use the active transformations of symmetries to construct the twists. As we will see in section~\ref{sec:star-def-symm}, this will be crucial when constructing twist deformations of \emph{gauge} theories. For the reader's convenience, in this section we recap some important facts about Drinfel'd twists, and we collect some observations on various notions of ``unimodularity conditions'', that  here are discussed simply from the point of view of Hopf algebras, but later will play an important role in the construction of the twisted gauge theories.

\subsection{Drinfel'd twists}
We refer to appendix~\ref{app:Drinf} for a short recap on Drinfel'd twists, following~\cite{drinfeld1983constant,Giaquinto:1994jx,Kulish2009}. A Drinfel'd twist $\mathcal F\in U(\mathfrak g)\otimes U(\mathfrak g)$ is an object that, in our case, is constructed out of elements of the universal enveloping algebra, and that we will also write as $\mathcal F=f^\alpha\otimes f_\alpha$, where summation over the index $\alpha$ is assumed. It admits an inverse that we will often denote just with a bar $\mathcal F^{-1}=\bar{\mathcal{F}}=\bar f^\alpha\otimes\bar f_\alpha$.
Thanks to the cocycle condition
\begin{equation}\label{eq:cocycle}
    (\mathcal F\otimes 1)(\Delta\otimes id)\mathcal F=(1\otimes \mathcal F)(id\otimes \Delta)\mathcal F,
\end{equation}
Drinfel'd twists may be used to twist a Hopf algebra $\mathcal H$. On the one hand, in fact, one has a twisted coproduct
\begin{equation}\label{eq:twisted-coproduct}
    \Delta_{\mathcal F}(\U)=\mathcal F\Delta(\U)\mathcal F^{-1},
\end{equation}
and on the other hand a twisted antipode
\begin{equation}\label{eq:SF}
    S_\mathcal{F}(\U)=V^{-1}S(\U)V,\qquad 
    V^{-1}=f^\alpha S(f_\alpha),\qquad
    V=S(\bar f^\alpha)\bar f_\alpha.
\end{equation}
Properties of Drinfel'd twists ensure that the universal enveloping algebra equipped with  $\Delta_{\mathcal F}$ and $S_\mathcal{F}$ is a twisted Hopf algebra $\mathcal H_{\mathcal F}$, meaning that the axioms of Hopf algebras are satisfied even in the presence of the twist.

We will be interested in Drinfel'd twists that are deformations of the identity, so that there is a deformation parameter $\xi$ and in the undeformed limit  $\lim_{\xi\to 0}\mathcal F=1\otimes 1$, so that the twisted Hopf algebra (together with its coproduct and antipode) reduces to the undeformed one. 
When the $\mathcal O(\xi)$ in the expansion of $\mathcal F$ is antisymmetric under the (graded) permutation $\tau$ of the two spaces, one says that the twist is ``$r$-symmetric'', and we may write
\begin{equation}
    \mathcal F=1\otimes1+\xi\ r+\mathcal O(\xi^2),\qquad r^{op}=\tau(r)=-r.
\end{equation}
The cocycle condition~\eqref{eq:cocycle} implies that $r$ is a solution of the classical Yang-Baxter equation (CYBE) written in~\eqref{eq:CYBE}.
Not all twists are $r$-symmetric, but it is always possible to find an equivalent twist that is. Two twists $\mathcal F,\mathcal F'$ are equivalent if they are related by 
\begin{equation}\label{eq:Fp-equiv}
    \mathcal F'=(w^{-1}\otimes w^{-1})\mathcal F\ \Delta(w),
\end{equation}
 where $w\in U(\mathfrak g)$ is  invertible.\footnote{It is easy to show that the two Hopf algebras obtained by $\mathcal{F}$ and $\mathcal{F}'$ are isomorphic to each other by the isomorphism $\phi:U(\mathfrak{g})\rightarrow U(\mathfrak{g}),~\phi(h)=w^{-1}hw$.} From compatibility with the cocycle and counital condition for both $\mathcal{F}$ and $\mathcal{F}'$ it follows that $w$ is a deformation of the identity as well
 \begin{equation}
     w=1+\mathcal O(\xi).
 \end{equation}
 To summarise, Drinfel'd twists continuously connected to the identity are organised in equivalent classes characterised by the corresponding classical $r$-matrix.

 Taking
 \begin{equation}\label{eq:RFF}
     \mathcal R=\mathcal F_{op}\mathcal F^{-1},
 \end{equation}
 one can in fact define a quantum $\mathcal R$-matrix that satisfies the quantum Yang-Baxter equation~\eqref{eq:YBE}.
 In fact, expanding in the deformation parameter one finds that the first non-trivial order is given by the classical $r$-matrix
 \begin{equation}
 \mathcal R=1-2\xi\ r+\mathcal O(\xi^2).    
 \end{equation}
 Notice that under an equivalence transformation as in~\eqref{eq:Fp-equiv}, the $\mathcal R$-matrix changes as
 \begin{equation}
     \mathcal R'=(w^{-1}\otimes w^{-1})\mathcal R(w\otimes w),
 \end{equation}
 because the undeformed coproduct satisfies $\Delta^{op}=\Delta$.
 Having a quantum $\mathcal R$-matrix, the twisted Hopf algebra is said to be quasi-triangular. It is in fact triangular because $\mathcal R$ satisfies the braiding unitarity condition $\mathcal R_{21}\mathcal R_{12}=1$. Obviously the triangularity condition remains in the undeformed limit, when the $\mathcal R$-matrix reduces to the identity.

\subsubsection{Unimodularity conditions}\label{sec:unim}
Given a quasi-triangular Hopf algebra with antipode $S$ and quantum $\mathcal R$, there is a so-called  Drinfel'd element
\begin{equation}
    u=m(S\otimes 1)\mathcal R_{op},
\end{equation}
that controls the square of the antipode
\begin{equation}
    S^2(h)=uhu^{-1}.
\end{equation}
One says that the quantum $\mathcal R$-matrix is unimodular if the Drinfel'd element is trivial
\begin{equation}
    u=1,
\end{equation}
which then also implies that the antipode squares to the identity $S^2=id$.

In our setup, we start in the undeformed case with a trivial $\mathcal R$-matrix $\mathcal R^{(0)}=1\otimes 1$ which is obviously unimodular. After the twist, the antipode is given by $S_{\mathcal F}$ defined in~\eqref{eq:SF}, and then the  unimodularity condition for the $\mathcal R$-matrix defined in~\eqref{eq:RFF} reads 
\begin{equation}
    u_\mathcal{F}=m(S_{\mathcal F}\otimes 1)\mathcal R_{op}=V^{-1}S(\bar f_\beta)S(f^\alpha)Vf_\alpha\bar f^{\beta}
    =V^{-1}S(\bar f_\beta)\bar f^{\beta}=V^{-1}S(V)\stackrel{!}{=}1,
    \end{equation}
    where we used an identity in~\eqref{eq:identities-V-gen}.
We see that this is only possible if we demand that $V=S(\bar f^\alpha)\bar f_\alpha$ defined in~\eqref{eq:SF} is invariant under the antipode
\begin{equation}
    S(V)=V.
\end{equation}
We will call this the ``$\mathcal R$-unimodularity condition''.  Obviously, this also implies that the twisted antipode squares to the identity
\begin{equation}
    S_{\mathcal F}^2(\U)=V^{-1}S(V)\ \U\ S(V^{-1})V=\U.
\end{equation}

At this point it is worth noticing that $V$ may be also written as
\begin{equation}
    V=m(S\otimes 1)\mathcal F^{-1}.
\end{equation}
Having a unimodularity condition for the $\mathcal R$-matrix, we may also introduce a ``$\mathcal F$-unimodularity condition'' for twists satisfying 
\begin{equation}
    V=m(S\otimes 1)\mathcal F^{-1}\stackrel{!}{=}1.
\end{equation}
Obviously, the $\mathcal F$-unimodularity condition $V=1$ is stronger than the $\mathcal R$-unimodularity condition $S(V)=V$.
Notice that the $\mathcal F$-unimodularity condition in fact implies that the twisted antipode coincides with the undeformed one
\begin{equation}
    S_{\mathcal F}=S,
\end{equation}
and then it obviously follows that the twist is unimodular also with respect to the twisted antipode $m(S_{\mathcal F}\otimes 1)\mathcal F^{-1}=1$. 

It is interesting to see how these unimodularity conditions change when considering an equivalent twist. 
When considering an equivalent twist $\mathcal F'$ as in~\eqref{eq:Fp-equiv}, we find
\begin{equation}\label{eq:mu1SFp}
        V'\equiv m(S\otimes 1)\mathcal F'^{-1}=S( w)Vw.
\end{equation}
See equation~\eqref{eq:mu1SFp-app} for the derivation.
Notice that if $V$ satisfies the $\mathcal R$-unimodularity condition $S(V)=V$, then $V'$ does as well, $S(V')=V'$. However, if $V$ satisfies the $\mathcal F$-unimodularity condition $V=1$, then in general $V'\neq 1$ does not, but it still satisfies the $\mathcal R$-unimodularity condition $S(V')=V'$. This is in fact consistent with the fact that a similar calculation to the previous one would give $u_{\mathcal F'}=V'{}^{-1}S(V')$.
We learn that the $\mathcal R$-unimodularity condition is preserved by the equivalence relation among Drinfel'd twists, while in general the $\mathcal F$-unimodularity condition is not.

Let us now see what happens in the case of an $r$-symmetric twist. One gets
\begin{equation}
    V=1-\xi \ r^{AB}\, (S(\X_A)\X_B-(-1)^{F(A)F(B)}S(\X_B)\X_A)+\mathcal O(\xi^2)=1+\xi\ r^{AB}\, f_{AB}{}^C\X_C+\mathcal O(\xi^2).
\end{equation}
Then, demanding the $\mathcal R$-unimodularity condition $S(V)=V$, at $\mathcal O(\xi)$ one gets the necessary condition
\begin{equation}\label{eq:r-unim}
    r^{AB}\, f_{AB}{}^C=0.
\end{equation}
We will call~\eqref{eq:r-unim} the $r$-unimodularity condition. Notice that demanding the $r$-unimodularity condition is enough to get the $\mathcal F$-unimodularity condition at least to the $\mathcal O(\xi)$ order.  Importantly, if a twist $\mathcal F$ is associated to a non-unimodular $r$-matrix, then no twist in its equivalence class will be $\mathcal R$-unimodular.

To make the above statements less abstract, we will now consider some examples.

\subsubsection{Examples: abelian twists}\label{sec:abel-ex}
Let us consider the simplest possible twists, the so-called abelian ones (see also section~\ref{sec:class}). To make the discussion as general as possible, we will take two commuting elements $a,b\in \mathfrak g$ so that $[a,b]=0$, and construct the Drinfel'd twist as
\begin{equation}
    \mathcal F=e^{\xi\, a\wedge b}.
\end{equation}
Physically important examples include the Groenewold-Moyal twist, where $a,b$ are translations $p_\mu$ of the Poincar\'e algebra.
Notice that this twist is $r$-symmetric, in fact $r=a\wedge b=-r^{op}$, and it trivially solves the CYBE. It is also $\mathcal F$-unimodular (and therefore $\mathcal R$- and $r$-unimodular) because, using the fact that all possible expressions commute with each other, one can check that $V=1$.

We may now construct a twist that is equivalent to the above one. We take
\begin{equation}
    w=e^{\xi\,  ab},
\end{equation}
so that 
\begin{equation}
\begin{aligned}
    \Delta(w)&= e^{\xi\, \Delta(a)\Delta(b)}=(e^{\xi\, ab}\otimes e^{\xi\, ab})e^{\xi\, a\otimes b}e^{\xi\, b\otimes a}.
\end{aligned}    
\end{equation}
Then, using~\eqref{eq:Fp-equiv}, the equivalent twist takes the form
\begin{equation}\label{eq:abel-equiv-tw}
    \mathcal F'=e^{2\xi\, a\otimes b}.
\end{equation}
Importantly, this twist is not $\mathcal F$-unimodular because
\begin{equation}
    V'=e^{2\xi\, ab}\neq 1.
\end{equation}
It is nevertheless $\mathcal R$-unimodular because $S(V')=V'$. Taking into account that $S(w)=w$, in fact we have $V'=S(w)Vw=w^2$.

\subsubsection{Examples: Jordanian twists}
Following~\cite{tolstoy2004chainsextendedjordaniantwists},  let us  consider the ``extended Jordanian superalgebra''\footnote{Compared to~\cite{tolstoy2004chainsextendedjordaniantwists}, we have set $t_\gamma=1/2$.}
\begin{equation}
    [h,e]=e,\qquad [h,f_\pm]=\frac12 f_\pm,\qquad \{f_+,f_-\}=e,
\end{equation}
where $h,e$ are even and $f_\pm$ are odd. It admits the classical Jordanian $r$-matrix 
\begin{equation}
    r^J=h\wedge e,
\end{equation}
that satisfies the CYBE. It does not satisfy the $r$-unimodularity condition, because  $m(S\otimes 1)r^J=-[h,e]=-e\neq0$. The corresponding twist will then also not be  $\mathcal F$-unimodular nor $\mathcal R$-unimodular. In fact, no twist in its equivalence class can be  $\mathcal F$-unimodular or $\mathcal R$-unimodular, otherwise we would get a contradiction with~\eqref{eq:mu1SFp}.

At the same time, this superalgebra admits also the extended Jordanian $r^{eJ}$-matrix
\begin{equation}
    r^{eJ}=h\wedge e-f_+\wedge f_-,
\end{equation}
that also satisfies the CYBE.
Notice that now this is unimodular because $m(S\otimes 1)r^{eJ}=-([h,e]-\{f_+,f_-\})=0$.
This means that if we construct an $r$-symmetric twist
\begin{equation}
    \bar{\mathcal F}^{eJ(1)}=1\otimes 1-\xi\ r^{eJ}+\mathcal O(\xi^2),
\end{equation}
this will be $\mathcal F$-unimodular, $m(S\otimes 1)\bar{\mathcal F}^{eJ(1)}=1$ at least to this order. We may now construct an equivalent twist $\mathcal F^{eJ(2)}=(\bar w\otimes \bar w)\mathcal F^{eJ(1)}\Delta( w)$ taking for example $w=1+\xi\ eh+\mathcal O(\xi^2)$. We obtain
\begin{equation}\label{eq:FeJ2}
    \bar{\mathcal F}^{eJ(2)}=1\otimes 1-\xi (2h\otimes e-f_+\wedge f_-)+\mathcal O(\xi^2),
\end{equation}
and
\begin{equation}
   V^{(2)} =m(S\otimes 1)\bar{\mathcal F}^{eJ(2)}=1-\xi(1-2h)e+\mathcal O(\xi^2),
\end{equation}
so that $\mathcal F^{eJ(2)}$ fails to be $\mathcal F$-unimodular, but we do satisfy the $\mathcal R$-unimodularity condition because $V^{(2)}=S(V^{(2)})$.
This is consistent with~\eqref{eq:mu1SFp} because one can check the relation  $V^{(2)}=S(w)V^{(1)}w$.

\subsection{Star products from active transformations}
Following~\cite{Drinfeld:1985rx}, one may use Drinfel'd twists to construct a non-commutative but associative ``star product'' of functions, that may be applied to the fields. To start, let us denote by $\mu$ the standard operation of multiplying  fields, which takes two of them and yields their product as output
\begin{equation}
    \mu(\Phi_1(x)\otimes\Phi_2(x))=\Phi_1(x)\Phi_2(x).
\end{equation}
It is important not to confuse this product with the one on the universal enveloping algebra that we denoted by $m$. In fact, $\mu$ is commutative $\Phi_1(x)\Phi_2(x)=\Phi_2(x)\Phi_1(x)$ while $m$ is not (in general, $\U\V\neq \V\U$).

We may now introduce a non-commutative product $\mu_{\hat{\mathcal F}}$, that we will simply denote with a star, by considering the composition of $\mu$ with the inverse twist. In this paper, we do this by taking the twist in the picture of active transformations, so that we will write
\begin{equation}
    \Phi_1(x)\hstar\Phi_2(x)\equiv \mu_{\hat{\mathcal F}}(\Phi_1\otimes \Phi_2)\equiv \mu(\hat{\mathcal F}\Phi_1\otimes \Phi_2).
\end{equation}
In this formula we are putting hats everywhere to remark that the twist and therefore the star product are defined in the active transformation picture. As already remarked, from a practical point of view, this means that if the abstract twist is given by
\begin{equation}\label{eq:abstract-twist}
    \mathcal F = f^\alpha(\X_1,\X_2,\ldots, \X_n)\otimes f_\alpha (\X_1,\X_2,\ldots, \X_n),
\end{equation}
where $\X_A$ are some generators of $\mathfrak g$, then 
\begin{equation}\label{eq:Fhat}
    \begin{aligned}
        \hat{\mathcal F} &= f^\alpha(\hat \X_1,\hat \X_2,\ldots, \hat \X_n)\otimes  f_\alpha(\hat \X_1,\hat \X_2,\ldots, \hat \X_n)\\
        &= (S\otimes S)(f^\alpha(\mathcal L^W_{\X_1},\mathcal L^W_{\X_2},\ldots, \mathcal L^W_{\X_n})\otimes f_\alpha(\mathcal L^W_{\X_1},\mathcal L^W_{\X_2},\ldots, \mathcal L^W_{\X_n})),
    \end{aligned}
\end{equation}
where we used the relation~\eqref{eq:hatX-LW} between $\hat\X$ and the Weyl-Lie derivative.
Another important remark is that the star product $\hstar$ is defined using $\hat{\mathcal F}$ rather than its inverse $\hat{\mathcal F}^{-1}$. The definition  above ensures that the star product $\hstar$ is associative
\begin{equation}
     \Phi_1\hstar(\Phi_2\hstar\Phi_3)
    =(\Phi_1\hstar\Phi_2)\hstar\Phi_3
    =\Phi_1\hstar\Phi_2\hstar\Phi_3
\end{equation}
as shown explicitly in~\eqref{eq:ass-star-pr}. To check this, one needs to use the cocycle condition for $\hat{\mathcal F}$ together with the implementation of the Leibniz rule for active transformations given in~\eqref{eq:UmuF-active}. Let us now comment on how these definitions compare to the ones that are normally employed in the literature.

\subsection{Comparison to other definitions of star products}\label{sec:comp-star}
Previous works on twist non-commutative deformations (see in particular~\cite{Aschieri:2005yw,Aschieri:2005zs,Aschieri:2006ye,Vassilevich:2006tc,Kontsevich:1997vb}) focused on twists of spacetime symmetries only (although see~\cite{Lunin:2005jy,Guica:2017mtd} for examples including also internal symmetries). The reason for the restriction of the classes of twists was of course consequence of the fact that the main motivation was to have a description of field theories defined on non-commutative spacetimes. As reviewed for example in~\cite{Szabo:2001kg}, an algebra of functions of non-commutative spacetime coordinates $\hat x^\mu$ (so that in general $[\hat x^\mu,\hat x^\nu]\neq 0$) is translated via the (inverse) Weyl map to a non-commutative algebra of functions that depend on standard (commutative) spacetime coordinates $x^\mu$, and with a multiplication that is given by a star product. Taking this perspective, the star product would be given by
\begin{equation}\label{eq:Lie-der-star}
    \Phi_1(x)\star \Phi_2(x) = \mu(\tilde{\mathcal F}^{-1}\Phi_1(x)\otimes \Phi_2(x)),
\end{equation}
with the twist written as
\begin{equation}
    \tilde{\mathcal F} = (\tilde{f}^\alpha(X_1,X_2,\ldots, X_n)\otimes \tilde f_\alpha(X_1,X_2,\ldots, X_n)),
\end{equation}
where $X_i=X_i^\mu\partial_\mu$ are the vectors that correspond to the symmetry transformations appearing in~\eqref{eq:abstract-twist}. In fact, in general, when taking that perspective the vectors $X_i$ that are used to construct the star product do not even need to be symmetries of the undeformed seed theory that is meant to be twisted. 

Importantly, the star product $\star$ in~\eqref{eq:Lie-der-star} is defined via the \emph{inverse} twist $\tilde{\mathcal F}^{-1}$. This is because Lie derivatives satisfy~\eqref{eq:UmuF-Lieder}, so that the derivation of associativity of the star product is slightly different compared to~\eqref{eq:ass-star-pr}, and the inverse cocycle condition should be used.

Notice that with the above definition of star product the twist acts in the same way on all fields, irrespectively if they are scalars, vectors, spinors, etc., because they are all  treated simply as functions of spacetime coordinates. A more refined construction is the one employed for example in~\cite{Aschieri:2009ky,Dimitrijevic:2011jg,Dimitrijevic:2014dxa} where the star product $\star$ is implemented via a twist written as
\begin{equation}\label{eq:Ft-LX}
    \tilde{\mathcal F} = (\tilde f^\alpha(\mathcal L_{\X_1},\mathcal L_{\X_2},\ldots, \mathcal L_{\X_n})\otimes \tilde f_\alpha(\mathcal L_{\X_1},\mathcal L_{\X_2},\ldots, \mathcal L_{\X_n})),
\end{equation}
where $\mathcal L_{\X_i}$ are the Lie derivatives for the vectors $X_i$. When dealing with vector fields, this construction can be understood as being equivalent to the previous one if working in the language of differential forms. In fact, by repackaging a vector field into the scalar quantity $A(x)=A_\mu(x)\dd x^\mu$, the Lie derivative reduces just to $X_i$.
Notice that~\eqref{eq:Ft-LX} differs from our implementation of the twist~\eqref{eq:Fhat} in two ways: first, the antipode does not appear in the definition, and second the Lie derivative rather than the Weyl-Lie derivative appears.

If we consider cases for which the Lie derivative and the Weyl-Lie derivative coincide (in particular when restricting to the Poincar\'e algebra, for which $\partial_\mu X^\mu=0$) then the difference in the two definitions is inconsequential, and it only leads to different conventions in how the star product is implemented. In order to match a star product constructed via the Lie derivative picture with a star product in the active transformation picture, we demand
\begin{equation}
    \Phi_1\star\Phi_2=\Phi_1\hat\star\Phi_2,
\end{equation}
where $\Phi_i$ denotes arbitrary fields. It follows immediately that
\begin{equation}\label{eq:Fh-Ft}
    \hat{\mathcal F}=(S\otimes S)\tilde{\mathcal F}^{-1}.
\end{equation}
It is possible to check that this relation between $\hat{\mathcal F}$ and $\tilde{\mathcal F}$ is compatible with the fact that they both satisfy the standard cocycle condition as in~\eqref{eq:cocycle}, where $\Delta^{op}=\Delta$ is assumed.

The discussion becomes more interesting when $\mathcal L^W\neq \mathcal L$ since, in general, in those cases the two recipes for constructing the star product yield different results. In section~\ref{sec:gauge-inv}, we will show that the star product $\hstar$ constructed with the twist $\hat{\mathcal F}$ in the picture of active transformations automatically gives rise to compatibility with gauge transformations. Using  the star product $\star$ constructed with the twist $\tilde{\mathcal F}$, instead, one may in general encounter issues to define gauge transformations. 

In~\cite{Meier:2023lku}  gauge theories deformed by Poincar\'e twists were constructed starting from the formulation of the gauge theory in terms of the language of differential forms. This strategy, as mentioned above, allows one to work with objects that transform as scalars under Poincar\'e and, together with the repackaging of vectors as $A(x)=A_\mu(x)\dd x^\mu$, fermionic fields were repackaged into scalar quantities by multiplying them by ``half-forms'' that compensated for their intrinsic spinorial transformations under Lorentz. When working in this adapted basis, all fields are rewritten in terms of scalar quantities only, and one effectively has $\mathcal L^W=\mathcal L= X^\mu\partial_\mu$. The formulation of~\cite{Meier:2023lku} and the one of this paper are then equivalent as in~\eqref{eq:Fh-Ft}. Similarly, in~\cite{Borsato:2025jre} twists were considered that together with  Poincar\'e generators included also the scale transformation of the conformal algebra. In that case, it was shown that  working with $\tilde{\mathcal F}$ written as in~\eqref{eq:Ft-LX} was not quite enough. First, one should construct a convenient basis of ``scaleless fields'' obtained by multiplying fields by powers of a dimensionful quantity $H$. Moreover, the action of symmetry generators needed to be non-trivial on $H$ too. We refer to~\cite{Borsato:2025jre} for more details. Effectively, when working in that basis one is still dealing with a twist constructed out of the vectors  $ X_i^\mu\partial_\mu$ only. Also the formulation of~\cite{Borsato:2025jre}, then,  is equivalent to the one of the present paper.

The constructions of~\cite{Meier:2023lku,Borsato:2025jre} have the disadvantage that one should first reformulate already the undeformed gauge theory in a convenient basis in order to then twist-deform it. Moreover, in those formulations, there are various realisations of the twist $\mathcal F$ that need to be worked out explicitly in various representations in order to check the consistency of the construction of the deformed gauge theories (see~\cite{Meier:2023lku,Borsato:2025jre} for details). Finally, it may be cumbersome to generalise that strategy to the case of twists of the more general superconformal algebra.\footnote{We refer to~\cite{StijnJulio} for important progress in this direction.} For these reasons, we view the star product $\hstar$ constructed out of the twist $\hat{\mathcal F}$ in the picture of active transformations as an easier implementation of the twist deformations, which reduces to previous constructions in the simpler setups that were considered so far.

We anticipate that the advantage of constructing the star product via \emph{active transformations} rather than just the Lie derivatives is that partial derivatives will always follow the Leibniz rule $\partial_\mu\left(\Phi_1\hstar\Phi_2\right)=\partial_\mu\Phi_1\hstar\Phi_2+\Phi_1\hstar\partial_\mu\Phi_2$, and this will make it straightforward to write down covariant derivatives under gauge transformations. In the usual picture of Lie derivatives, instead, the Leibniz rule for partial derivatives may change in general, see for example the case of the $\kappa$-deformation in~\cite{Wess:2006cm}.
Let us remind that an additional point in this paper is that we will relax the $\mathcal F$-unimodularity condition imposed since the results of~\cite{Aschieri:2009ky}, because we will show that in order to have consistency with gauge invariance it will be enough to require the $\mathcal R$-unimodularity condition. We will then also prove the planar equivalence theorem in the more general setup.

\section{Star deformations of field and gauge theories}\label{sec:star-theories}

\subsection{Star deformations based on symmetries}\label{sec:star-def-symm}
When constructing star deformations of field theories, the typical recipe (which is followed also in this paper) is to replace everywhere the ordinary product with the star product
\begin{equation}
    \Phi_1(x)\cdot\Phi_2(x)\to \Phi_1(x)\hstar\Phi_2(x).
\end{equation}
In order to construct the twisted theories, the important assumption that we need in this paper is that the undeformed action  is invariant under the symmetries that are used to construct the twist. In particular, this means that the Lagrangian density $\mathscr L$ will be invariant under the symmetry transformations $\X$, possibly up to a total derivative
\begin{equation}
    \hat\X(\mathscr L)=\text{total derivative}.
\end{equation}
If the twist is constructed out of conformal transformations, we will then assume that the Lagrangian density is a scalar of dimension $\Delta_{\mathscr L}=d$, so that
\begin{equation}
    \hat \X(\mathscr L)=-X^\mu\partial_\mu\mathscr L-\frac{\Delta_{\mathscr L}}{d}\partial_\mu X^\mu\mathcal L=-\partial_\mu(X^\mu\mathscr L).
\end{equation}
When working with twists of the Poincar\'e algebra only, for example, one can relax the above requirement and work also with Lagrangian terms of different dimensions, as long as they respect relativistic invariance.
Demanding compatibility with internal symmetries and (off-shell) supersymmetries, instead, means that the Lagrangian density will be simply invariant under the transformations, with no total derivatives expected.

Let us remark that in this section the whole construction is classical, and then it is enough to assume that the symmetries of the action hold at the classical level. In section~\ref{sec:quant-plan} we will move to the quantum case, and then we will have to assume that the symmetries appearing in the twist still hold at the quantum level, i.e.~there are no anomalies.

When dealing with gauge theories, the above simple recipe of replacing the ordinary product with the twisted one may entail complications related to the gauge transformations. These complications are consequence of the fact that, depending on the definition of the star product that is used, partial derivatives may not follow the usual Leibniz rule; when that is not the case, gauge transformations of derivatives of fields take a complicated expression, and it becomes challenging to define covariant derivatives. See for example~\cite{Borsato:2025jre,Dimitrijevic:2011jg,Meier:2023lku} for discussions on this point.

These issues disappear when working with the star product $\hstar$ constructed out of active transformations. As already remarked, active transformations commute with partial derivatives, and therefore it follows that partial derivatives still satisfy the Leibniz rule also in the presence of the star product
\begin{equation}
    \partial_\mu(\Phi_1(x)\hat\star\Phi_2(x)) = \partial_\mu \Phi_1(x)\hat\star\Phi_2(x)+\Phi_1(x)\hat\star\partial_\mu\Phi_2(x).
\end{equation}
This simple observation makes it straightforward to write down star-deformed  gauge theories, because we can really just follow the usual recipe of replacing the ordinary product with the star product. This ensures that we can define  covariant derivatives in a straightforward way, see appendix~\ref{app:star-gauge-tr}. 

To conclude, to construct the star-deformed field theories we will replace the ordinary product with the hatted star product. 
\begin{equation}
    S=\int d^4x\ \mathscr L\qquad\implies\qquad S^{\hat \star}=\int d^4x\ \mathscr L^{\hat \star},\qquad
    \text{where }\mathscr L^{\hat \star}=\mathscr L|_{\cdot\to \hat\star},
\end{equation}
where $\mathscr L$ is the  Lagrangian density of the undeformed model. 
As we will see later, demanding gauge invariance of the action will however require an additional condition on the twist.
Before going there, though, we will point out some facts that are valid for more general star-deformed field theories, and then we will discuss the particular case of \emph{gauge} theories.

\subsection{Twisted symmetries}\label{sec:twisted-symm}

In general, the actions deformed by the star product are not any more invariant under the original superconformal  or internal symmetries. Nevertheless, there is a notion of twisted symmetries under which the actions are still invariant. In particular, one can prove that if the original Lagrangian density $\mathscr L$ is invariant under $\hat \X$ up to a total derivative, then the same is true also for $\mathscr L^{\hstar}$
\begin{equation}
    \hat\X(\mathscr L)=\text{total derivative}\qquad\implies\qquad
    \hat\X(\mathscr L^{\hstar})=\text{total derivative}.
\end{equation}
The reason why the symmetries are said to be twisted after the deformation is that their action on products of fields gets modified compared to the undeformed case. 
An advantage of the formulation of the present paper, though, is that star products of fields have the same tensorial transformation properties as their undeformed counterparts. It is for this reason that the invariance of the twisted action immediately follows from the invariance of the undeformed one.

To see this, let us start from the fact that, before introducing any twist, symmetry transformations  follow the Leibniz rule, because taking for example two fields and $\hat\X\in\mathfrak g$ one has
\begin{equation}
    \hat\X(\Phi_1(x)\Phi_2(x))=\mu(\Delta(\hat\X)(\Phi_1(x)\otimes \Phi_2(x))),
\end{equation}
with $\Delta(\hat\X)=\hat\X\otimes 1+1\otimes \hat\X$ .
After the twist, one has instead
\begin{equation}\label{eq:Xmu-twist}
\begin{aligned}
         \hat\X(\Phi_1(x)\hat\star\Phi_2(x))&=\hat\X\mu(\hat{\mathcal F}( \Phi_1(x)\otimes \Phi_2(x)))\\
         &=\mu(\hat{\mathcal F}\Delta(\hat\X)( \Phi_1(x)\otimes \Phi_2(x))),
\end{aligned}
\end{equation}
where~\eqref{eq:UmuF-active} was used. The above reveals the twisted nature of the symmetry transformations.
In fact, the relations in~\eqref{eq:Xmu-twist} are consistent with the usual formulas for the action of symmetries in the presence of star products, see e.g.~\cite{Meier:2023lku,Aschieri:2009zz}. When we use the composition rules of active transformations discussed in section~\ref{sec:comp} and the relation to the Weyl-Lie derivative~\eqref{eq:hatU-LWSU} we get
\begin{equation}\label{eq:Xofstar}
    \hat\X(\Phi_1(x)\hat\star\Phi_2(x))=\mu\left(\mathcal L^W_{(S\otimes S)\Delta(\X)}\mathcal L^W_{(S\otimes S)\mathcal F}(\Phi_1(x)\otimes \Phi_2(x))\right).
\end{equation}
It is immediate to see, then, that from an operational point of view we first act with the Weyl Lie derivative with respect to the twist, and only then we implement the symmetry transformation via  $\mathcal L^W_{(S\otimes S)\Delta(\X)}$.

This simple observation has deep consequences. To start, take for example the case of internal symmetries, so that $\hat\X\Phi^i=u^i{}_j\Phi^j$. In the undeformed theory, the action of the symmetry generator on the product of two fields is
\begin{equation}
      \hat\X(\Phi^i\Phi^j)=(u^i{}_k\delta^j{}_l+\delta^i{}_ku^j{}_l)\Phi^k\Phi^l.
\end{equation}
Using~\eqref{eq:Xmu-twist}, we conclude that even in the presence of the twist the star product of two fields inherits the above tensorial nature under the symmetry transformation
\begin{equation}
      \hat\X(\Phi^i\hat\star\Phi^j)=\mu(\hat{\mathcal{F}}(u^i{}_k\Phi^k\otimes \Phi^j+u^j{}_l\Phi^j\otimes \Phi^l))
      =(u^i{}_k\delta^j{}_l+\delta^i{}_ku^j{}_l)\Phi^k\hat\star\Phi^l.
\end{equation}
At this point, to argue the twisted symmetry of the deformed action, we simply remind that our initial assumption is to start from an undeformed action that is invariant under a set of symmetries. This means that, even if the Lagrangian is constructed out of fields that have non-trivial transformation properties under the symmetry under consideration, for example $\hat\X\Phi^i=u^i{}_j\Phi^j$ for internal symmetries, the indices of these fields must be contracted to give rise to singlets of that symmetry. An example may be a collection of $N$ scalar fields $\phi^i$ with kinetic term $\delta_{ij}\partial^\mu \phi^i\partial_\mu\phi^j$, which is invariant under $O(N)$ if the fields transform as $\hat\X\phi^i=u^i{}_j\phi^j$ with $u_{ij}=\delta_{ik}u^k{}_j$ antisymmetric matrices. Then, in the presence of the star product, the kinetic term will be $\delta_{ij}\partial^\mu \phi^i\hstar\partial_\mu\phi^j$, and it will still be invariant under $O(N)$ 
\begin{equation}
      \hat\X(\delta_{ij}\partial^\mu \phi^i\hstar\partial_\mu\phi^j)=\delta_{ij}(u^i{}_k\delta^j{}_l+\delta^i{}_ku^j{}_l)\partial^\mu \phi^k\hstar\partial_\mu\phi^l=0,
\end{equation}
for the same reason as in the undeformed case.

Even though the explicit calculations are more complicated, the same logic applies also for spacetime transformations, as also in that case star products of spacetime tensors maintain the tensorial properties of their undeformed counterparts. We refer to appendix~\ref{app:tens-spt} for the explicit calculations. In particular, given that under the symmetry transformation the star product of tensors changes itself as a tensor as written in~\eqref{eq:sptXofstar}, the contraction of the spacetime indices and the assignement of the correct conformal weight to construct a singlet as in the undeformed case will imply that  also the deformed Lagrangian changes just by a total derivative. The case of off-shell supersymmetry transformations formally works like that of internal symmetries.

Although we only discussed the case when we star multiply two fields, the above observations can obviously be iterated to any number of star products.

Although in the previous discussion we focused on the Lagrangian density and on elements of $\mathfrak g$, we may generalise it to a more general function $\mathcal W$  of the fields and their derivatives that  is invariant under an element $\hat\U$ of the universal enveloping algebra up to a total derivative. Then, also the function $\mathcal W^{\hat\star}=\mathcal W|_{\cdot\to \hat\star}$ obtained by replacing the ordinary product with the hatted star product will be invariant under $\hat\U$ up to a total derivative
\begin{equation}\label{eq:tot-der-tw}
    \hat\U(\mathcal W)=\text{total derivative}\qquad\implies\qquad \hat\U(\mathcal W^{\hat\star})=\text{total derivative}.
\end{equation}
This observation will be useful later.

\subsection{Equivalent twists and field redefinitions}\label{sec:equiv-red}
In this section we point out that star products $\hstar$ and $\hstar'$ constructed out of equivalent twists $\mathcal F$ and $\mathcal F'$ satisfying~\eqref{eq:Fp-equiv} are related by field redefinitions. 
In fact, we may use $w$ to construct the field redefinition
\begin{equation}
    \Phi_i(x)=\hat w^{-1}\left(\Phi'_i(x)\right)=\mathcal{L}_{S(w^{-1})}(\Phi'_i(x)),
\end{equation}
which is then both linear in the fields and  a deformation of the trivial redefinition. We have 
\begin{equation}
    \begin{aligned}
        \Phi_1\hstar\Phi_2&=\mu(\hat{\mathcal F}(\Phi_1\otimes \Phi_2))=\mu(\mathcal L_{(S\otimes S){\mathcal F}}(\mathcal{L}_{S(w^{-1})}\otimes \mathcal{L}_{S(w^{-1})})(\Phi'_1\otimes \Phi'_2))\\
        &=\mu((\hat w^{-1}\otimes\hat w^{-1})\hat{\mathcal F}(\Phi'_1\otimes \Phi'_2))
        =\mu((\hat w^{-1}\otimes \hat w^{-1})\hat{\mathcal F}\Delta(\hat w)\Delta(\hat w^{-1})(\Phi'_1\otimes \Phi'_2))\\
        &=\hat w^{-1}\mu((\hat w^{-1}\otimes \hat w^{-1})\hat{\mathcal F}\Delta(\hat w)(\Phi'_1\otimes \Phi'_2))\\
        &=\hat w^{-1} (\Phi'_1\hstar'\Phi'_2),
    \end{aligned}
\end{equation}
where $\hat {\mathcal{F}}'$ is an equivalent twist to $\hat{\mathcal{F}}$ as in~\eqref{eq:Fp-equiv}. We see that the field redefinition induced by $w$ carries over to the star product of two fields, and in fact to an arbitrary number of star products thanks to the associativity of $\hstar$. In particular, a consequence of this is the transformation of the Lagrangian density
\begin{equation}
    \mathcal L^{\hstar}(\Phi_i)=\hat w^{-1}(\mathcal L^{\hstar'}(\Phi'_i)).
\end{equation}
At this point, we may assume that $w$ is of the form
\begin{equation}
    w=1+\U,\qquad\qquad \U\in U(\mathfrak g)\setminus\{1\},
\end{equation}
so that the piece proportional to the identity operator is normalised to 1.\footnote{It is not possible to have $w=f+\U$, where $f$ is any arbitrary function of the deformation parameter other than $f=1$. The reason is the normalization condition of the twist: $(\epsilon\otimes id)(\mathcal{F}^{-1})=1$. By demanding the transformed twist to satisfy the same condition, we get
\begin{equation}
    \begin{aligned}
        1\stackrel{!}{=}(\epsilon\otimes id)(\mathcal{F'}^{-1})&=(\epsilon\otimes id)(\Delta(w^{-1}))(\epsilon\otimes id)(\mathcal{F}^{-1})(\epsilon\otimes id)(w\otimes w)\\
        &=w^{-1}1\epsilon(w)w=\epsilon(w).
    \end{aligned}
\end{equation}
Hence, $w=1+\U$, with $\U\in U(\mathfrak{g})/\{1\}$.} The element $\hat\U$, therefore, is a symmetry that leaves the Lagrangian density invariant, possibly up to total derivatives, and this is enough to conclude that the actions constructed with $\hstar$ and $\hstar'$ are in fact equal
\begin{equation}\label{eq:S=Sp}
    \int d^4x \ \mathscr L^{\hstar}(\Phi_i)=\int d^4x \ \mathscr L^{\hstar'}(\Phi'_i).
\end{equation}

\subsection{Cyclic integration and $\mathcal R$-unimodularity condition}\label{sec:cyclic}
As we will see in the next section, in order to construct the star deformations of the gauge theories, we will need to require that the star product, albeit being non-commutative, is cyclic under integration. This means that if $\mathcal W^{\hat\star}_1,\mathcal W^{\hat\star}_2$ are two functions of the fields involving star products, then
\begin{equation}\label{eq:cycl-int}
    \int d^dx \ \mathcal W^{\hat\star}_1\ \hat\star\  \mathcal W^{\hat\star}_2=\int d^dx \ \mathcal W^{\hat\star}_2\ \hat\star \ \mathcal W^{\hat\star}_1.
\end{equation}
In the above we are assuming, as usual, that integrals of total derivatives are zero.

As we are about to see, not all twists will give rise to cyclic star products. Interestingly, we can link cyclicity to the $\mathcal R$-unimodularity condition of section~\ref{sec:unim}. In fact, first we can write
\begin{equation}\label{eq:L1sL2}
\begin{aligned}
    \int d^dx \ \mathcal W^{\hat\star}_1\ \hat\star\  \mathcal W^{\hat\star}_2&= \int d^dx \ \hat{ f}^\alpha( \mathcal W^{\hat\star}_1)\ \hat{ f}_\alpha( \mathcal W^{\hat\star}_2)=\int d^dx \    \mathcal W^{\hat\star}_1\ (\hat{ f}_\alpha S(\hat{ f}^\alpha)\mathcal W^{\hat\star}_2)\\
    &=\int d^dx \   \mathcal W^{\hat\star}_1\ (S(\hat V^{-1})\mathcal W^{\hat\star}_2),
\end{aligned}
\end{equation}
where  we used the identity~\eqref{eq:IBP-new} that implements integration by parts for $\hat\U=\hat{ f}^\alpha$, and we recognised the object $S(\hat V^{-1})$ defined in~\eqref{eq:SF}. Importantly, the fact that the ``integration by parts'' implemented by~\eqref{eq:IBP-new} only gives rise to extra total derivatives that we don't explicitly write, is related to the assumption that we are deforming a Lagrangian that is symmetric.

Importantly, in the above expression a star product has been removed at the cost of acting with $S(\hat V^{-1})$ on $\mathcal W^{\hat\star}_2$. Because $\mathcal W^{\hat\star}_1$ and $S(\hat V^{-1})( \mathcal W^{\hat\star}_2)$ are now multiplied by an ordinary product, the order in which they appear under the integral does not matter any more.

We obviously get a similar expression if we started from the star product in the other order, just with the indices 1 and 2 swapped;  if  we further use again the identity~\eqref{eq:IBP-new}, now for $\hat\U=S(\hat V^{-1})$, we get
\begin{equation}
\begin{aligned}
    \int d^dx \ \mathcal W^{\hat\star}_2\ \hat\star\  \mathcal W^{\hat\star}_1&=\int d^dx \  \mathcal W^{\hat\star}_2(S(\hat V^{-1}) \mathcal W^{\hat\star}_1)=\int d^dx \  (S(\hat V^{-1}) \mathcal W^{\hat\star}_1)\ \mathcal W^{\hat\star}_2\\
    &=\int d^dx \   \mathcal W^{\hat\star}_1\ (\hat V^{-1}\mathcal W^{\hat\star}_2).
\end{aligned}
\end{equation}
We conclude that the  cyclicity of the star-product under integration written in~\eqref{eq:cycl-int} is equivalent to the $\mathcal R$-unimodularity condition
\begin{equation}\label{eq:cycl-cond-tw}
    S(\hat V)=\hat V.
\end{equation}
It is nice to see that if a twist gives rise to a cyclic star product, then thanks to~\eqref{eq:mu1SFp} all twists in its equivalence class will give rise to cyclic star products. This is in fact consistent with the discussion of section~\ref{sec:equiv-red}.

Let us point out that the $\mathcal F$-unimodularity condition  $V=1$  first introduced in Ref.~\cite{Aschieri:2009ky} is certainly sufficient for cyclicity, but it actually requires  a stronger property for integration of star products, since it implies that one star product can be removed under integration and replaced by an ordinary one
\begin{equation}
    \int d^dx \ \mathcal W^{\hat\star}_1\ \hat\star\ \mathcal W^{\hat\star}_2=\int d^dx \ \mathcal W^{\hat\star}_1 \ \ \mathcal W^{\hat\star}_2.
\end{equation}
This of course implies also cyclicity under integration.\footnote{Importantly, this does not mean that the star-product completely disappears under integration: when more than one star-product appears, only one may be replaced by the ordinary one, having the option to choose which one, e.g.
\begin{equation}
        \int d^dx\ \mathcal W^{\hat\star}_1\hat\star\mathcal W^{\hat\star}_2\hat\star\mathcal W^{\hat\star}_3\hat\star\ldots=\int d^dx\ \mathcal W^{\hat\star}_1(\mathcal W^{\hat\star}_2\hat\star\mathcal W^{\hat\star}_3\hat\star\ldots)=\int d^dx\ (\mathcal W^{\hat\star}_1\hstar\mathcal W^{\hat\star}_2)(\mathcal W^{\hat\star}_3\hat\star\ldots).
\end{equation}
}
We conclude that the $\mathcal F$-unimodularity condition of~\cite{Aschieri:2009ky} is unnecessarily strong, and we will use the $\mathcal R$-condition~\eqref{eq:cycl-cond-tw} instead. In fact, the $\mathcal F$-unimodularity condition of~\cite{Aschieri:2009ky} would exclude from the construction twists that, although not being $\mathcal F$-unimodular themselves, are equivalent to $\mathcal F$-unimodular twists. Examples of this kind are the abelian twist in~\eqref{eq:abel-equiv-tw} or the extended Jordanian twist in~\eqref{eq:FeJ2}. Our cyclicity condition, instead, is more general and allows us to construct star gauge theories also for those twists. Moreover, if a twist satisfies our cyclicity condition then all twists in its equivalence class will as well, in contrast to what happens for the stronger condition of \cite{Aschieri:2009ky} which is not left invariant by the similarity transformation in \eqref{eq:Fp-equiv}.

Before continuing, let us comment on the fact that we can generalise the usual integration by parts of equation~\eqref{eq:IBP-new} to the twisted setup as
\begin{equation}\label{eq:IBP-twist}
    \hat\U(\mathcal W^{\hat\star}_1)\hstar\mathcal W^{\hat\star}_2=\mathcal W^{\hat\star}_1 \hstar(S(\hat\U)(\mathcal W^{\hat\star}_2))+\text{total derivative},\qquad 
     \quad \forall \hat\U\in U(\mathfrak g) .
\end{equation}
Notice that the undeformed antipode appears. The proof uses \eqref{eq:IBP} for $\hat\U$ directly
\begin{equation}
    \begin{aligned}
        \int d^dx\ \hat\U(\mathcal W^{\hat\star}_1)\hstar\mathcal W^{\hat\star}_2
    &=\int \mathrm{d}^dx~\mu( \hat f^\alpha\otimes\hat f_\alpha) (\hat\U\otimes 1)(\mathcal{W}_1^{\hstar}\otimes\mathcal{W}_2^{\hstar})\\
    &=\int \mathrm{d}^dx~ \mu( \hat f^\alpha\otimes\hat f_\alpha) (1\otimes S(\hat\U_{(2)}))\Delta(\hat\U_{(1)})(\mathcal{W}_1^{\hstar}\otimes\mathcal{W}_2^{\hstar})\\
    &=\int \mathrm{d}^dx~ \hat\U_{(1)}\mu( \hat f^\alpha\otimes\hat f_\alpha) (1\otimes S(\hat\U_{(2)}))(\mathcal{W}_1^{\hstar}\otimes\mathcal{W}_2^{\hstar})\\
    &=\int \mathrm{d}^dx~ \mu( \hat f^\alpha\otimes\hat f_\alpha) (1\otimes S(\hat\U))(\mathcal{W}_1^{\hstar}\otimes\mathcal{W}_2^{\hstar})+\text{tot. derivative}\\
    &=\int \mathrm{d}^dx~ \mathcal W^{\hat\star}_1\hstar S(\hat\U)(\mathcal W^{\hat\star}_2).
    \end{aligned}
\end{equation}
One may also get a twisted version of integration by parts, with the twisted antipode $S_{\mathcal F}$, when swapping the application of $\hat\U$ and of the star product
\begin{equation}
\begin{aligned}
    \int\mathrm{d}^dx~\hat\U(\hat f^\alpha \mathcal{W}^{\hstar}_1)\hat f_\alpha \mathcal{W}^{\hstar}_2&=\int\mathrm{d}^dx~\hat\U( \mathcal{W}^{\hstar}_1)\hat f_\alpha S(\hat f^\alpha)\mathcal{W}^{\hstar}_2\\
    &=\int\mathrm{d}^dx~\hat\U( \mathcal{W}^{\hstar}_1)\hat V^{-1}(\mathcal{W}^{\hstar}_2)\\
    &=\int\mathrm{d}^dx~ \mathcal{W}^{\hstar}_1\hat V^{-1}(S(\hat\U)\mathcal{W}^{\hstar}_2)\\
    &=\int\mathrm{d}^dx~ \mathcal{W}^{\hstar}_1\hat V^{-1}S(\hat\U)\hat V\hat V^{-1}\mathcal{W}^{\hstar}_2\\
    &=\int\mathrm{d}^dx~\hat f^\alpha \mathcal{W}^{\hstar}_1S_{\mathcal{F}}(\hat\U)(\hat f_\alpha \mathcal{W}^{\hstar}_2).
\end{aligned}
\end{equation}

\subsection{Gauge invariant actions}\label{sec:gauge-inv}
We now have all the ingredients to check the gauge invariance of the actions for the star-deformed gauge theories. Let us consider the gauge transformation of the kinetic term of the gauge fields. We get
\begin{equation}
     i\int d^dx \ \tr([\epsilon\stackrel{\hat\star}{,}F_{\mu\nu}]\, \hat\star \, F^{\mu\nu}+F_{\mu\nu}\hat\star[\epsilon\stackrel{\hat\star}{,}F^{\mu\nu}])=
     i\int d^dx \ \tr[\epsilon\stackrel{\hat\star}{,} F_{\mu\nu} \hat\star F^{\mu\nu}]=0,
\end{equation}
which vanishes thanks to cyclicity under integration.
Notice that cyclicity under integration is necessary to check the gauge invariance only of  terms involving adjoint fields, and it is not needed for fields in the fundamental representation of the gauge group. In fact, considering for example the kinetic term of a complex scalar field transforming as $\delta_\epsilon\phi=i\epsilon\hat\star\phi$ (which implies $\delta_\epsilon\phi^\dagger=-i\phi^\dagger\hat\star\epsilon$) one has that
\begin{equation}
    \delta_\epsilon\int d^dx\ D_\mu\phi^\dagger\hat\star D^\mu\phi=i\int d^dx(D_\mu\phi^\dagger\hat\star \epsilon\hat\star D^\mu\phi-D_\mu\phi^\dagger\hat\star \epsilon\hat\star D^\mu\phi)=0,
\end{equation}
without the need to appeal to cyclicity under integration. Even when dealing with matter fields in the fundamental representations, however, the gauge fields are still in the adjoint representation, and for this reason cyclicity under integration is nevertheless a necessary requirement. In the following we will therefore assume that the twist  satisfies the $\mathcal R$-unimodularity condition~\eqref{eq:cycl-cond-tw} so that we can always use cyclicity under integration.

As a final remark, we point out that the usual simple way of describing the recipe to construct the twist deformations---i.e.~to replace the ordinary product with the star product as said in section~\ref{sec:star-def-symm}---may be subtle in some cases. For example, when twist-deforming a $U(1)$ gauge theory one should really write $ F_{\mu\nu} = \partial_\mu A_\nu-\partial_\nu A_\mu-i[A_\mu\stackrel{\hat\star}{,}A_\nu]$ in order to ensure gauge invariance of the action; on the one hand, the last term with the star commutator in $F_{\mu\nu}$ is non-trivial because of the star-product, but on the other hand, strictly speaking, it does not come from replacing the ordinary product with the star product in the undeformed $F_{\mu\nu}$, because in the undeformed limit the commutator trivialises. Overall, what one needs  to ensure consistency with gauge invariance is that each term that in the undeformed Lagrangian density is gauge covariant remains such even after the prescription to introduce the twist. In some cases, like in the example of the $U(1)$ $F_{\mu\nu}$, one may need to complement the naive recipe with extra terms.
One may expect similar subtleties to appear for example in the construction of deformations involving twists by supercharges.\footnote{We thank Stijn van Tongeren for useful discussions on this point} Also in that case one should identify a prescription to twist that ensures that gauge covariance of each term in the Lagrangian density is maintained. We refer to~\cite{StijnJulio} for more details on this discussion. 

\section{Quantisation and planar equivalence theorem}\label{sec:quant-plan}
We will now move to the quantisation of the deformed gauge theories. We will start with a brief section to set up our conventions and to prove a couple of identities that will be useful, in the following subsection, to prove the planar equivalence theorem.

\subsection{Preliminaries on quantisation}\label{sec:quant}
When quantising the twist-deformed gauge theories, we will have in mind the path integral formulation. This is convenient for various reasons, and in this case it helps to sweep under the rug certain issues. In particular, certain twist deformations may introduce higher time derivatives of the fields; in fact, this happens when a symmetry generator entering the twist corresponds to a vector $X^\mu$ with a timelike component. In the path integral formulation, having time derivatives up to order $N$ means that the boundary conditions in the path integral must specify the $i$-th derivatives $\partial_t^i\Phi(t_i),\partial_t^i\Phi(t_f)$ at the initial and final time for $i=0,\ldots,N-1$. We will nevertheless never see these details appearing.

To make the presentation simple, we will now discuss the case of a scalar field, since this is enough to identify the effects of the twist deformation. The generalisation to fermionic spinors and vectors will be immediate.
We will work with the usual definition of the generating functional
\begin{equation}\label{eq:gen-funct}
    \mathcal Z[J]=\int \mathcal D\phi e^{i(S^{\hstar}+\int d^4x J(x)\phi(x))},
\end{equation}
where the action $S^{\hstar}$ is defined via the star product. Notice that the coupling to the source $J$ is via the ordinary product. To start, we may consider the deformation of the action of the Klein-Gordon field
\begin{equation}
    \begin{aligned}
        S^{\hstar}&=\frac12\int d^4x\ (\partial_\mu\phi\hstar\partial^\mu\phi-m^2\phi\hstar\phi)=\frac12\int d^4x\ (\partial_\mu\phi \hat V^{-1}(\partial^\mu\phi)-m^2\phi \hat V^{-1}(\phi)).
    \end{aligned}
\end{equation}
Defining the propagator of the undeformed theory as
\begin{equation}
    \Delta_F(x) = \int \frac{d^4p}{(2\pi)^4}\ \frac{i}{p^2-m^2+i\epsilon}\ e^{-ip\cdot x},\quad
    \implies\quad
    (\Box+m^2)\Delta_F(x)=-i\delta^{(4)}(x),
\end{equation}
one can  make minor modifications to the usual considerations to obtain a useful expression for the generating functional of the twist deformation of the free theory. In particular, redefining $\phi(x)=\phi'(x)+i\int d^4y\Delta_F(x-y)\hat V(J(y))$ one gets\footnote{If we had  instead defined the generating functional with $J\hstar \phi=V^{-1}(J)\phi$, we would have redefined $\phi(x)=\phi'(x)+i\int d^4y\Delta_F(x-y)J(y)$ and obtained in the exponent $-\frac12\int d^4x d^4y\ \hat V^{-1}(J(x))\Delta_F(x-y)J(y)$. Then the  propagator would be $  \Delta^{\hstar}_F(x,y)=V_x^{-1}\Delta_F(x-y)$. However, in this scenario, correlators would be generated by derivatives with respect to $V^{-1}J(x)$. By simply redefining the source $J'=V^{-1}(J)$, this is  exactly equivalent to starting with a source term $\int\mathrm{d}^4x J'(x)\phi(x)$ as done in the main text.}
\begin{equation}
    \mathcal Z[J]=\mathcal Z[0]\exp\left(-\frac12\int d^4x d^4y\ J(x)\Delta_F(x-y)\hat V(J(y))\right).
\end{equation}
Importantly, the operator $\hat V$ acting on $J(y)$ is inherited from the action of $\hat V$ on $\phi$. That means that the way the elements of the universal enveloping algebra will be represented when acting on the source $J$ (and later on the propagator) will be the same as when acting on the corresponding field, which in this case is $\phi$. Because we are integrating in $d^4y$, we can do integration by parts and write
\begin{equation}
    \mathcal Z[J]=\mathcal Z[0]\exp\left(-\frac12\int d^4x d^4y\ J(x)(\hat V_y\Delta_F(x-y))J(y)\right),
\end{equation}
where we used $S(\hat V)=\hat V$ and we added a subindex $y$ to remind ourselves that we are acting with $\hat V$ on the second leg of the propagator. 
Looking at the above expression, it is natural to define the propagator of the deformed theory as
\begin{equation}
   \Delta^{\hstar}_F(x,y)\equiv\hat V_y\Delta_F(x-y),
\end{equation}
where  the notation makes it clear  that now the twisted propagator is not necessarily of difference form. Notice that when one imposes the stronger $\mathcal F$-unimodularity condition $\hat V=1$, the propagator remains undeformed as in~\cite{Meier:2023lku}.
We take the above definition for the twisted propagator because then
\begin{equation}
    \braket{0| T\phi(x_1)\phi(x_2)|0}\equiv -\left.\left(\frac{\delta}{\delta J(x_1)}\frac{\delta}{\delta J(x_2)}\mathcal Z[J]\right)\right|_{J=0}=\Delta^{\hstar}_F(x,y).
\end{equation}
Moreover, it is still true, as in the undeformed setup, that the deformed propagator is a Green's function for the differential operator implementing the equations of motion. In fact, when calculating the variation of the action we have
\begin{equation}
    \delta S^{\hstar}=-\int d^{4} x\ \delta\phi \hat V^{-1}(\Box+m^2)\phi,
\end{equation}
so that the equations of motion are
\begin{equation}
    \hat V^{-1}(\Box+m^2)\phi=0\qquad\implies\qquad (\Box+m^2)\phi=0,
\end{equation}
because $\hat V^{-1}$ is invertible. 
Adding a test function $\varphi$ to make the calculation with distributions more clear, we have
\begin{equation}
\begin{aligned}
    \varphi(x,y)\hat V_x^{-1}(\Box_x+m^2)\Delta^{\hstar}_F(x,y)&= \varphi(x,y)\hat V_x^{-1}(\Box_x+m^2)\hat V_y\Delta_F(x-y)=-i\varphi(x,y)\hat V_x^{-1}\hat V_y\delta^{(4)}(x-y)\\
    &=-i\hat V_y\hat V_x^{-1}(\varphi(x,y))\delta^{(4)}(x-y)=-i\varphi(x,y)\delta^{(4)}(x-y),
\end{aligned}
\end{equation}
where we used integration by parts and $S(\hat V)=\hat V$.

At this point, it is useful to highlight a couple of identities that are satisfied by the propagator. They are essentially consequence of the Ward identities that hold in the undeformed setup. In particular, given an $N$-point correlation function
\begin{equation}
    \braket{\Phi_1(x_1)\Phi_2(x_2)\ldots\Phi_N(x_N)}\equiv\braket{0|T\Phi_1(x_1)\Phi_2(x_2)\ldots\Phi_N(x_N)|0},
\end{equation}
where $T$ denotes time ordering, one may write Ward identities for a global symmetry implemented by $\hat\X\in\mathfrak g$ as
\begin{equation}\label{eq:Ward-Hopf}
    \Delta^{(N)}(\hat\X)\braket{\Phi_1(x_1)\Phi_2(x_2)\ldots\Phi_N(x_N)}=0,
\end{equation}
where $\Delta^{(N)}$ is the $N$-fold coproduct defined in appendix~\ref{app:Hopf}, so that $\Delta^{(N)}(\hat\X)=\sum_{k=0}^{N-1}1^{\otimes k}\otimes \hat\X\otimes 1^{\otimes (N-k-1)}=\hat\X\otimes 1^{\otimes (N-1)}+1\otimes \hat\X\otimes 1^{\otimes (N-2)}\ldots+1^{\otimes (N-1)}\otimes\hat\X$. See appendix~\ref{app:Ward} for the rewriting of Ward identities in the above language.
A direct consequence of this is that the propagator of the undeformed theory is invariant under $\hat\X$, which reads
\begin{equation}  
\Delta(\hat\X)\braket{\Phi_1(x_1)\Phi_2(x_2)}=\braket{\hat\X(\Phi_1(x_1))\Phi_2(x_2)}+\braket{\Phi_1(x_1)\hat\X(\Phi_2(x_2))}=0.
\end{equation}
Importantly, this implies
\begin{equation} \label{eq:inv-prop}
    \hat \X_x\Delta_F(x-y)=S(\hat \X_y)\Delta_F(x-y),
\end{equation}
where the subscripts on the symmetry generators indicate the leg on which they act.
In~\cite{Meier:2023lku} this was used for the Poincar\'e symmetry only, while here we extend it to the whole (super)conformal symmetry, whenever that applies.

Generalising the above identity to the whole universal enveloping algebra one may write the symmetry invariance of the propagator as
\begin{equation}
    \hat \U_x \hat\V_x\hat\V'_y\Delta_F(x-y)=S(\hat \U_y)\hat\V_x\hat\V'_y\Delta_F(x-y),
\end{equation}
where, to be more general, we allowed for the presence of other ``spectator elements'' $\hat\V_x,\hat\V'_y\in U(\mathfrak g)$ acting on the propagator. Notice that these spectator elements act innermost, while the symmetry invariance of the propagator is implemented on the outermost element $\hat \U$. The reason for this is that, when realising active transformations as explicit Weyl-Lie derivatives, $\mathcal L^W_U$ will act innermost. In that case, we are then allowed to use the symmetry invariance implemented as the differential equation satisfied by the propagator.

 A first consequence of the symmetry invariance is that the twisted propagator is symmetric under the exchange of the two spacetime points. In fact,
\begin{equation}
   \Delta^{\hstar}_F(x,y)=\hat V_y\Delta_F(x-y)=S(\hat V_x)\Delta_F(x-y)=\hat V_x\Delta_F(x-y)=\Delta^{\hstar}_F(y,x),
\end{equation}
where we used also the $\mathcal R$-unimodularity condition.

We also point out that the explicit expression for an element $\hat \U\in U(\mathfrak g)$ acting on the second leg of the twisted propagator is
\begin{equation}
    \hat \U_y  \Delta^{\hstar}_F(x,y)=\hat \U_y  \Delta^{\hstar}_F(x,y)=\hat V_y\hat \U_y  \Delta_F(x,y),
\end{equation}
where $\hat V_y$ acts outermost. In fact, using the definition of the twisted propagator
\begin{equation}
    \begin{aligned}
    \hat \U_y  \Delta^{\hstar}_F(x,y)
    &=-\hat \U_y
    \left.\left(\frac{\delta}{\delta J(x)}\frac{\delta}{\delta J(y)}\exp\left(-\frac12\int d^4z_1 d^4z_2\ J(z_1)(\hat V_{z_2}\Delta_F(z_1-z_2))J(z_2)\right)\right)\right|_{J=0}\\
    &=\int d^4z_1 d^4z_2\ \hat V_{z_2}\Delta_F(z_1-z_2)\ \delta^{(4)}(z_1-x)\ \hat \U_y\delta^{(4)}(z_2-y)\\
    &=\int d^4z_1 d^4z_2\ \hat V_{z_2}\Delta_F(z_1-z_2)\ \delta^{(4)}(z_1-x)\ S(\hat \U)_{z_2}\delta^{(4)}(z_2-y)\\
    &=\int d^4z_1 d^4z_2\ \hat V_{z_2}\hat \U_{z_2}\Delta_F(z_1-z_2)\ \delta^{(4)}(z_1-x)\ \delta^{(4)}(z_2-y)\\
    &=\hat V_y\hat \U_y  \Delta_F(x,y).
    \end{aligned}
\end{equation}
In fact, the above expression may be proved also by using the previous properties of the propagator
\begin{equation}
    \hat \U_y  \Delta^{\hstar}_F(x,y)=\hat \U_y  \hat V_x\Delta_F(x,y)= \hat V_x\hat \U_y \Delta_F(x,y)
    = S(\hat V)_y\hat \U_y \Delta_F(x,y)= \hat V_y\hat \U_y \Delta_F(x,y).
\end{equation}
With the above, we can then prove an identity that later will be useful for the proof of the planar equivalence theorem. In particular,  when both legs of the  twist act on the twisted propagator, one gets back the untwisted one. In fact,
\begin{equation}\label{eq:FDeltastar=Delta}
\begin{aligned}
    \hat {{\mathcal F}}_{xy}\Delta^{\hstar}_F(x,y)
    &=\hat{ f}^\alpha_x\hat V_y\hat{ f}_{\alpha y}\Delta_F(x-y)
    =S(\hat{ f}^\alpha_y)\hat V_y\hat{ f}_{\alpha y}\Delta_F(x-y)\\
    &
    =\Delta_F(x-y),
\end{aligned}
\end{equation}
where in the last step we used one of the identities in~\eqref{eq:identities-V-gen}.

Although these considerations were made for the case of the scalar field, one may generalise them to the case of fermionic spinors and gauge vectors. In particular, one will have a twisted propagator that is equal to the undeformed one dressed by $\hat V$ acting on one of the two legs. Moreover, Ward identities will still imply the symmetry invariance of the propagator.

At this point we have all the necessary ingredients to prove the planar equivalence theorem.

\subsection{The planar equivalence theorem}
Loosely speaking, the planar equivalence theorem states that for all planar Feynman diagrams of the twisted theory, the effects of the twist cancel in the internal structure of the diagram and remain non-trivial only on the external legs. Moreover, the way the twist acts on the external legs mimics what happens on fundamental vertices, and in particular it leads to cyclicity of those legs and invariance of the diagram under the symmetries of the undeformed theory. This theorem was originally proven in \cite{FILK199653} for the Groenewold-Moyal deformation and generalized to twists in the Poincar\'e algebra in the Lie-derivative picture in \cite{Meier:2023lku}.

To make the above statements more precise, we will build diagrams of increasing complexity. We will start with fundamental vertices where the validity of the planar equivalence theorem is immediate; we will then use them to construct tree-level diagrams and extend the theorem to that case; finally we will include diagrams with loops and show that the theorem applies also in that case.

Our proof will closely follow the one of~\cite{Meier:2023lku}, generalising it to the case of twists beyond the Poincar\'e algebra (and beyond the inclusion of only scale transformations as in~\cite{Borsato:2025jre}) and assuming only $S(\hat V)=\hat V$ rather than the stronger $\hat V=1$. In our explanation, we will use the example of a scalar field, for simplicity, but later we will comment on why the proof applies to more general fields.

\subsubsection{Fundamental interaction terms}
Let us assume that the undeformed theory has a polynomial interaction of order $k$. Knowing that in the presence of the deformation we have to apply the twist to each of the fields in the product in different ways, we introduce an index $i=1,\ldots,k$ that allows us to distinguish between the different fields in the product. Moreover, we note that we can formally write
\begin{equation}
    \int d^dx\ \phi_1(x)\phi_2(x)\cdots\phi_k(x)=\int d^dx\ \mu^{(k)}\left(\bigotimes_{i=1}^k\phi_i(x)\right),
\end{equation}
where $\mu^{(k)}$ is the $k$-fold product allowing us to multiply $k$ fields associatively, and $\bigotimes_{i=1}^k\phi_i(x)=\phi_1(x)\otimes\cdots\otimes \phi_k(x)$.
Diagrammatically, we may represent the  integrand of the fundamental interaction term as a vertex as one may do with Feynman diagrams.

Importantly,  the  integrand of the interaction term  becomes  a contribution to the amplitude with $k$ external legs in perturbation theory only after we contract the fields $\phi_i(x)$ to ``external fields'' $\phi_i(z_i)$ that appear in the correlation function $\braket{\phi_1(z_1)\phi_2(z_2)\ldots\phi_k(z_k)}$. This contraction is of course what gives rise to Feynman propagators $\Delta_F(z_i-x)$. 

To present the proof of the theorem, instead of working with Feynman diagrams we will first work with the corresponding objects where, like above, the external fields located at points $z_i$ are not yet contracted with the fields located at the points $x_i$. We will still call these objects ``diagrams'', as done above. We will do so not only for the fundamental interaction terms, but also when building more complicated tree-level and loop diagrams.  At the end, we will show what happens when performing the final contractions.

\begin{figure}[h]
\begin{center}
\includegraphics[width=0.7\textwidth]{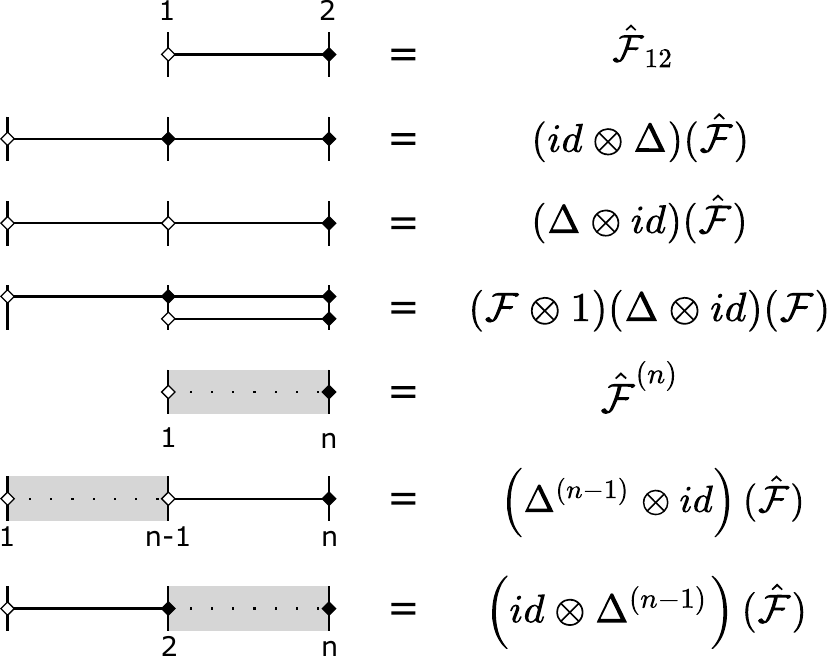}
\caption{The Feynman rules representing the Drinfel'd twist. All twists appearing in a given Feynman diagram act on individual legs, here represented by vertical lines. Diagrammatically, we represent the twist via  lines (that here are horizontal) connecting the individual propagator lines. Here, unfilled squares denote the action of $\hat f^\alpha$ while filled squares denote the action of $\hat f_\alpha$. When extending their actions  to several legs by means of the coproduct, we will use the notation as in the figure. By grey areas we denote the possibility of having several legs in between.}
\label{fig:TwistRules}
\end{center}
\end{figure}

\begin{figure}[h]
\begin{center}
\includegraphics[width=0.7\textwidth]{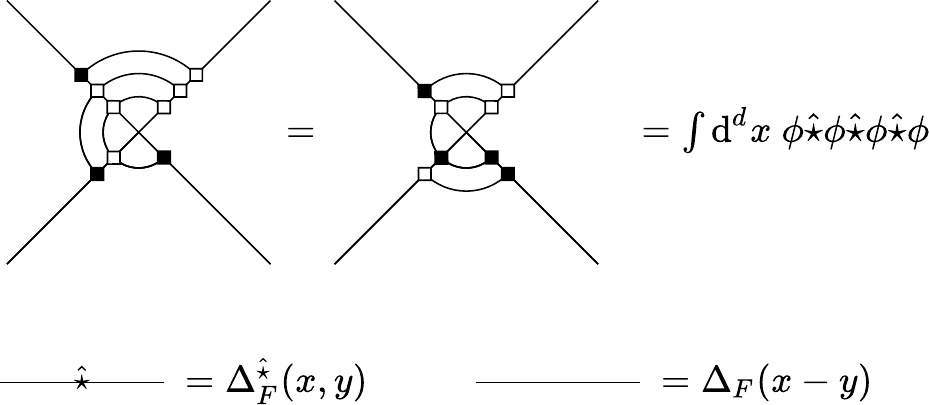}
\caption{The Feynman rules for the deformed theory. The star products entering at a given vertex are denoted by twist lines, which satisfy the cocycle condition and hence lead to different choices of assigning twist lines, which are related by associativity of the star product. The deformed propagator is denoted by a $\hat\star$ sitting on a line. This additional $\hat\star$ can be understood as the operator $\hat V$ acting on the undeformed propagator. Taking into account that the propagator remains symmetric in its arguments, we prefer to use a symmetric diagrammatic representation and have the operator $\hat\star$ sitting in the middle of a given line. For completeness, by an ordinary line, we denote the undeformed propagator.}
\label{fig:FeynmanRules}
\end{center}
\end{figure}

In the presence of the deformation, the above interaction term becomes 
\begin{equation}
\begin{aligned}
    \int d^dx\ \phi_1(x)\hstar\phi_2(x)\hstar\cdots\hstar\phi_k(x)&=\int d^dx\ \mu^{(k)}\left(\hat{{\mathcal F}}^{(k)}\ \bigotimes_{i=1}^k\phi_i(x)\right),
\end{aligned}
\end{equation}
where  $\hat{{\mathcal F}}^{(k)}$ is the $k$-fold twist defined in~\eqref{eq:n-F}. To represent this, we will modify the Feynman rules to account for the additional twists acting on the individual legs of a vertex. In particular,  following \cite{Meier:2023lku}, we introduce twist lines that connect different legs. On them,  we will use filled and unfilled boxes to indicate the two different spaces in which the twist is acting; cf. Figure \ref{fig:TwistRules} for a diagrammatic definition of these twist lines. With this additional structure, we can give the fundamental Feynman rules in Figure \ref{fig:FeynmanRules} to build Feynman diagrams.

At this point we can finally note that there are three important properties satisfied by the  integrand of the twisted interaction terms:
\begin{enumerate}

    \item 
Going from the undeformed to the deformed theory amounts just to ``dressing'' the  integrand of the undeformed interaction term with the ``$k$-fold twist''  
\begin{equation}
    \bigotimes_{i=1}^k\phi_i(x)\qquad\to\qquad
    \hat{{\mathcal F}}^{(k)}\bigotimes_{i=1}^k\phi_i(x).
\end{equation}

\item 
The interaction terms of the undeformed theory are, by assumption, invariant under the action of elements $\U$ of the universal enveloping algebra. As proved in section~\ref{sec:twisted-symm}, the symmetry remains even in the presence of the deformation, so that we may write
\begin{equation}
    \hat\U\left(\phi_1(x)\hstar\phi_2(x)\hstar\cdots\hstar\phi_k(x)\right)\simeq 0,
\end{equation}
which more explicitly is
\begin{equation}
    \mu^{(k)}\left(  \hat{{\mathcal F}}^{(k)}\Delta^{(k)}(\hat\U)\bigotimes_{i=1}^k\phi_i(x)\right)\simeq 0.
\end{equation}
Here we use the symbol $\simeq$  rather than $=$, because  in general the result, instead of vanishing, could be a total derivative, so that the invariance may hold only after integration. From the above identity it also follows the ``integration by parts'' identity
\begin{equation}
    \hat\U_1\left(\phi_1(x)\hstar\phi_2(x)\hstar\cdots\hstar\phi_k(x)\right)\simeq S(\hat\U)_{2\cdots k}\left(\phi_1(x)\hstar\phi_2(x)\hstar\cdots\hstar\phi_k(x)\right),
\end{equation}
where the subindices on the elements of the universal enveloping algebra indicate the fields on which they act. More explicitly, the above equation is
\begin{equation}\label{eq:inv-vert}
    \mu^{(k)}\left(   \hat{{\mathcal F}}^{(k)}\ (\hat\U\otimes id_{k-1})\bigotimes_{i=1}^k\phi_i(x)\right)\simeq \mu^{(k)}\left(  \hat{{\mathcal F}}^{(k)}\ (id\otimes \Delta^{(k-1)}(S(\hat\U))\bigotimes_{i=1}^k\phi_i(x)\right).
\end{equation}
In fact, on the right-hand-side we are simply using~\eqref{eq:UmuF-active}.

Of course, the above invariance is related to the fact that one may replace one star product by the ordinary product, at the cost of introducing the action of $V^{-1}$. For example, eliminating the last star product one may write 
\begin{equation}
    \phi_1(x)\hstar\phi_2(x)\hstar\cdots\hstar\phi_k(x)\simeq\phi_1(x)\hstar\cdots\hstar\phi_{k-1}(x)\ \hat V^{-1}(\phi_k(x)), 
\end{equation}
which may also be written as
\begin{equation}\label{eq:inv-vert-V}
    \mu^{(k)}\left(  \hat{{\mathcal F}}^{(k)}\bigotimes_{i=1}^k\phi_i(x)\right)\simeq\mu^{(k)}\left( (\hat{{\mathcal F}}^{(k-1)}\otimes \hat V^{-1})\bigotimes_{i=1}^k\phi_i(x)\right).
\end{equation}

\item 
The final property is that the integrand of the twisted interaction term can be taken to be cyclic
\begin{equation}
     \phi_1(x)\hstar\phi_2(x)\hstar\cdots\hstar\phi_k(x)\simeq \phi_k(x)\hstar\phi_1(x)\hstar\phi_2(x)\hstar\cdots\hstar\phi_{k-1}(x)
\end{equation}
or equivalently
\begin{equation}
   \mu^{(k)}\left(  \hat{{\mathcal F}}^{(k)}\bigotimes_{i=1}^k\phi_i(x)\right) \simeq  \mu^{(k)}\left( \hat{{\mathcal F}}^{(k)}\left(\phi_k(x)\otimes \bigotimes_{i=1}^{k-1}\phi_i(x)\right)\right).
\end{equation}
Here we use again the symbol $\simeq$ because in general the integrand of the twisted interaction term may not be cyclic. The cyclicity property will however hold after taking the integral, also thanks to the $\mathcal R$-unimodularity condition $S(\hat V)=\hat V$. In fact,  all interaction terms appear under integration.

\end{enumerate}

In the rest of this section we will not use the symbol $\simeq$ any more, we will simply write $=$.

\subsubsection{Tree-level diagrams}
We now want to use the fundamental interaction terms of the twisted theory to construct all possible tree-level diagrams. We will start by considering an interaction of order $k+1$ at position $x$ and one of order $l+1$ at position $y$, so that by contracting the two interaction terms by one propagator one obtains a diagram with $k+l$ legs. We will write $\phi_1(x),\ldots,\phi_{k+1}(x)$ and $\phi_{ 1}(y),\ldots,\phi_{ l+1}(y)$.

\begin{figure}[h]
\begin{center}
\includegraphics[width=0.9\textwidth]{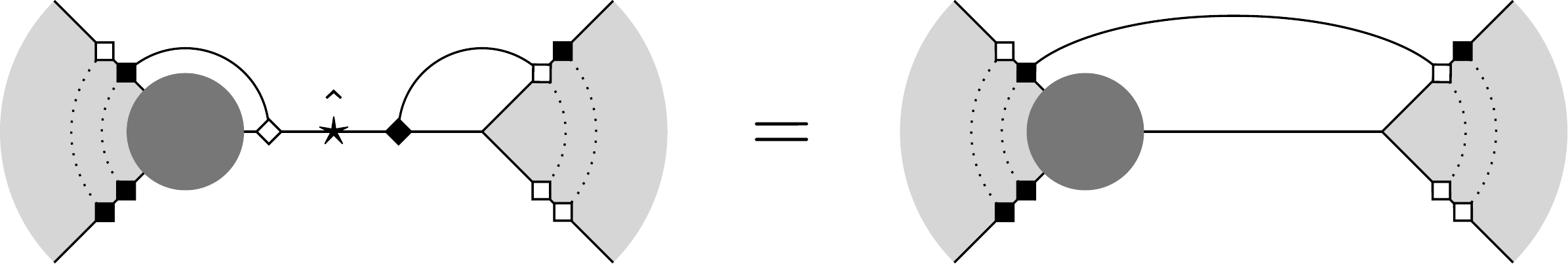}
\caption{Adding a vertex to an existing planar diagram. On the left-hand-side of the equation, adding a new vertex to an already existing planar diagram results in two series of star products: one located at the new vertex (on the right) and one on the old diagram (on the left). Furthermore, the contraction between the two fields leads to a deformed propagator connecting both. Light-grey areas denote arbitrary many propagators, and dotted twist lines denote the action of the twist on all lines in between. As described in the main text, it can be shown that this is equal to twists only acting on the external fields in an undeformed diagram, as given on the right-hand-side of the picture.}
\label{fig:AddVertex}
\end{center}
\end{figure}

\paragraph{Dressing of the uncontracted fields by the twist.}
Because the interaction terms are cyclic, without loss of generality we may consider the diagram obtained by contracting the last fields in each list,~i.e. $\phi_{k+1}(x)$ and $\phi_{ l+1}(y)$. 
The statement that we will now prove is that this diagram is given just by the following expression
\begin{equation}\label{eq:fusion-vertices}
    \phi_1(x)\hstar\cdots\hstar \phi_{k}(x)\hstar\phi_{ 1}(y)\hstar\cdots\hstar\phi_{ l}(y) \ \Delta_F(x-y).
\end{equation}
In other words, we simply have the star product of the uncontracted fields multiplying the \emph{undeformed} propagator, i.e.~the one that would appear in the undeformed theory when contracting $\phi_{k+1}(x)$ and $\phi_{ l+1}(y)$, cf. Figure \ref{fig:AddVertex}.

To prove this, it is convenient to start with each of the two interaction terms written as in~\eqref{eq:inv-vert-V}, i.e.~where a star product is  replaced by $V^{-1}$, and we do it so that $\phi_{k+1}(x)$ and $\phi_{ l+1}(y)$ are singled out. Although it does not correspond to an interaction term, we will first show the proof for the simpler case $k=l=1$, because it contains the essential points of the argument. We have
\begin{equation}
    \begin{aligned}
         \phi(x)\phi(y)\hat V_x^{-1}\, \hat V_y^{-1}\, \Delta^{\hstar}_F(x,y)
        =&\phi(x)\phi(y) \hat V_y^{-1}\, \Delta_F(x-y)\\
        =&\phi(x)\phi(y) \hat{ f}^\alpha_yS(\hat{ f}_{\alpha y})\, \Delta_F(x-y)\\
       =&\phi(x)\phi(y) S(\hat{ f}^\alpha_x)S(\hat{ f}_{\alpha y})\, \Delta_F(x-y)\\
        =&\hat{ f}^\alpha_x\phi(x)\hat{ f}_{\alpha y}\phi(y) \, \Delta_F(x-y)\\
        =&\phi(x)\hstar\phi(y) \ \Delta_F(x-y),
    \end{aligned}
\end{equation}
where we  used the definition of the twisted propagator, the definition of $V^{-1}$,  the invariance of the propagator to rewrite $\hat { f}^\alpha_y$ as $S(\hat { f}^\alpha_x)$,  the symmetry invariance~\eqref{eq:inv-vert} both in $x$ and $y$, and  finally the definition of the star product.
In the more general case one has
\begin{equation}
    \begin{aligned}
        &\mu^{(k+l)}\left( \left(\hat{{\mathcal F}}^{(k)}\otimes \hat{{\mathcal F}}^{(l)}\right)\bigotimes_{i=1}^{k}\phi_i(x)\bigotimes_{i=1}^{l}\phi_i(y)\right)\hat V_x^{-1}\, \hat V_y^{-1}\, \Delta^{\hstar}_F(x-y)\\
        =&\mu^{(k+l)}\left( \left(\hat{{\mathcal F}}^{(k)}\otimes \hat{{\mathcal F}}^{(l)}\right)\bigotimes_{i=1}^{k}\phi_i(x)\bigotimes_{i=1}^{l}\phi_i(y)\right) \hat V_y^{-1}\, \Delta_F(x-y)\\
        =&\mu^{(k+l)}\left( \left(\hat{{\mathcal F}}^{(k)}\otimes \hat{{\mathcal F}}^{(l)}\right)\bigotimes_{i=1}^{k}\phi_i(x)\bigotimes_{i=1}^{l}\phi_i(y)\right) \hat { f}^\alpha_yS(\hat { f}_{\alpha y})\, \Delta_F(x-y)\\
        =&\mu^{(k+l)}\left( \left(\hat{{\mathcal F}}^{(k)}\otimes \hat{{\mathcal F}}^{(l)}\right)\bigotimes_{i=1}^{k}\phi_i(x)\bigotimes_{i=1}^{l}\phi_i(y)\right) S(\hat { f}^\alpha_x)S(\hat { f}_{\alpha y})\, \Delta_F(x-y)\\
        =&\mu^{(k+l)}\left(\left(\hat{{\mathcal F}}^{(k)}\otimes \hat{{\mathcal F}}^{(l)}\right)\left(\Delta^{(k)}(\hat { f}^\alpha)\otimes \Delta^{(l)}(\hat { f}_\alpha)\right)\bigotimes_{i=1}^{k}\phi_i(x)\bigotimes_{i=1}^{l}\phi_i(y)\right)  \, \Delta_F(x-y)\\
        =&\mu^{(k+l)}\left(  \left(\hat{{\mathcal F}}^{(k)}\otimes \hat{{\mathcal F}}^{(l)}\right)\left(\Delta^{(k)}\otimes \Delta^{(l)}\right)(\hat{{\mathcal F}})\ \bigotimes_{i=1}^{k}\phi_i(x)\bigotimes_{i=1}^{l}\phi_i(y)\right) \, \Delta_F(x-y)\\
        =&(\phi_1(x)\hstar\cdots\hstar \phi_{k}(x))\hstar(\phi_{ 1}(y)\hstar\cdots\hstar\phi_{ l}(y)) \ \Delta_F(x-y).
    \end{aligned}
\end{equation}
Compared to the previous calculation, now when using the symmetry invariance~\eqref{eq:inv-vert}  in $x$ and $y$ we need to use the $k$-fold and $l$-fold coproduct; moreover, in the last step, we used~\eqref{eq:Fn-FkFl} to rewrite the $(k+l)$-fold twist and to conclude that we are left with the star product acting on all uncontracted fields at the same time.

We have therefore proved the first point of the planar equivalence theorem for this tree-level diagram, namely the dressing of the undeformed result by the relevant twist.

\begin{figure}[h]
\begin{center}
\includegraphics[width=0.9\textwidth]{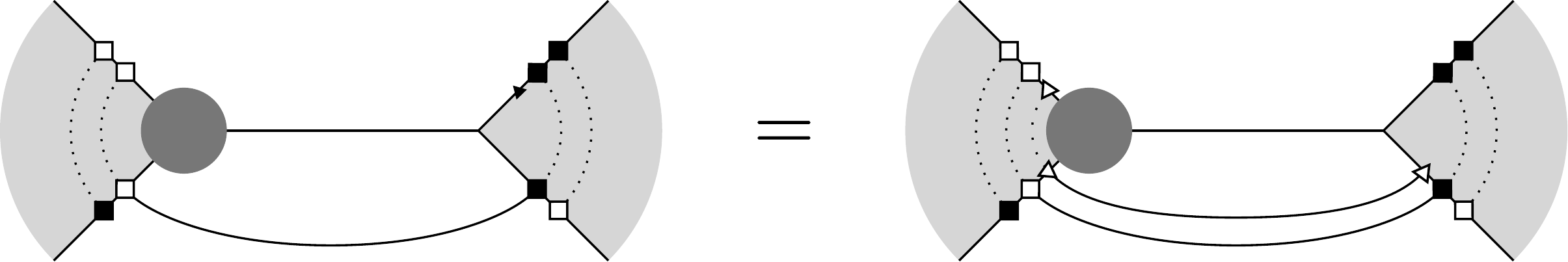}
\caption{Invariance of the deformed diagram. A given element of $U(\mathfrak{g})$ (here denoted by a black triangle) acting on one leg of the new extended Feynman diagram is equivalent to acting with the antipode of that element (here denoted by a white triangle) on all other legs instead.}
\label{fig:Invariance}
\end{center}
\end{figure}

\paragraph{Symmetry invariance.}
The second property appearing in the theorem is the symmetry invariance. In particular, we can prove   the identity
\begin{equation}
    \begin{aligned}
    &\hat\U_l\left(\phi_1(x)\hstar\cdots\hstar \phi_{k}(x)\hstar\phi_{ 1}(y)\hstar\cdots\hstar\phi_{ l}(y)\right) \ \Delta_F(x-y)\\
    =&S(\hat\U)_{1\cdots k1\cdots l-1}\left(\phi_1(x)\hstar\cdots\hstar \phi_{k}(x)\hstar\phi_{ 1}(y)\hstar\cdots\hstar\phi_{ l}(y)\right)\ \Delta_F(x-y),
    \end{aligned}
\end{equation}
or more explicitly
\begin{equation}
    \begin{aligned}
        &\mu^{(k+l)}\left( \hat{{\mathcal F}}^{(k+l)}\left(id^{(k)}\otimes id^{(l-1)}\otimes \hat\U\right)\bigotimes_{i=1}^{k}\phi_i(x)\bigotimes_{i=1}^{l}\phi_i(y)\right)\ \Delta_F(x-y)\\
        =&\mu^{(k+l)}\left( \hat{{\mathcal F}}^{(k+l)}\left(\Delta^{(k+l-1)}(S(\hat\U))\otimes id\right)\bigotimes_{i=1}^{k}\phi_i(x)\bigotimes_{i=1}^{l}\phi_i(y)\right)\ \Delta_F(x-y),
    \end{aligned}
\end{equation}
see Figure \ref{fig:Invariance} for a diagrammatical expression.

Notice two important points. First, the propagator plays the role of a spectator, since the elements of the universal enveloping algebra are not acting on it, despite the fact that it does depend on the coordinates $x,y$. Second, when moving the action of $\hat\U$ to the rest of the fields, this works irrespectively of the coordinate dependence; in this example, on the left-hand-side of the equation, $\hat\U$ acts only on a field depending on $y$, but on the right-hand-side of the equation $S(\hat\U)$ acts on all remaining fields depending on $y$ \emph{and} on all fields depending on $x$. 

As before, the essence of the proof may be understood in the simple case when $k=l=1$. Then
\begin{equation}
    \begin{aligned}
        &\mu\left( \hat{{\mathcal F}}\left(id\otimes \hat\U\right)\left(\phi(x)\otimes \phi(y)\right)\right)\ \Delta_F(x-y)\\
        =&\mu\left( \hat{{\mathcal F}}\left(\phi(x)\otimes \phi(y)\right)\right)\ S(\hat\U)_{y}\Delta_F(x-y)\\
        =&\mu\left( \hat{{\mathcal F}}\left(\phi(x)\otimes \phi(y)\right)\right)\ \hat\U_{x}\Delta_F(x-y)\\
        =&\mu\left( \hat{{\mathcal F}}\left(S(\hat\U)\otimes id\right)\left(\phi(x)\otimes \phi(y)\right)\right)\ \Delta_F(x-y),
    \end{aligned}
\end{equation}
where we first used integration by parts in $y$, then we used the invariance of the propagator, and then we used integration by parts in $x$.
For general $k,l$ the idea of the calculation is the same, although one needs to pay attention to all the fields on which the symmetry elements are acting. One has
\begin{equation}
    \begin{aligned}
        &\mu^{(k+l)}\left(\hat{{\mathcal F}}^{(k+l)} \left(id^{(k)}\otimes id^{(l-1)}\otimes \hat\U\right)\bigotimes_{i=1}^{k}\phi_i(x)\bigotimes_{i=1}^{l}\phi_i(y)\right)\ \Delta_F(x-y)\\
        =&\mu^{(k+l)}\left(\hat{{\mathcal F}}^{(k+l)} \left(id^{(k)}\otimes \Delta^{(l-1)}(S(\hat\U)_{(1)})\otimes id\right)\bigotimes_{i=1}^{k}\phi_i(x)\bigotimes_{i=1}^{l}\phi_i(y)\right)\ S(\hat\U)_{(2)y}\Delta_F(x-y)\\
        =&\mu^{(k+l)}\left( \hat{{\mathcal F}}^{(k+l)}\left(id^{(k)}\otimes \Delta^{(l-1)}(S(\hat\U)_{(1)})\otimes id\right)\bigotimes_{i=1}^{k}\phi_i(x)\bigotimes_{i=1}^{l}\phi_i(y)\right)\ \hat\U_{(2)x}\Delta_F(x-y)\\
        =&\mu^{(k+l)}\left( \hat{{\mathcal F}}^{(k+l)}\left(\Delta^{(k)}(S(\hat\U_{(2)}))\otimes \Delta^{(l-1)}(S(\hat\U)_{(1)})\otimes id\right)\bigotimes_{i=1}^{k}\phi_i(x)\bigotimes_{i=1}^{l}\phi_i(y)\right)\ \Delta_F(x-y)\\
        =&\mu^{(k+l)}\left( \hat{{\mathcal F}}^{(k+l)}\left(\Delta^{(k+l-1)}(S(\hat\U))\otimes id\right)\bigotimes_{i=1}^{k}\phi_i(x)\bigotimes_{i=1}^{l}\phi_i(y)\right)\ \Delta_F(x-y).
    \end{aligned}
\end{equation}
The explanation of the steps in the first lines is as in the simplified $k=l=1$ case; now, however, we also have an additional final step where, using that the undeformed coproduct is co-commutative ($\Delta=\Delta_{op}$), we  notice that the expression appearing there can be understood as the $(k+l-1)$-fold coproduct.

\begin{figure}[h]
\begin{center}
\includegraphics[width=0.9\textwidth]{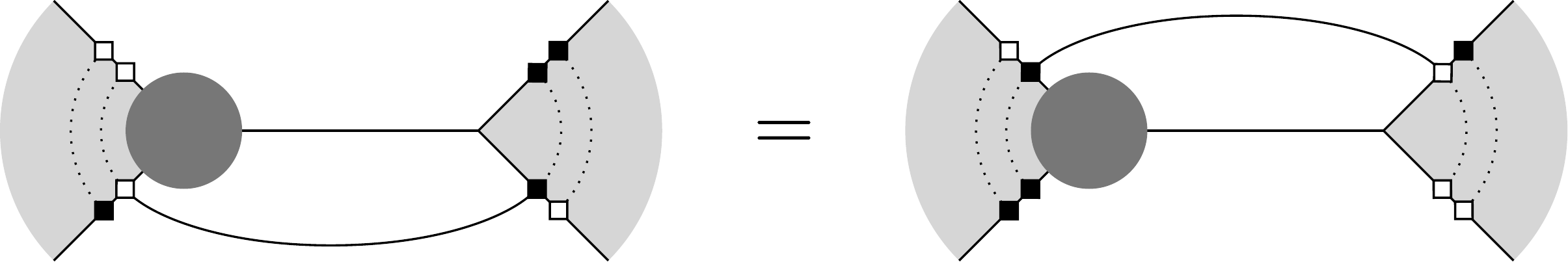}
\caption{Cyclicity of the deformed diagram. A given orientation of the star product of the new extended Feynman diagram can be changed cyclically.}
\label{fig:Cyclicity}
\end{center}
\end{figure}

\paragraph{Cyclicity.}
 The third and last property to prove is cyclicity, for example 
\begin{equation}
    \begin{aligned}
    &\phi_1(x)\hstar\cdots\hstar \phi_{k}(x)\hstar\phi_{ 1}(y)\hstar\cdots\hstar\phi_{ l}(y) \ \Delta_F(x-y)\\
    =&\phi_{ l}(y)\hstar\phi_1(x)\hstar\cdots\hstar \phi_{k}(x)\hstar\phi_{ 1}(y)\hstar\cdots\hstar\phi_{ l- 1}(y) \ \Delta_F(x-y),
    \end{aligned}
\end{equation}
see Figure \ref{fig:Cyclicity} for a diagrammatic expression.
At first sight this is not obvious, because one should remember that the cyclicity property of the star product involves integration by parts that give rise to $\hat V^{-1}$, and in this case one may naively expect the propagator $\Delta_F(x-y)$ to interfere with the calculation. However, thanks to the symmetry invariance property just proved, we know that we can essentially treat the propagator as a spectator. 
Then the proof of cyclicity follows in the same way as in the case of the classical action
\begin{equation}
    \begin{aligned}
         & \left(\phi_1(x)\hstar\cdots\hstar \phi_{k}(x)\hstar\phi_{ 1}(y)\hstar\cdots\hstar\phi_{ l-1}(y) \right)\hstar\phi_{ l}(y)\ \Delta_F(x-y)\\
        =&\mu\left(\hat { f}^\alpha \left(\phi_1(x)\hstar\cdots\hstar \phi_{k}(x)\hstar\phi_{ 1}(y)\hstar\cdots\hstar\phi_{ l-1}(y) \right)\otimes \hat { f}_\alpha\phi_{ l}(y)\right)\ \Delta_F(x-y)\\
        =&\mu\left( \left(\phi_1(x)\hstar\cdots\hstar \phi_{k}(x)\hstar\phi_{ 1}(y)\hstar\cdots\hstar\phi_{ l-1}(y) \right)\otimes \hat { f}_\alpha S(\hat { f}^\alpha) \phi_{ l}(y)\right)\ \Delta_F(x-y)\\
        =&\mu\left( \left(\phi_1(x)\hstar\cdots\hstar \phi_{k}(x)\hstar\phi_{ 1}(y)\hstar\cdots\hstar\phi_{ l-1}(y) \right)\otimes \hat { f}^\alpha S(\hat { f}_\alpha)\phi_{ l}(y)\right)\ \Delta_F(x-y)\\
        =&\mu\left( \hat { f}_\alpha\left(\phi_1(x)\hstar\cdots\hstar \phi_{k}(x)\hstar\phi_{ 1}(y)\hstar\cdots\hstar\phi_{ l-1}(y) \right)\otimes \hat { f}^\alpha\phi_{ l}(y)\right)\ \Delta_F(x-y)\\
        =&\mu\left(\hat { f}^\alpha\phi_{ l}(y)\otimes \hat { f}_\alpha\left(\phi_1(x)\hstar\cdots\hstar \phi_{k}(x)\hstar\phi_{ 1}(y)\hstar\cdots\hstar\phi_{ l-1}(y) \right)\right)\ \Delta_F(x-y)\\
         =&\phi_{ l}(y) \hstar\left(\phi_1(x)\hstar\cdots\hstar \phi_{k}(x)\hstar\phi_{ 1}(y)\hstar\cdots\hstar\phi_{ l- 1}(y)\right)\ \Delta_F(x-y).
    \end{aligned}
\end{equation}
Here we used the definition of the star product and the usual rewriting of the twist; then we used the symmetry invariance already proved, then we noticed that $\hat { f}_\alpha S(\hat { f}^\alpha)=\hat { f}^\alpha S(\hat { f}_\alpha)$ thanks to the $\mathcal R$-unimodularity condition; then we used symmetry invariance again; then we used the commutativity of $\mu$; and finally we applied the definition of the star product.

Alternatively, another way to prove cyclicity is to show that even in the presence of the propagator one can replace one star product by $\hat V^{-1}$,~i.e.
\begin{equation}\label{eq:star-x-y-V-Delta}
    \begin{aligned}
         & \phi_1(x)\hstar\cdots\hstar \phi_{k}(x)\hstar\phi_{ 1}(y)\hstar\cdots\hstar\phi_{ l}(y) \ \Delta_F(x-y)\\
         =&\phi_1(x)\hstar\cdots\hstar \phi_{k}(x)\hstar\phi_{ 1}(y)\hstar\cdots\hstar\phi_{ l- 1}(y)\ \hat V^{-1}(\phi_{ l}(y)) \ \Delta_F(x-y).
    \end{aligned}
\end{equation}
The proof follows again the same logic, and it goes as follows. If, as done before, we order fields in the tensor product such that we first have the $k$ fields evaluated in $x$ and then the $l$ fields evaluated in $y$, then we have
\begin{equation}\label{eq:proof-star-x-y-V-Delta}
    \begin{aligned}
         & \phi_1(x)\hstar\cdots\hstar \phi_{k}(x)\hstar\phi_{ 1}(y)\hstar\cdots\hstar\phi_{ l}(y) \ \Delta_F(x-y)\\
         =&\mu^{(k+l)}\left(\left(\hat {{\mathcal F}}^{(k+l-1)}\otimes 1\right)\left(\Delta^{(k+l-1)}\otimes 1\right)(\hat {{\mathcal F}})\bigotimes \phi_i\right)\Delta_F(x-y)\\
         =&\mu^{(k+l)}\left(\left(\hat {{\mathcal F}}^{(k+l-1)}\otimes 1\right)\left(\Delta^{(k)}(\hat { f}^\alpha_{(1)})\otimes \Delta^{(l-1)}(\hat { f}^\alpha_{(2)})\otimes \hat { f}_\alpha\right)\bigotimes \phi_i\right)\Delta_F(x-y)\\
         =&\mu^{(k+l)}\left(\left(\hat {{\mathcal F}}^{(k+l-1)}\otimes 1\right)\left(id^{(k)}\otimes \Delta^{(l-1)}(\hat { f}^\alpha_{(2)})\otimes \hat { f}_\alpha\right)\bigotimes \phi_i\right)\, S(\hat { f}^\alpha_{(1)x})\Delta_F(x-y)\\
         =&\mu^{(k+l)}\left(\left(\hat {{\mathcal F}}^{(k+l-1)}\otimes 1\right)\left(id^{(k)}\otimes \Delta^{(l-1)}(\hat { f}^\alpha_{(2)})\otimes \hat { f}_\alpha\right)\bigotimes \phi_i\right)\, \hat { f}^\alpha_{(1)y}\Delta_F(x-y)\\
         =&\mu^{(k+l)}\left(\left(\hat {{\mathcal F}}^{(k+l-1)}\otimes 1\right)\left(id^{(k)}\otimes id^{(l-1)}\otimes \hat { f}_\alpha S(\hat { f}^\alpha)\right)\bigotimes \phi_i\right)\, \Delta_F(x-y)\\
         =&\phi_1(x)\hstar\cdots\hstar \phi_{k}(x)\hstar\phi_{ 1}(y)\hstar\cdots\hstar\phi_{ l- 1}(y)\ \hat V^{-1}(\phi_{ l}(y)) \ \Delta_F(x-y).
    \end{aligned}
\end{equation}
In the above proof we first used the rewriting of the $n$-fold twist as in~\eqref{eq:n-F}; then we used the fact that we can write $\Delta^{(k+l-1)}=(\Delta^{(k)}\otimes\Delta^{(l-1)})\Delta$; then we performed integration by parts in $x$, acting then on the propagator; the we used invariance of the twist; finally, we performed integration by parts in $y$ after realising that having $\hat { f}^\alpha_{(1)y}$ acting on the propagator and $\Delta^{(l-1)}(\hat { f}^\alpha_{(2)})$ on the first $l-1$ fields can again be understood as $\Delta^{(l)}(\hat { f}^\alpha)$ acting on the tensor product of all these objects, with the propagator standing first.

\paragraph{More general tree-level diagrams.}
We may now consider more contractions of the same interaction terms, or even contractions with other diagrams to construct the most general tree-level diagrams. We can immediately conclude that also the most general tree-level diagrams will satisfy the three properties of the planar equivalence theorem. In fact, to prove that the planar equivalence theorem holds for just one  contraction, we used properties of the fundamental interaction terms; but we also proved that the resulting diagrams do satisfy the same properties. This means that the proof may be iterated for contractions of more complicated diagrams, and then the planar equivalence theorem holds for all possible tree-level diagrams.

In particular, in a given tree-level diagram, we may collect all propagators coming from contractions and connecting internal vertices into a function $I_n$, where $n$ stands for the number of internal vertices. Importantly, $I_n$ will be the same as in the undeformed setup, because we already proved that $\Delta_F(x-y)$ and not $\Delta_F^{\hstar}(x-y)$ appears. A tree-level diagram with $k$ external legs will then be of the form
\begin{equation}\label{eq:tree-level-diagr}
    I_n \cdot\phi_1(x_1)\hstar\phi_1(x_2)\hstar\cdots\hstar \phi_{k}(x_k),
\end{equation}
where in general the uncontracted fields may be evaluated at different points. This object, then, is manifestly the natural twisted version of the undeformed one, and it will be invariant under the symmetries and cyclic as a consequence of the theorem.

\subsubsection{Loop diagrams}
\begin{figure}[h]
\begin{center}
\includegraphics[width=0.7\textwidth]{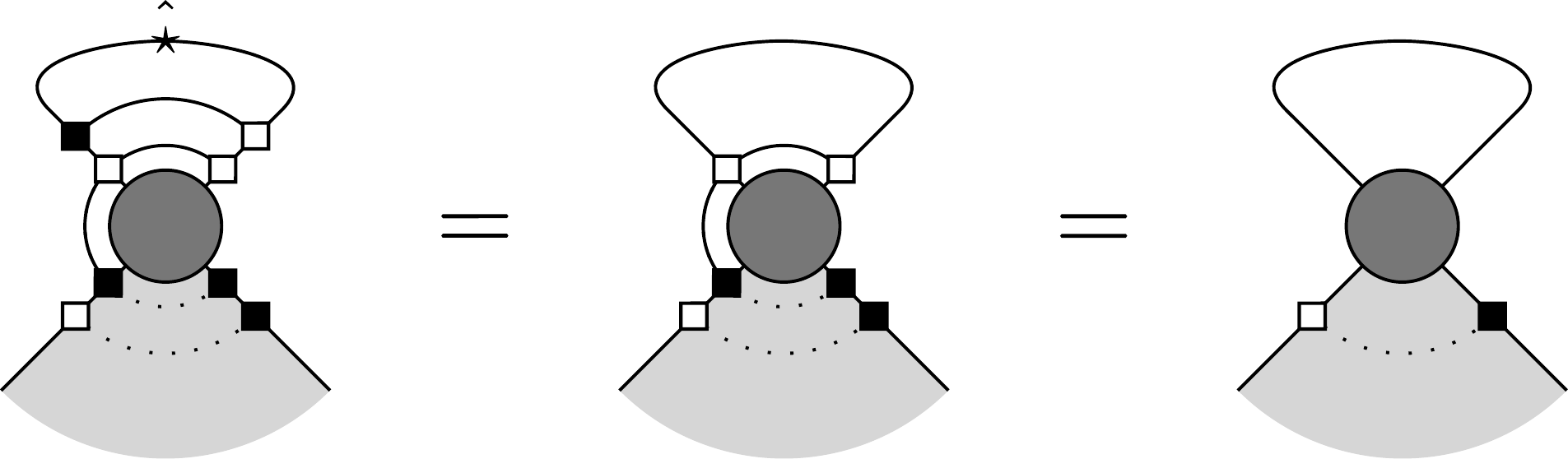}
\caption{Contracting two neighbouring fields in a given planar Feynman diagram leads to the left-hand-side expression. Using the relations involving the deformed propagator and the twist leads to the right-hand-side.}
\label{fig:CloseLoops}
\end{center}
\end{figure}

Loop diagrams are built by connecting legs via propagators. 
We suppose to start from a tree-level diagram of the form~\eqref{eq:tree-level-diagr}, albeit with $k+2$ legs, so that by contracting two of them we are left with $k$ legs. Because we only want to construct planar diagrams, we can only contract adjacent fields. To start, we  take for example the diagram obtained by contracting the last two fields $\phi_{k+1}(x),\phi_{k+2}(x)$.
We will now prove that after the contraction we simply have
\begin{equation}\label{eq:loop-diagr}
    I_n \cdot\phi_1(x_1)\hstar\phi_1(x_2)\hstar\cdots\hstar \phi_{k}(x_k)\cdot \Delta_F(x_{k+1}-x_{k+2}),
\end{equation}
meaning that once again we have just the undeformed expression (notice the undeformed propagator $\Delta_F(x_{k+1}-x_{k+2})$ and the internal structure $I_n$) dressed with the relevant twist on the uncontracted fields, see also Figure \ref{fig:CloseLoops} for a diagrammatic expression. 

To prove the above statement, it is  convenient to  rewrite the tree level diagram by first grouping the first $k$ fields and then the last two
\begin{equation}
    \begin{aligned}
        &I_n \cdot\left(\phi_1(x_1)\hstar\phi_1(x_2)\hstar\cdots\hstar \phi_{k}(x_k)\right)\hstar \left(\phi_{k+1}(x_{k+1})\hstar \phi_{k+2}(x_{k+2})\right)\\
        =&I_n \cdot\left(\phi_1(x_1)\hstar\phi_1(x_2)\hstar\cdots\hstar \phi_{k}(x_k)\right)\ \hat V^{-1} \left(\phi_{k+1}(x_{k+1})\hstar \phi_{k+2}(x_{k+2})\right)\\
       =&I_n \cdot\left(\phi_1(x_1)\hstar\phi_1(x_2)\hstar\cdots\hstar \phi_{k}(x_k)\right)\ \mu \left(\hat{\mathcal F}\Delta(\hat V^{-1})(\phi_{k+1}(x_{k+1})\otimes \phi_{k+2}(x_{k+2}))\right).
    \end{aligned}
\end{equation}
After preparing the tree-level diagram as above, we may now contract the last two fields. Given that the first part of the expression $I_n \cdot\left(\phi_1(x_1)\hstar\phi_1(x_2)\hstar\cdots\hstar \phi_{k}(x_k)\right)$ plays the role of spectator in the calculation, we will ignore it and simply focus on the last factor. After the contraction we have
\begin{equation}
    \begin{aligned}
         \hat {{\mathcal F}}_{k+1,k+2}\ \hat V^{-1}_{(1)k+1}\hat V^{-1}_{(2)k+2}\ \Delta^{\hstar}_F(x_{k+1}-x_{k+2})
        &=\hat V^{-1}_{(1)k+1}\hat V^{-1}_{(2)k+2}\ \Delta_F(x_{k+1}-x_{k+2})\\
        &=S(\hat V^{-1}_{(1)})_{k+2}\hat V^{-1}_{(2)k+2}\ \Delta_F(x_{k+1}-x_{k+2})\\
        &=\epsilon(\hat V^{-1})_{k+2}\ \Delta_F(x_{k+1}-x_{k+2})\\
        &=\Delta_F(x_{k+1}-x_{k+2}).
    \end{aligned}
\end{equation}
Here we first used the identity~\eqref{eq:FDeltastar=Delta}; then we used the invariance of the twist to bring the action of the element from $k+1$ to $k+2$; then, after noticing that $S(\hat V^{-1}_{(1)})\hat V^{-1}_{(2)}=m(S\otimes 1)\Delta(\hat V^{-1})$, we used the antipode relation; and finally we used $\epsilon(\hat V^{-1})=1$ which is a consequence of the fact that we consider twists that are deformations of the identity.

We have just proved the first point of the planar equivalence theorem.  The proofs for the remaining two properties follow the same logic as above. First, symmetry invariance implies
\begin{equation}
    \begin{aligned}
        &\hat \U_k\left(I_n \cdot\phi_1(x_1)\hstar\phi_1(x_2)\hstar\cdots\hstar \phi_{k}(x_k)\cdot \Delta_F(x_{k+1}-x_{k+2})\right)\\
        =&S(\hat \U)_{1\cdots k-1}\left(I_n \cdot\phi_1(x_1)\hstar\phi_1(x_2)\hstar\cdots\hstar \phi_{k}(x_k)\cdot \Delta_F(x_{k+1}-x_{k+2})\right).
    \end{aligned}
\end{equation}
In fact, singling out the product between the fields and the propagator $\Delta_F(x_{k+1}-x_{k+2})$ one has
\begin{equation}
    \begin{aligned}
        &\hat \U_k\left(I_n \cdot\phi_1(x_1)\hstar\phi_1(x_2)\hstar\cdots\hstar \phi_{k}(x_k)\cdot \Delta_F(x_{k+1}-x_{k+2})\right)\\
        =&I_n \cdot S(\hat \U)_{(1)1\cdots k-1}\left(\phi_1(x_1)\hstar\phi_1(x_2)\hstar\cdots\hstar \phi_{k}(x_k)\right)\cdot S(\hat \U)_{(2)k+1,k+2}\Delta_F(x_{k+1}-x_{k+2}),
    \end{aligned}
\end{equation}
but if we focus on the last factor we have
\begin{equation}
    \begin{aligned}
        S(\hat \U)_{(2)k+1,k+2}\Delta_F(x_{k+1}-x_{k+2})
        &=S(\hat \U)_{(2)(1)k+1}S(\hat \U)_{(2)(2)k+2}\Delta_F(x_{k+1}-x_{k+2})\\
        &=S(S(\hat \U)_{(2)(1)})_{k+2}S(\hat \U)_{(2)(2)k+2}\Delta_F(x_{k+1}-x_{k+2})\\
        &=\epsilon(S(\hat \U)_{(2)})_{k+2}\Delta_F(x_{k+1}-x_{k+2}),
    \end{aligned}
\end{equation}
where we first used~\eqref{eq:Umu}, then the invariance of the propagator and finally the antipode condition, after noticing that $S(S(\hat \U)_{(2)(1)})S(\hat \U)_{(2)(2)}=m(S\otimes 1)\Delta(S(\hat \U)_{(2)})$.
At this point, we notice that $S(\hat\U)_{(1)}\epsilon(S(\hat\U)_{(2)})=(id\otimes\epsilon)\Delta(S(\hat\U))=S(\hat\U)$ as $\epsilon$ is the counit of the Hopf algebra. This then proves the statement. 

Thanks to the fact that, as we just proved, the propagator appears to be  a spectator when doing integrations by parts, it is straightforward to prove cyclicity following exactly the same steps as before
\begin{equation}
    \begin{aligned}
         &I_n \cdot\phi_1(x_1)\hstar\phi_1(x_2)\hstar\cdots\hstar \phi_{k}(x_k)\cdot \Delta_F(x_{k+1}-x_{k+2})\\
         =& I_n \cdot\hat { f}^\alpha \left(\phi_1(x_1)\hstar\phi_1(x_2)\hstar\cdots\hstar \phi_{k-1}(x_{k-1})\right)\ \hat { f}_\alpha(\phi_{k}(x_k))\cdot \Delta_F(x_{k+1}-x_{k+2})\\
         =& I_n \cdot \hat { f}^\alpha S(\hat { f}_\alpha)\left(\phi_1(x_1)\hstar\phi_1(x_2)\hstar\cdots\hstar \phi_{k-1}(x_{k-1})\right)\cdot \phi_{k}(x_k)\cdot \Delta_F(x_{k+1}-x_{k+2})\\
         =& I_n \cdot \hat { f}_\alpha S(\hat { f}^\alpha)\left(\phi_1(x_1)\hstar\phi_1(x_2)\hstar\cdots\hstar \phi_{k-1}(x_{k-1})\right)\cdot \phi_{k}(x_k)\cdot \Delta_F(x_{k+1}-x_{k+2})\\
         =& I_n \cdot  \hat { f}_\alpha\left(\phi_1(x_1)\hstar\phi_1(x_2)\hstar\cdots\hstar \phi_{k-1}(x_{k-1})\right) \hat { f}^\alpha( \phi_{k}(x_k))\cdot \Delta_F(x_{k+1}-x_{k+2})\\
         =&I_n \cdot\phi_{k}(x_k)\hstar\phi_1(x_1)\hstar\phi_1(x_2)\hstar\cdots\hstar \phi_{k-1}(x_{k-1})\cdot \Delta_F(x_{k+1}-x_{k+2}),
    \end{aligned}
\end{equation}
where we used the definition of the twist, integration by parts and the $\mathcal R$-unimodularity condition.

To conclude, also when contracting legs to form a loop one preserves the properties valid in the planar equivalence theorem. This obviously can be iterated to have as many loops as desired. All the propagators resulting from these loops can then be collected into the internal part $I_n$ which, as previously observed,  remains undeformed.

In the above calculations,  we have considered the contraction of the last two fields appearing in the star product. Importantly, thanks to the cyclicity property, this implies that the result will be the same if we contract two \emph{neighbouring} fields $\phi_i,\phi_{i+1}$. If, instead, one contract fields that are not nearest-neighbour inside the star product, then in general the properties of the theorem will not necessarily apply. In particular, non-planar diagrams can be constructed only by contracting fields that are not nearest-neighbours, so in general for non-planar diagrams the properties of the theorem are not expected. This justifies the name of ``planar equivalence theorem''.

\subsubsection{Performing the final contractions with the external fields}
So far, we have proved that a generic diagram with uncontracted fields $\phi_i(x_i)$ is of the form
\begin{equation}\label{eq:gen-diagr}
   a_n^{\hstar}(x_1,\ldots,x_k;I_n)\equiv I_n \cdot\phi_1(x_1)\hstar\phi_1(x_2)\hstar\cdots\hstar \phi_{k}(x_k),
\end{equation}
where $I_n$ collects all the internal structure with propagators that is identical to the undeformed result, and where the star product of the uncontracted fields still preserves the symmetry invariance and cyclicity. At this point, to obtain a Feynman diagram, one wants to contract the fields $\phi_i(x_i)$ with the external fields $\phi_i(z_i)$ that appear in the correlation function being studied.
If we were in the undeformed case, the Wick contractions between $a_n(x_1,\ldots,x_k;I_n)$ and the external fields would leave
\begin{equation}
    A_n(z_1,\ldots,z_k;I_n)\equiv I_n \cdot \Delta_F(x_1-z_1)\cdots \Delta_F(x_k-z_k).
\end{equation}
We will now prove that in the presence of the twist the corresponding expression is
\begin{equation}\label{eq:tw-Feyn-diag}
\begin{aligned}
    A_n^{\hstar}(z_1,\ldots,z_k;I_n)&= I_n\cdot \mu^{(k)}\left(\hat {\bar{\mathcal F}}_{op}^{(k)}\bigotimes_{i=1}^k\Delta_F(x_i-z_i)\right)
\end{aligned}
\end{equation}
where $\hat {\bar{\mathcal F}}_{op}^{(k)}$ is the inverse of the ``opposite twist'' defined in~\eqref{eq:n-Fop}, and in the above expression it is understood that it acts on the external points $z_1,\ldots,z_k$.
In fact, contracting~\eqref{eq:gen-diagr} with external fields evaluated at $z_1,\ldots,z_k$, we obtain
\begin{equation}
    \begin{aligned}
        A_n^{\hstar}(z_1,\ldots,z_k;I_n)&=I_n \cdot\mu^{(k)}\left(\hat {{\mathcal F}}^{(k)}_{x_1\cdots x_k}\bigotimes_{i=1}^k\Delta^{\hstar}_F(x_i-z_i)\right)\\
        &=I_n \cdot\mu^{(k)}\left( \hat {{\mathcal F}}^{(k)}_{x_1\cdots x_k}(\bigotimes_{i=1}^k\hat V_{z_i})\bigotimes_{i=1}^k\Delta_F(x_i-z_i)\right)\\
        &=I_n \cdot\mu^{(k)}\left(S^{\otimes k}(\hat {{\mathcal F}}^{(k)}_{z_1\cdots z_k})(\bigotimes_{i=1}^k\hat V_{z_i}) \bigotimes_{i=1}^k\Delta_F(x_i-z_i)\right)\\
        &=I_n \cdot\mu^{(k)}\left(\Delta^{(k)}(\hat { V})_{z_1\cdots z_k}\hat {\bar{\mathcal F}}^{(k)}_{op,z_1\cdots z_k}\bigotimes_{i=1}^k\Delta_F(x_i-z_i)\right)\\
        &=I_n \cdot\mu^{(k)}\left(\Delta^{(k)}(S(\hat {V}))_{x_1\cdots x_k}\hat {\bar{\mathcal F}}^{(k)}_{op,z_1\cdots z_k}\bigotimes_{i=1}^k\Delta_F(x_i-z_i)\right)\\
        &=I_n \cdot\mu^{(k)}\left(\hat {\bar{\mathcal F}}^{(k)}_{op,z_1\cdots z_k}\bigotimes_{i=1}^k\Delta_F(x_i-z_i)\right),
    \end{aligned}
\end{equation}
where we used the definition of the twisted propagator, the symmetry invariance of the undeformed propagator to move the action of the twist from $x_i$ to $z_i$, the  identity~\eqref{eq:VSF=FopV-n-inverse}, and finally the symmetry invariance of the uncontracted diagrams, that allow us to conclude that $\Delta^{(k)}(\hat { V})$ has a trivial action, i.e.~as the identity operator.

This concludes our discussion. As shown in~\eqref{eq:tw-Feyn-diag}, a planar Feynman diagram of the twisted theory is obtained by taking the corresponding Feynman diagram of the undeformed theory and dressing it with $\hat {\bar{\mathcal F}}_{op}$, the inverse of the opposite twist,  acting on the external legs $z_i$.

\subsubsection{Final remarks on the planar equivalence theorem}
For simplicity of notation and of the presentation, we discussed the proof of the planar equivalence theorem for the case of scalar fields. However, the same theorem holds also for more generic fields, and the proof carries over exactly in the same way. In fact, if we were not working with twists in the picture of active transformations, we would need to worry about constructing an adapted basis of fields where the calculations formally work as in the case of the scalar fields, as done in~\cite{Meier:2023lku,Borsato:2025jre}. In the active picture employed in this paper, instead, the twist already satisfies the same properties irrespectively of the fields it acts on, and then the generalisation of the proof is straightforward.

Let us now consider twists involving the dilatation generator $D$ or the special conformal transformations $k_\mu$. When acting on fields, they will produce expressions where the scaling dimensions $\Delta$ appear. When considering the planar equivalence theorem we are dealing with the theory at the quantum level, and it is then natural to wonder if the scaling dimensions that appear when acting with the (inverse opposite) twist on the external legs of the Feynman diagrams are the classical ones or the ones that include the quantum corrections. We note that the manipulations that are necessary to prove the planar equivalence theorem make use of the symmetry invariance of the propagator coming from the free part of  the theory, not necessarily of the correlation functions of the interacting theory. As such, the Drinfel'd twists constructed out of $D$ and $k_\mu$ can be taken to be built with the \emph{classical} counterpart of these symmetry generators. Notice that constructing twists built out of the quantum counterpart of these symmetry generators would be cumbersome even in perturbation theory. On the one hand, when studying the spectral problem, the anomalous dimension to be determined should appear also in the twist that constructs the two-point function, possibly leading to recursivity; on the other hand, when considering the application of the twist on the product of $n$ fields, one should also take into account that these products may not be eigenstates of the quantum $D$, for example, thus complicating a lot the construction. In this sense, it is reassuring that we can construct the twists in terms of the classical symmetries only. This is in fact consistent with twists of the spin-chain  of $AdS_5/CFT_4$ that involve $D$ and that were considered  in~\cite{Beisert:2005if}.

When dealing with star-products arising from supersymmetric versions of Drinfel'd twists, one should keep in mind the assumptions explained in section~\ref{sec:susy-symm}. First, we claim that our discussion includes all twists of $\mathcal N=1$ or $\mathcal N=2$ supersymmetric gauge theories, because in all possible cases we can (formally) have linearly realised off-shell supersymmetry (possibly with the inclusion of an infinite number of auxiliary fields in the case of $\mathcal N=2$ hypermultiplets). In all these cases, in general, one should pay attention to the fact that the inverse opposite twist $\hat {\bar{\mathcal F}}_{op}^{(k)}$ acting on the external legs of the Feynman diagrams may generate auxiliary fields even when the starting point only contains physical ones. Apart from this remark, the validity of the planar equivalence theorem is ensured.

In the case of twist deformations of $\mathcal N=4$ super Yang-Mills, our construction is still valid for most twists. However, one may consider also twists that cannot be embedded into just two copies of supersymmetry. Examples of this kind of twist are some extended Jordanian solutions, see the classification in section~\ref{sec:class}. A direct consequence of this is that not all the supercharges participating in the construction of the Drinfel'd twist will be simultaneously linearly realised in an off-shell formulation.  These twists will escape our construction unless a way is found to have a simultaneous linear and off-shell realisation for all the supersymmetry transformations appearing in the twists.

\section{Classification of possible twists}\label{sec:class}
In this section we provide a rough classification and some examples of possible twists that may be applied to field and gauge theories to generate non-commutative deformations. We do not claim to provide the full classification of all possible twists, since the purpose of this section is simply to show that there is a rich list of possibilities.\footnote{For the sake of the discussion, we will assume that the spacetime has dimension four.}

Given that equivalence classes of twists $\mathcal F$ are identified by the corresponding classical $r$-matrices, we will construct the classification by listing possible solutions $r$ of the classical Yang-Baxter equation (CYBE). Obviously, according to the assumptions of this paper, $r$ should solve the CYBE on $\mathfrak{g}$, where $\mathfrak{g}$ is the (super)algebra of global symmetries of the undeformed field theory. We will also write the expression for a representative twist $\mathcal F$ in the cases when we know that.

\subsection{Abelian}
The class of abelian twists is the simplest one. It is characterised by $r$-matrices of the form
\begin{equation}
    r=r^{AB}a_A\wedge a_B,\qquad r^{AB}=-r^{BA},\qquad [a_A,a_B]=0,\ \forall A,B,
\end{equation}
which of course justifies the name abelian. In fact, if all generators making up the $r$-matrix commute with each other, than the CYBE is trivially solved.
A corresponding ($r$-symmetric) twist is given simply by
\begin{equation}
    \mathcal F=e^{\xi \ r},
\end{equation}
where $\xi$ is a deformation parameter. Given that the generators $a_A$ may be rescaled without affecting the statement that this is an abelian Drinfel'd twist, this is in fact a multi-parametric twist. 

Given that the elements that compose them all commute with each other, abelian twists satisfy the $r$-unimodularity condition. The above twist $\mathcal{F}$ is also $\mathcal F$-unimodular, but we remind that in section~\ref{sec:abel-ex} we pointed out that one may consider other equivalent twists that are not $\mathcal F$-unimodular.

To construct abelian twists, it is then enough to identify abelian subalgebras of $\mathfrak g$.
In the context of AdS/CFT, abelian twists correspond to (sequences of) TsT transformations of the string background~\cite{Osten:2016dvf}.

\subsubsection{Internal symmetries}
Internal symmetries are normally organised into a compact algebra, and then calling $h_i,\ i=1,\ldots,n$ the corresponding Cartan elements, the most general $r$-matrix (up to Lie-algebra automorphisms) is
\begin{equation}
    r=r^{ij}h_i\wedge h_j,
\end{equation}
corresponding then to a twist with $n(n-1)/2$ parameters $r^{ij}=-r^{ji}$. Twists of this kind, i.e.~constructed out of Cartan elements only, are normally referred to as ``Drinfel'd-Reshetikhin twists''.
An example is the $\beta$-deformation of Lunin and Maldacena~\cite{Lunin:2005jy}, that is a reinterpretation of the Leigh-Strassler deformation~\cite{Leigh:1995ep}. Obviously, when twisting via internal symmetries, the star-product will not generate spacetime non-commutativity: the result of the deformation will be an ordinary (commutative) field theory, albeit with deformed interaction terms.

\subsubsection{Spacetime symmetries}
The family of possibilities grows with the amount of spacetime symmetries available; in fact one may also consider abelian twists outside the class of Drinfel'd-Reshetikhin because they may be built out of elements that are not Cartan. When dealing with theories that are invariant under translations $p_\mu$, one can construct twists with $r$-matrix
\begin{equation}
    r=\theta^{\mu\nu}\ p_\mu\wedge p_\nu.
\end{equation}
The above is in fact the well-known Groenewold-Moyal deformation~\cite{Moyal:1949sk,Groenewold:1946kp,Douglas:2001ba,Szabo:2001kg}. There are subtleties when switching on timelike directions $\theta^{0i}\neq 0$ in the deformation~\cite{Gomis:2000zz,Seiberg:2000gc}, but we will not deal with these issues here.

If the undeformed theory is in fact Poincar\'e invariant, then there are also Lorentz generators $J_{\mu\nu}$ and another abelian option is
\begin{equation}
    r=\epsilon^{\mu\nu\rho\sigma}\ J_{\mu\nu}\wedge J_{\rho\sigma},
\end{equation}
which is essentially just the $r$-matrix $J_{01}\wedge J_{23}$ of~\cite{Meier:2023kzt}. Other options appear when pairing a Lorentz generator and a translation with different indices, for example
\begin{equation}
    r= J_{12}\wedge p_3,\qquad\text{or}\qquad r= J_{01}\wedge p_2,\qquad\text{or}\qquad r= J_{12}\wedge p_0.
\end{equation}

When the available symmetries are enlarged to the conformal algebra, one can take advantage of the fact that the dilatation operator $D$ commutes with Lorentz generators, and work with
\begin{equation}
    r=\xi^{\mu\nu}\ D\wedge J_{\mu\nu},
\end{equation}
where $\xi^{\mu\nu}$ are the six possible parameters.
Moreover, taking advantage of the automorphism swapping translations with special conformal transformations\footnote{This dualisation relates the corresponding deformed theories  by the inversion symmetry that exchanges $p_\mu$ and $k_\mu$.} $p_\mu\leftrightarrow k_\mu$ and changing the sign of rescalings $D\to -D$, one can ``dualise'' previous $r$-matrices to obtain
\begin{equation}
    r=\tilde\theta^{\mu\nu}\ k_\mu\wedge k_\nu,\qquad
    r= J_{\mu\nu}\wedge k_\rho,\ \rho\neq \mu\nu.
\end{equation}

\subsubsection{Mixed twists}
Since spacetime and internal symmetries commute, one can construct twists of the form
\begin{equation}
    r=X\wedge a,
\end{equation}
where $X$ is any spacetime symmetry and $a$ any internal one. An example of this is the $r$-matrix $p_-\wedge Q$ of~\cite{Guica:2017mtd} for the dipole deformation, where $p_-$ is a light-cone momentum and $Q$ an $R$-symmetry charge. Yet another option is the angular-dipole deformation $J_{23}\wedge Q$ of~\cite{Meier:2025tjq}.

\subsubsection{Supercharges}

In principle, one may exploit anticommuting supercharges to construct abelian twists only out of fermionic generators.\footnote{In the Euclidean setting, such twists are possible and were considered in~\cite{Seiberg:2003yz}.} For example, taking the Poincar\'e supercharges $Q_\alpha^I$ with $I=1,\ldots,\mathcal N$ and $\alpha$ being a Lorentz spinor index, one may consider
\begin{equation}
    r=r_{IJ}^{\alpha\beta}\ Q_\alpha^I\wedge Q_\beta^J,
\end{equation}
where $r_{IJ}^{\alpha\beta}$ are parameters and $\wedge$ is graded, so that it is actually the symmetrised tensor product in this case. Similarly, one may construct solutions of the kind $\bar Q\wedge\bar Q$ with the complex conjugate supercharges $\bar Q_{\dot \alpha}^I$.

In a superconformal theory like $\mathcal N=4$ super Yang-Mills, together with the Poincar\'e supercharges $Q,\bar Q$ one has also the conformal supercharges $S,\bar S$, so that thanks to the anticommutation relations one can construct also the $r$-matrices of the form $S\wedge S,\bar S\wedge \bar S,Q\wedge\bar S,\bar Q\wedge S$.

In all these cases, the twist will spoil the reality condition of the action. This can be understood simply by noticing that if we are working with $r\sim Q\wedge Q$, for example, then after complex conjugation this will give rise to terms of the form $\bar Q\wedge \bar Q$. 

A natural question is whether it is possible to identify abelian $r$-matrices that are constructed only out of supercharges and that are compatible with reality conditions. One can check that  it is not possible to identify pairs of anticommuting \emph{real} supercharges. An intuitive explanation of this is that real supercharges are essentially a combination of $Q$ and $\bar Q$, for example, which have non-trivial anticommutation relations between themselves. Therefore, all purely fermionic abelian twists necessarily spoil the reality condition of the action.

\subsection{Non-abelian}
Together with the abelian ones, interesting options are also the non-abelian twists. Given that compact algebras only admit abelian classical $r$-matrices, non-abelian solutions will be possible only if the non-compact algebra of spacetime symmetries participates.

\subsubsection{Almost abelian}
A class that may be considered is that of ``almost abelian twists''~\cite{vanTongeren:2016eeb}. They are characterised by abelian building blocks that are ``subordinate'' to each other~\cite{Borowiec:2008se} and that, when put together, entail a non-abelian structure. In the case of rank-4 $r$-matrices, they may be written as
\begin{equation}
    r=r_1+r_2,\qquad r_1=a\wedge b,\qquad r_2=c\wedge d,\qquad[a,b]=[c,d]=0,
\end{equation}
and they satisfy the property that $c,d$ are symmetries of $r$, i.e.
\begin{equation}
    (\mathrm{ad}_c\otimes 1+1\otimes \mathrm{ad}_c)r=(\mathrm{ad}_d\otimes 1+1\otimes \mathrm{ad}_d)r=0.
\end{equation}
They may be understood as the composition of two successive abelian twists, so that 
\begin{equation}
    \mathcal F=e^{\xi \ r_2}e^{\xi \ r_1}.
\end{equation}
Despite  $r_1,r_2$ being abelian on their own, when put together they give rise to a non-abelian structure because some of the generators of one $r$-matrix do not commute with some of the other. It is for this reason that almost abelian twists can be understood as non-commuting compositions of abelian twists, because $\mathcal F\neq e^{\xi \ r_1}e^{\xi \ r_2}$.

Being compositions of abelian twists, they satisfy the $r$-unimodularity condition.

At rank-4, the options that can be embedded into the conformal algebra were classified in~\cite{Borsato:2016ose} up to inner automorphisms. The amount of available $r$-matrices indicated in that paper actually doubles if we use the automorphism $p_\mu\leftrightarrow k_\mu, D\to -D$. To summarise the classification, we now proceed to list the relevant non-abelian algebras with its non-trivial commutation relations, and the enumeration of the corresponding classical $r$-matrices that appear in Table 2 of~\cite{Borsato:2016ose}:
\begin{itemize}
    \item $\mathfrak h_3\oplus \mathbb R$ algebra: $[c,a]=b$, with classical $r$-matrices $1,2,3$ that only require Poincar\'e invariance, plus the $r$-matrix number $4$ that requires conformal symmetry.
    \item $\mathfrak r_{3,-1}\oplus \mathbb R$ algebra: $[c,a]=a,\ [c,b]=-b$, with classical $r$-matrices $6,7$ that only require Poincar\'e invariance, plus the $r$-matrix $5$ that also requires invariance under $D$.
    \item $\mathfrak r'_{3,0}\oplus \mathbb R$ algebra: $[c,a]=-b,\ [c,b]=a$, with classical $r$-matrices $8,\ldots,14$ that only require Poincar\'e invariance.
\end{itemize}
We notice that some of these $r$-matrices have extra parameters, so that the family is very rich.

Given that the element $d$ is central in all the $r$-matrices of this class, it may be replaced by an element of internal, rather than spacetime, symmetries. When doing this, the number of possibilities is even higher, and in fact there is also one option in $\mathfrak r_{3,-1}\oplus \mathbb R$~\cite{Borsato:2016ose}
\begin{equation}
    r=(D+2J_{03})\wedge T+p_1\wedge(p_0+p_3),
\end{equation}
that is possible when $T$ is an internal symmetry, but not when it is a spacetime symmetry.

Going to rank-6, there is at least the almost abelian $r$-matrix
\begin{equation}
    r=p_0\wedge p_1+p_2\wedge p_3+J_{01}\wedge J_{23},
\end{equation}
with twist
\begin{equation}
    \mathcal F=e^{\xi\ J_{01}\wedge J_{23}}e^{\xi\ (p_0\wedge p_1+p_2\wedge p_3)}.
\end{equation}

One may check that it is not possible to embed a rank-8 $r$-unimodular $r$-matrix in the conformal algebra, and that means that the highest rank is 6.

\subsubsection{$\mathfrak{n}_4$ algebra}
Going beyond the class of almost abelian $r$-matrices, we encounter the class of the $\mathfrak{n}_4$ algebra characterised by
\begin{equation}
    r=a\wedge b+c\wedge d,\qquad [a,c]=-b,\qquad [b,c]=d,
\end{equation}
and corresponding to the $r$-matrices $15,16,17$ of Table 2 of~\cite{Borsato:2016ose}, which only require Poincar\'e invariance.
We refer to appendix A of~\cite{Meier:2023lku} for a discussion on the construction of the corresponding twists.

\subsubsection{Extended Jordanian}
Extended Jordanian $r$-matrices are solutions of the CYBE of the form~\cite{tolstoy2004chainsextendedjordaniantwists}
\begin{equation}
    r=h\wedge e-\sum_{i=1}^N e_{-i}\wedge e_{+i},
\end{equation}
where the (super)algebra elements satisfy
\begin{equation}
    [h,e]=e,\qquad [e_{\pm i},e]=0,\qquad
    [h,e_{\pm i}]=\left(\tfrac12\pm \xi_i \right)e_{\pm i},\ \xi_i\in \mathbb C,\qquad
    [[e_k,e_l]]=\delta_{k,-l}e,
\end{equation}
where $k>l$ and $[[.,.]]$ denotes the graded commutator. When $N=0$ one recovers the usual Jordanian solution $r=h\wedge e$. If we write $N=N_0+N_1$, where $N_0$ is the number of bosonic pairs of $e_\pm i$ elements and $N_1$ that of fermionic ones, then $r$-unimodularity implies~\cite{Borsato:2016ose}
\begin{equation}
    N_1=N_0+1.
\end{equation}
The first unimodular Jordanian examples in AdS/CFT were considered in~\cite{vanTongeren:2019dlq}, and the classification of unimodular extended Jordanian $r$-matrices for the $\mathfrak{psu}(2,2|4)$ superalgebra relevant for the duality between the $AdS_5\times S^5$ superstring and $\mathcal N=4$ super Yang-Mills was given in~\cite{Borsato:2022ubq}. The interested reader is invited to consult section 2 of~\cite{Borsato:2022ubq}, where the solutions are organised by the rank of the bosonic piece.
In particular, considering only the bosonic $r$-matrices that admit a unimodular extension, we may have the following observations:
\begin{itemize}
    \item 
In the case of rank-2 (see Table 1 of~\cite{Borsato:2022ubq}) there are 6 families\footnote{We talk about families of solutions because the results presented in the tables of~\cite{Borsato:2022ubq} depend on parameters that give rise to multi-parametric twists.} of solutions, two of which only require Poincar\'e invariance (number 1 when $b=-1$ and number 3 of the table); moreover, in order to be able to have a unimodular extension, for the first 5 families it is enough to start with a theory with just $\mathcal N=1$ supersymmetry, while for the sixth one at least $\mathcal N=2$ is needed.
\item In the case of rank-4, there are 6 more families, two of which only require Poincar\'e (number 7 when $b=-1$ and number 10 of the table); moreover, in order to be able to have a unimodular extension, the undeformed theory should have at least $\mathcal N=2$ supersymmetry.
\item In the case of rank-6 (see equations (2.11)-(2.14)), there are 2 more families, one of which  only requires Poincar\'e (number 13 when $b=-1$); moreover, in order to be able to have a unimodular extension, the undeformed theory should have the full $\mathcal N=4$ supersymmetry.\footnote{In principle $\mathcal N=3$ is enough, but we assume to have a Lagrangian theory.}
\end{itemize}
The corresponding formulas for the extended Jordanian twists $\mathcal F$ may be found in~\cite{tolstoy2004chainsextendedjordaniantwists}.

\section{Conclusions}\label{sec:concl}

In this paper we presented a very general construction to build deformations of field and gauge theories via non-commutative star-products induced by Drinfel'd twists. One important ingredient is that the Drinfel'd twists are constructed out of elements of the symmetry algebra of the undeformed theory, interpreted as \emph{active} symmetry transformations. A direct consequence of this setup is that partial derivatives follow the usual Leibniz rule even in the presence of the star-product, making it therefore straightforward to construct covariant derivatives. Importantly, we also explained that all Drinfel'd twist deformations of gauge theories  appeared in the literature so far are captured by our construction; on the one hand, ours is an equivalent formulation of those deformations, and on the other hand it is a significant extension of the number of deformations that can be considered. We refer to section~\ref{sec:class} for a rough classification of the possible twists depending on the symmetry algebra of the original undeformed theory.

We also explained that the cyclicity of the star-product under integration---a requirement that is essential for maintaining the gauge invariance even in the twisted setup---is equivalent to what we call the ``$\mathcal R$-unimodularity condition'', which is weaker than the condition imposed since the work of~\cite{Aschieri:2009ky}. In this sense, the correct identification of the unimodularity condition allows us to consider twist deformations that otherwise would be left out from the construction. As observed also in~\cite{Meier:2023lku} for the case of Poincar\'e twists, what is very interesting is that the unimodularity condition of the classical $r$-matrix that here is necessary to have gauge invariance, already appeared in the context of AdS/CFT as being the condition to have  valid type II solutions for the Yang-Baxter deformations of the supergravity backgrounds~\cite{Borsato:2016ose}.\footnote{While in the case of the $AdS_5\times S^5$ background the unimodularity of the $r$-matrix is a necessary and sufficient condition for remaining within type II supergravity, for other seed backgrounds it may be only sufficient, so that the condition can be relaxed~\cite{Wulff:2018aku,Hronek:2020skb}.} It is nice to observe, then, that the very same condition controls the consitency of the construction both on the side of the deformed gauge theory and on that of the dual deformed supergravity background.

We discussed several aspects of the twist-deformed gauge theories, including the fact that global symmetries get in general twisted, because their implementation at the level of star-products of fields involves the twist. Moreover, we proved the planar equivalence theorem in the most general setup of our construction. In particular, we showed that in the deformed theories planar Feynman diagrams maintain the original undeformed internal structure, and the effect of the twist is only on the external legs. This  extends the results known since the work of~\cite{FILK199653} in the case of Groenewold-Moyal, and later demonstrated in~\cite{Meier:2023lku} to be valid for Poincar\'e twists.

Compared to previous approaches, we must limit the construction to  twists built out of \emph{symmetries} of the undeformed field theory. While this might seem a limitation, it is in fact a very natural strategy, and it is in line with the fact that deformations like the Groenewold-Moyal one are of that kind (because translation invariance for the gauge theories that are studied is always assumed). The option to twist by means of symmetries actually offers a rich set of possibilities, because one can have twists involving also internal symmetries. This is natural also from the point of view of AdS/CFT, see in particular~\cite{Lunin:2005jy}. 

Our construction provides field theories with a non-commutative star product that live on ordinary (commutative) spacetimes. When considering twists involving spacetime symmetries, it would be interesting to understand the formal map that relates these to field theories on non-commutative spacetimes. Such a map should probably involve going from the active picture of symmetry transformations (where only the fields change) to the passive picture (where only the spacetime coordinates change), so as to induce a non-trivial star-product for coordinates.

Let us point out that our construction relies on the assumption that the symmetry transformations are realised linearly and off-shell. This comment is particularly important in the case of supersymmetry transformations, that may appear in twists of superalgebras relevant for supersymmetric gauge theories. While we can cover several twists of this kind, those that cannot be embedded into $\mathcal N\leq 2$ supersymmetry escape our construction. This issue is related to the fact that there is no off-shell formulation for $\mathcal N=4$ superspace. While it would be interesting to understand how to include twists that require more than two copies of supersymmetry, this question is beyond the scope of our paper.

Related to these comments, it would be very interesting to understand what happens when trying to integrate out auxiliary fields or fixing gauge degrees of freedom in the presence of star-products with Drinfel'd twists featuring supersymmetry transformations. In general, these operations imply that supercharges may not be realised linearly and off-shell any more, so one may expect important difficulties to arise when writing the twisted theories only in terms of the physical degrees of freedom. We refer to~\cite{StijnJulio} for more detailed discussions on these points.

Having provided a construction for the twist-deformed gauge theories, a natural next step now is to  understand how to build  gauge-invariant operators and compute their correlation functions. We expect that the construction of gauge invariant operators will entail a generalisation of the idea of~\cite{Gross:2000ba}. Moreover, it would be particularly interesting to do this for twist deformations of $\mathcal N=4$ super Yang-Mills, and explicitly show the direct relation to the expected Drinfel'd twist deformations of the spin-chain description that underlies the planar spectral problem.

Beyond the planar limit, in general one should expect the possibility of UV/IR mixing for the twisted gauge theories as for example shown in \cite{Minwalla:1999px} in the Groenewold-Moyal case. However, the family of possibilities is so rich that it may be worth scanning them to identify possible examples that do not feature UV/IR mixing.

For the phenomenology of quantum gravitational effects, a potentially important role is played by the $\kappa$-Minkowski spacetime~\cite{Kowalski-Glikman:2004fsz}. This can be realised also as a Drinfel'd twist with classical $r$-matrix of Jordanian type $D\wedge p_0$~\cite{Dimitrijevic:2011jg,Dimitrijevic:2014dxa}. While it is true that this $r$-matrix is not unimodular, one can in principle promote it to an extended Jordanian $r$-matrix that does satisfy the unimodularity condition~\cite{vanTongeren:2019dlq,Borsato:2022ubq}, as long as a suitable pair of supercharges is identified. In this sense, our construction opens the possibility to construct a twist deformation of gauge theories that should be equivalent to placing them on the $\kappa$-Minkowski non-commutative spacetime, thus overcoming the issues of~\cite{Dimitrijevic:2011jg,Dimitrijevic:2014dxa}. We remark, however, that in order for this to work the original undeformed theories should have at least $\mathcal N=2$ superconformal symmetry.

Although in this paper the goal was to construct non-commutative deformations via Drinfel'd twists, another very interesting possibility is that of constructing $q$-deformations of gauge theories. In particular, an explicit construction of the $q$-deformation of $\mathcal N=4$ super Yang-Mills would provide the obvious candidate for the holographic dual of the ``$\eta$-deformation'' of $AdS_5\times S^5$ of~\cite{Delduc:2013qra}. In fact, it is likely that some ideas in this paper  may be helpful to understand how to construct $q$-deformations as well; for example, it may be natural to work in the active picture of symmetry transformations also in that case. Moreover, thanks to the experience from the case of Drinfel'd twist deformations and the knowledge of the unimodularity conditions arising for the $\eta$-deformation of $AdS_5\times S^5$~\cite{Borsato:2016ose,Hoare:2018ngg}, it is natural to expect that some sort of unimodularity condition should appear also for $q$-deformations of gauge theories. It would be very interesting to understand in detail how that construction would work.

\section*{Acknowledgements}
We thank Stijn van Tongeren and Julio Cabello Gil for very useful discussions.

This work was supported by the grants RYC2021-032371-I (funded by the European Union ``NextGenerationEU''/PRTR and by MCIN/AEI/10.13039/501100011033), by 2023-PG083 (with reference code ED431F 2023/19 funded by Xunta de Galicia), by PID2023-152148NB-I00 (funded by AEI-Spain), by the Mar\'ia de Maeztu grant CEX2023-001318-M (funded by MICIU/AEI /10.13039/501100011033), by the CIGUS Network of Research Centres, and by the European Union.

\appendix

\section{Conformal transformations of  fields}\label{app:symm}
In this appendix we collect the explicit expressions of symmetry transformations of various fields of interest to construct field theories.

\subsection{Primary scalar fields}
We may define a primary scalar field $\phi(x)$ as a field satisfying
\begin{equation}
    \X_{tot}(\phi(x))=-\frac{\Delta_\phi}{d}\partial_\mu X^\mu\ \phi(x).
\end{equation}
 Given that for Poincar\'e transformations $\partial_\mu X^\mu=0$, in that case the reader will recognise the usual definition of a scalar field for the Poincar\'e group as $\phi'(x')=\phi(x)$.
From~\eqref{eq:tot-pas-act} it follows that the active transformation of a primary scalar field is 
\begin{equation}
    \hat\X(\phi(x))=-X^\mu\partial_\mu\phi(x)-\frac{\Delta_\phi}{d}\partial_\mu X^\mu\ \phi(x).
\end{equation}
For a primary scalar field we have that the Weyl-Lie derivative is
\begin{equation}
    \mathcal L^W_{\X}\phi(x)=X^\mu\partial_\mu\phi(x)+\frac{\Delta_\phi}{d}\partial_\mu X^\mu\ \phi(x),
\end{equation}
and then
\begin{equation}
     \hat\X(\phi(x))=\mathcal L^W_{-\X}\phi(x).
\end{equation}

\subsection{Covariant/contravariant vector and tensor fields}

 Considering the covariant vector field $A_\mu(x)$, one may generalise the above discussion by defining the scalar field $A(x)$ 
\begin{equation}
    A(x)\equiv A_\mu(x)\dd x^\mu,
\end{equation}
where $\dd x^\mu$ is a basis of differential forms. Being a scalar field, its total transformation is 
\begin{equation}
    \X_{tot}(A(x))=-\frac{\Delta_A-1}{d}\partial_\mu X^\mu\ A(x),
\end{equation}
where we assign the weight $w=\Delta_A-1$ because we take into account that $A_\mu$ has scaling dimension $\Delta_A$ and $\dd x^\mu$ has scaling dimension $-1$.\footnote{This is necessary also to match with conventions of~\cite{Freedman:2012zz}, see their section 15.3.}
Knowing that 
\begin{equation}
    \dd x'{}^\mu\approx\dd x^\mu+\dd X^\mu=\dd x^\mu+\dd x^\nu\partial_\nu X^\mu,
\end{equation}
we can compute the passive transformation as
\begin{equation}
\begin{aligned}
        \check \X(A(x)) &= A_\mu(x')\dd x'{}^\mu-A_\mu(x)\dd x^\mu\\
        &= (X^\nu\partial_\nu A_\mu(x)+\partial_\mu X^\nu A_\nu (x))\dd x^\mu.
\end{aligned}
\end{equation}
It follows that the active transformation is
\begin{equation}
    \begin{aligned}
        \hat \X(A(x)) &=  \X_{tot}(A(x)) -\check \X(A(x)) \\
        &= -\left(X^\nu\partial_\nu A_\mu(x)+\partial_\mu X^\nu A_\nu (x)+\frac{\Delta_A-1}{d}\partial_\nu X^\nu\ A_\mu(x)\right)\dd x^\mu.
    \end{aligned}
\end{equation}
Remembering that active transformations leave coordinates unchanged, we may write
\begin{equation}
    \hat \X(A(x))=\hat \X(A_\mu(x))\dd x^\mu.
\end{equation}
After defining the Weyl-Lie derivative as
\begin{equation}
    \mathcal L^W_\X A_\mu(x) = X^\nu\partial_\nu A_\mu(x)+\partial_\mu X^\nu A_\nu (x)+\frac{\Delta_A-1}{d}\partial_\nu X^\nu\ A_\mu(x),
\end{equation}
we may therefore conclude
\begin{equation}
    \hat \X(A_\mu(x)) =\mathcal L^W_{-\X} A_\mu(x).
\end{equation}

Similar considerations can be made for a contravariant vector field $B^\mu(x)$. Given that $B(x)\equiv B^\mu(x)\partial_\mu$ transforms like a scalar field, we write
\begin{equation}
    \X_{tot}(B(x))=-\frac{\Delta_B+1}{d}\partial_\mu X^\mu\ B(x),
\end{equation}
where now we use that $\partial_\mu$ has dimension 1.
At this point, using that 
\begin{equation}
    \partial'_\mu\equiv \frac{\partial}{\partial x'{}^\mu}=\frac{\partial x^\nu}{\partial x'{}^\mu}\frac{\partial}{\partial x^\nu} \approx-\partial_\mu X^\nu\partial_\nu,
\end{equation}
we have
\begin{equation}
\begin{aligned}
        \check \X(B(x)) &= B^\mu(x')\partial'_\mu-B^\mu(x)\partial_\mu\\
        &= (X^\nu\partial_\nu B_\mu(x)-\partial_\nu X^\mu B^\nu (x))\partial_\mu.
\end{aligned}
\end{equation}
We can then take
\begin{equation}
    \mathcal L^W_\X B^\mu(x) = X^\nu\partial_\nu B^\mu(x)-\partial_\nu X^\mu B^\nu (x)+\frac{\Delta_B+1}{d}\partial_\nu X^\nu\ B^\mu(x),
\end{equation}
so that 
\begin{equation}
    \hat \X(B^\mu(x)) =\mathcal L^W_{-\X} B^\mu(x).
\end{equation}
Symmetry transformations of $A^\mu(x)=\eta^{\mu\nu}A_\nu(x)$ and $B_\mu(x)=\eta_{\mu\nu}B^\nu(x)$ are obtained simply by raising/lowering indices in the above formulas with the Minkowski metric. In particular, using the conformal Killing equation
\begin{equation}\label{eq:conf-Kill-eq}
    \partial_\mu X^\nu+\partial^\nu X_\mu=\frac{2}{d}\, \delta_\mu^\nu\, \partial_\rho X^\rho,
\end{equation}
one can recast the active transformation (i.e.~the Weyl-Lie derivative) of $A^\mu$ as that of a contravariant field, and that of $B_\mu$ as that of a covariant field
\begin{equation}
    \begin{aligned}
         \hat \X(A^\mu(x)) &= -X^\nu\partial_\nu A^\mu(x)-\partial^\mu X_\nu A^\nu (x)-\frac{\Delta_A-1}{d}\partial_\nu X^\nu\ A^\mu(x)\\
         &= -X^\nu\partial_\nu A^\mu(x)+\partial_\nu X^\mu A^\nu (x)-\frac{\Delta_A+1}{d}\partial_\nu X^\nu\ A^\mu(x)\\
         \hat \X(B_\mu(x)) &= -X^\nu\partial_\nu B_\mu(x)+\partial^\nu X_\mu B_\nu (x)-\frac{\Delta_B+1}{d}\partial_\nu X^\nu\ B_\mu(x)\\
         &= -X^\nu\partial_\nu B_\mu(x)-\partial_\mu X^\nu B_\nu (x)-\frac{\Delta_B-1}{d}\partial_\nu X^\nu\ B_\mu(x).
    \end{aligned}
\end{equation}
The shifts by $\pm 2$ of the scaling dimensions are justified by the fact that the scaling dimension of the (inverse) metric participates in these formulas.

Obviously, transformation rules for tensors $T_{\mu_1\cdots \mu_m}^{\nu_1\cdots \nu_n}(x)$ with $m$ covariant and $n$ contravariant indices are easily obtained from the above formulas
\begin{equation}\label{eq:WeylLieTensor}
\begin{aligned}
\mathcal L^W_\X T_{\mu_1\cdots \mu_m}^{\nu_1\cdots \nu_n}(x) &= X^\rho\partial_\rho T_{\mu_1\cdots \mu_m}^{\nu_1\cdots \nu_n}(x)
+\sum_{i=1}^m\partial_{\mu_i} X^\rho T_{\mu_1\cdots \mu_{i-1}\rho\mu_{i+1}\cdots \mu_m}^{\nu_1\cdots \nu_n} (x)\\
&
-\sum_{j=1}^n\partial_\rho X^{\nu_j} T_{\mu_1\cdots \mu_m}^{\nu_1\cdots \nu_{j-1}\rho\nu_{j+1}\cdots \nu_n} (x)+\frac{\Delta_T+n-m}{d}\partial_\rho X^\rho\ T_{\mu_1\cdots \mu_m}^{\nu_1\cdots \nu_n}(x).
\end{aligned}
\end{equation}

\subsection{Spinor fields}
Restricting for simplicity to the case of $d=4$ spacetime dimensions, for Dirac spinor fields $\Psi(x)$  we have~\cite{Freedman:2012zz}
\begin{equation}
    \X_{tot}(\Psi(x))=-\frac{1}{4}\partial^\mu X^\nu\gamma_{\mu\nu}\Psi(x)-\frac{\Delta_{\Psi}}{4}\, \partial_\mu X^\mu\, \Psi(x),
\end{equation}
where $\gamma_{\mu\nu}=\frac12[\gamma_\mu,\gamma_\nu]$, and the first term implements the intrinsic transformation of the Dirac spinor. 
It follows that 
\begin{equation}
    \hat\X(\Psi(x))=\mathcal L^W_{-\X}\Psi(x),
\end{equation}
if we define
\begin{equation}
    \mathcal L^W_{\X}\Psi(x)=X^\mu\partial_\mu \Psi(x)+\frac{1}{4}\partial^\mu X^\nu\gamma_{\mu\nu}\Psi(x)+\frac{\Delta_{\Psi}}{4}\, \partial_\mu X^\mu\, \Psi(x).
\end{equation}
For the Dirac conjugate fermion $\bar\Psi(x)=\Psi(x)^\dagger\gamma^0$,  the transformation is
\begin{equation}
    \mathcal L^W_{\X}\bar\Psi(x)=X^\mu\partial_\mu \bar\Psi(x)-\frac{1}{4}\partial^\mu X^\nu\bar\Psi(x)\gamma_{\mu\nu}+\frac{\Delta_{\Psi}}{4}\, \partial_\mu X^\mu\, \bar\Psi(x).
\end{equation}
These transformations can be written also for Weyl spinors. When writing gamma matrices in the Weyl representation as 
\begin{equation}
    \gamma^\mu = \left(\begin{array}{cc}
      \mathbf 0   & \sigma^\mu \\
         \bar \sigma^\mu&\mathbf  0 
    \end{array}\right),
\end{equation}
we have that a left Weyl fermion $\psi(x)$ transforms as
\begin{equation}
    \begin{aligned}
        &\X_{tot}(\psi(x))=-\frac{1}{2}\partial^\mu X^\nu\sigma_{\mu\nu}\psi(x)-\frac{\Delta_{\psi}}{4}\, \partial_\mu X^\mu\, \psi(x),\\
        &\hat\X(\psi(x))=\mathcal L^W_{-\X}\psi(x),\\
        &\mathcal L^W_{\X}\psi(x)=X^\mu\partial_\mu \psi(x)+\frac{1}{2}\partial^\mu X^\nu\sigma_{\mu\nu}\psi(x)+\frac{\Delta_{\psi}}{4}\, \partial_\mu X^\mu\, \psi(x),
    \end{aligned}
\end{equation}
while a right Weyl fermion $\chi(x)$ transforms as
\begin{equation}
    \begin{aligned}
        &\X_{tot}(\chi(x))=-\frac{1}{2}\partial^\mu X^\nu\bar\sigma_{\mu\nu}\chi(x)-\frac{\Delta_{\chi}}{4}\, \partial_\mu X^\mu\, \chi(x),\\
        &\hat\X(\chi(x))=\mathcal L^W_{-\X}\chi(x),\\
        &\mathcal L^W_{\X}\chi(x)=X^\mu\partial_\mu \chi(x)+\frac{1}{2}\partial^\mu X^\nu\bar\sigma_{\mu\nu}\chi(x)+\frac{\Delta_{\chi}}{4}\, \partial_\mu X^\mu\, \chi(x),
    \end{aligned}
\end{equation}
where
\begin{equation}
    \sigma^{\mu\nu}=\frac{1}{4}(\sigma^\mu\bar\sigma^\nu-\sigma^\nu\bar\sigma^\mu),\qquad\qquad
    \bar\sigma^{\mu\nu}=\frac{1}{4}(\bar\sigma^\mu\sigma^\nu-\bar\sigma^\nu\sigma^\mu).
\end{equation}

\subsection{Partial derivatives of fundamental fields}\label{sec:descendantFields}
Already in undeformed field theories, there are derivatives of fundamental fields appearing in kinetic terms in the action. In the deformed theory, we need to understand how active transformations change for partial derivatives of certain fields. Here we give an alternative argument to the one of the main text to the fact that partial derivatives commute with active transformations. Let us consider the derivative of the active transformation of a scalar field $\phi$ with scaling dimension $\Delta_\phi$, i.e.
\begin{equation}\label{eq:partialXPhi}
    \begin{aligned}
        \partial_\mu(\hat\X\phi)&=-\partial_\mu(\mathcal{L}^W_X(\phi))=-\partial_\mu(X^\nu\partial_\nu(\phi)+\frac{\Delta_\phi}{d}\partial_\nu X^\nu\phi)\\
        &=-X^\nu\partial_\nu(\partial_\mu\phi)+\left[X^\nu\partial_\nu,\partial_\mu\right]\phi-\frac{\Delta_\phi}{d}\partial_\nu X^\nu\partial_\mu\phi-\frac{\Delta_\phi}{d}\partial_\nu \partial_\mu X^\nu \phi\\
        &=-X^\nu\partial_\nu(\partial_\mu\phi)-\partial_\mu X^\nu\partial_\nu\phi-\frac{\Delta_\phi}{d}\partial_\nu X^\nu\partial_\mu\phi-\frac{\Delta_\phi}{d}\partial_\nu (\partial_\mu X^\nu) \phi.
    \end{aligned}
\end{equation}
On the other hand, $\partial_\mu\phi$ is a conformal descendant of the primary field $\phi$, and hence transforms as
\begin{equation}\label{eq:XPartialPhi}
    \begin{aligned}
        \hat\X(\partial_\mu\phi)=-X^\nu\partial_\nu(\partial_\mu\phi)-\partial_\mu X^\nu\partial_\nu\phi-\frac{\Delta_\phi}{d}\partial_\nu X^\nu\partial_\mu\phi-\frac{\Delta_\phi}{d}\partial_\nu (\partial_\mu X^\nu) \phi.
    \end{aligned}
\end{equation}
Note that \eqref{eq:partialXPhi} is identical to \eqref{eq:XPartialPhi}, and then the active transformation commutes with partial derivatives and we have
\begin{equation}
    \hat\X(\partial_\mu\phi)=\partial_\mu(\hat\X\phi).
\end{equation}
The property can be similarly extended to vector and fermionic fields as well, and also to derivatives of arbitrary tensor fields.

\subsection{Composition of active transformations}\label{app:CompTF}
In order to identify representations of the universal enveloping algebra of active field transformations, we have to investigate compositions of active transformations built from the Lie algebra of symmetries itself. As argued in the main text already, the representation of active transformations on fundamental fields is given via the Weyl-Lie derivative. Furthermore, products of such transformations are given by swapping their order when represented as Weyl-Lie derivatives. We will give an alternative proof for this behaviour in this section.

Let us start by considering a primary scalar field. Under consecutive conformal transformations generated by
\begin{equation}
    x''^\mu=x'^\mu+Y^\mu(x')\qquad x'^\mu=x^\mu+X^\mu(x),
\end{equation}
a scalar field transforms as
\begin{equation}\label{eq:activeTransDouble}
\begin{aligned}
    \phi''(x'')&=\phi'(x')-\frac{\Delta_\phi}{d}\partial'_{\mu}Y^\mu(x')\phi'(x')\\
    &=\phi(x)-\frac{\Delta_\phi}{d}\partial_\mu X^\mu(x)\phi(x)-\frac{\Delta_\phi}{d}\partial_\mu Y^\mu(x)\phi(x)-\frac{\Delta_\phi}{d}X^\nu(\partial_\mu\partial_\nu Y^\nu)\phi(x)+\frac{\Delta^2}{d^2}\partial_\mu X^\mu\partial_\nu Y^\nu \phi. 
\end{aligned}
\end{equation}
On the other hand, we have
\begin{equation}\label{eq:doubleFullTrans}
    \begin{aligned}
        \phi''(x'')=\phi'(x')+Y^\mu(x')\partial'_\mu\phi'(x')+[\hat\Y\phi'](x'),
    \end{aligned}
\end{equation}
where the vector field $Y$ evaluated at the position $x'$ can be evaluated at $x$ as $Y(x')=Y(x)+\left[X,Y\right]$ while $[\hat\Y\phi](x')$ denotes the field $\hat\Y\phi$ evaluated at the position $x'$ and hence,
\begin{equation}
\begin{aligned}
    [\hat\Y\phi'](x)&=\hat\Y\phi'(x')+\left[\hat\X,\hat\Y\right]\phi(x)\\
    &=\hat\Y\hat\X\phi(x)+\hat\Y(X^\mu\partial_\mu\phi(x))+\left[\hat\X,\hat\Y\right]\phi(x).
\end{aligned}
\end{equation}
As a result, \eqref{eq:doubleFullTrans} becomes
\begin{equation}
    \begin{aligned}
        \phi''(x'')&=\phi(x)+Y^\mu\partial_\mu\phi(x)+\hat\Y\phi(x)+X^\mu\partial_\mu\phi(x)+\hat\X\phi(x)+\left[X,Y\right]^\mu\partial_\mu\phi(x)+\left[\hat\X,\hat\Y\right]\phi(x)\\&+Y^\mu\partial_\mu(X^\nu\partial_\nu\phi(x)+\hat\X\phi(x))+\hat\Y(X^\nu\partial_\nu\phi(x)+\hat\X\phi(x))\\
        &=\phi-\frac{\Delta_\phi}{d}(\partial_\mu Y^\mu+\partial_\mu X^\mu)\phi-\frac{\Delta_\phi}{d}\partial_\mu\left[X,Y\right]^\mu\phi-Y^\mu\partial_\mu(\frac{\Delta_\phi}{d}\partial_\nu X^\nu\phi)\\
        &+ X^\mu\partial_\mu(\hat\Y\phi)+\hat\Y\hat\X\phi\\
        &=\phi-\frac{\Delta_\phi}{d}(\partial_\mu Y^\mu+\partial_\mu X^\mu)\phi-\frac{\Delta_\phi}{d}(X^\mu\partial_\mu\partial_\nu Y^\nu-Y^\mu\partial_\mu\partial_\nu X^\nu)\phi\\
        &-\frac{\Delta_\phi}{d}\partial_\nu X^\nu Y^\mu\partial_\mu\phi-\frac{\Delta}{d}Y^\mu\partial_\mu\partial_\nu X^\nu\phi- X^\mu\partial_\mu(Y^\nu\partial_\nu\phi)-\frac{\Delta}{d}X^\mu\partial_\mu(\partial_\nu Y^\nu\phi)+\hat\Y\hat\X\phi\\
        &=\phi-\frac{\Delta_\phi}{d}(\partial_\mu Y^\mu+\partial_\mu X^\mu)\phi\\
        &-\frac{\Delta}{d}\left(\partial_\mu X^\mu Y^\nu\partial_\nu\phi+X^\mu\partial_\mu\partial_\nu Y^\nu\phi+X^\mu\partial_\mu(Y^\nu\partial_\nu\phi)\right).
    \end{aligned}
\end{equation}
Comparing the latter with \eqref{eq:activeTransDouble} results in
\begin{equation}
    \begin{aligned}
        \hat\Y\hat\X\phi(x)=\mathcal{L}_{-X}^W\mathcal{L}_{-Y}^W(\phi(x)).
    \end{aligned}
\end{equation}

\section{Hopf algebras and their representations}\label{app:Hopf}

\subsection{Definitions, axioms and notation}
A Hopf algebra $\mathcal H$ is a bialgebra with an antipode, satisfying some compatibility conditions. In our context, we will apply this mathematical structure to the case of the universal enveloping algebra $U(\mathfrak g)$ of the Lie (super)algebra $\mathfrak g$, and we will specify formulas to that case to make clearer examples. Being $\mathcal H$ a bialgebra, it has at the same time the structure of an algebra and a coalgebra. 

First, being an algebra, there is a product $m:\mathcal H\otimes \mathcal H\to \mathcal H$ that is associative, $m(id\otimes m)=m(m\otimes id)$, where $id$ is the identity operation in $\mathcal H$. When $\U,\V,\mathbb W\in U(\mathfrak g)$, this is the standard multiplication $m(\U\otimes \V)=\U\cdot \V=\U\V$, so that associativity is $\U(\V\mathbb W)=(\U\V)\mathbb W$. It also has a unit map $\iota:\mathbb C\to \mathcal H$ satisfying $m(\iota\otimes id)=m(id\otimes \iota)=id$ that may be interpreted simply as the statement that there exists a unit element $1\in \mathcal H$ such that $m(1\otimes \U)=m(\U\otimes 1)=\U, \forall \U\in \mathcal H$, as can be seen by assigning $\iota(1_{\mathbb C})=1_{\mathcal H}$ (and linearly for other elements of $\mathbb C$). 

Being a coalgebra, it has similar but ``dual'' properties.  First, there is a coproduct $\Delta: \mathcal H\to\mathcal H\otimes \mathcal H$ that is coassociative, $(id\otimes \Delta)\Delta=(\Delta\otimes id)\Delta$.  In the case of $U(\mathfrak g)$, one may take the ``primitive coproduct'' $\Delta(\X)=\X\otimes 1+1\otimes \X$ for $\X\in\mathfrak g$ and $\Delta(1)=1\otimes 1$. Then one satisfies the previous condition $(id\otimes \Delta)\Delta(\U)=(\Delta\otimes id)\Delta(\U)$. It also has a counit map $\epsilon:\mathcal H\to \mathbb C$ satisfying $(\epsilon\otimes id)\Delta=(id\otimes \epsilon)\Delta=id$, so that in the case of $U(\mathfrak g)$ one has $\epsilon(1_{\mathcal H})=1_{\mathbb C}$ and $\epsilon(\X)=0$ if $\X\in \mathfrak g$. For the two structures to be compatible, the following conditions must also be satisfied
\begin{equation}
     \epsilon\iota=id,\qquad
    \epsilon m=\epsilon\otimes\epsilon,\qquad
    \Delta\iota=\iota\otimes\iota,\qquad
   \Delta m=(m\otimes m)(id\otimes\tau\otimes id)(\Delta\otimes \Delta),
\end{equation}
where $\tau$ is the (graded) permutation map $\tau(\X\otimes \Y)=(-1)^{F(\X)F(\Y)}\Y\otimes \X$ where $F(\X)=0$ for $\X$ even (bosonic) and $F(\X)=1$ for $\X$ odd (fermionic).
The first condition above tells us that the unit and counit maps are the inverse of each other. The following two conditions state that $\epsilon(\X\Y)=\epsilon(\X) \epsilon(\Y)$ and $\Delta(1)=1\otimes 1$. The last property may be understood as the statement that $\Delta$ is an algebra homomorphism, $\Delta(\X\Y)=\Delta(\X)\cdot\Delta(\Y)$, where $(a\otimes b)\cdot(c\otimes d)=(-1)^{F(b)F(c)}ac\otimes bd$.

Finally, to have a Hopf algebra one needs an antipode $S:\mathcal H\to \mathcal H$ satisfying the so-called antipode condition $m(id\otimes S)\Delta=m(S\otimes id)\Delta=\iota\epsilon$. In the case of $U(\mathfrak g)$ one has $S(1)=1$ and $S(\X)=-\X$ for $\X\in\mathfrak g$. The antipode on a generic element of $U(\mathfrak g)$ may be found by knowing that it is an antihomomorphism $S(\X\Y)=S(\Y)S(\X)$. More abstractly, this can be written as $S\circ m=m((S\otimes S)\tau)$, and its dual identity is $\Delta\circ S=\tau\circ(S\otimes S)\circ\Delta$.

Given that the product $m$ is associative, it makes sense to define the operation multiplying $n$ elements,  $m^{(n)}:\mathcal H^{\otimes n}\to \mathcal H$. We may then have the recursive definition $m^{(n)}=m(m^{(n-1)}\otimes id)$ with $m^{(2)}=m$ which would correspond to the bracketing of the product as $(\ldots((\U_1\U_2)\U_3)\ldots)\U_n$. For example, $m^{(4)}=m(m\otimes 1)(m\otimes 1\otimes 1)$. Of course other expressions corresponding to different bracketings are possible, but they are all equivalent due to associativity. If $n=k+l$, a useful way to think of it may be $m^{(n)}=m\circ(m^{(k)}\otimes m^{(l)})$. Similarly, we may define the $n$-fold coproduct $\Delta^{(n)}:\mathcal H\to\mathcal H^{\otimes n}$ as $\Delta^{(n)}=(\Delta\otimes 1^{\otimes (n-2)})\Delta^{(n-1)}$ where $\Delta^{(2)}=\Delta$. For example, $\Delta^{(3)}=(\Delta\otimes 1)\Delta$ and $\Delta^{(4)}=(\Delta\otimes 1\otimes 1)(\Delta\otimes 1)\Delta$. If $n=k+l$, we also have $\Delta^{(n)}=(\Delta^{(k)}\otimes \Delta^{(l)})\Delta$.

\vspace{12pt}

Taking into account that elements  $\mathcal X\in\mathcal H\otimes \mathcal H$ may be complicated (because they may be infinite sums of more elementary elements), we will often use notation like $\mathcal X=x^\alpha\otimes x_\alpha$, where summation over the index $\alpha$ is assumed. For example, we will write the Drinfel'd twist as $\mathcal F=f^\alpha\otimes f_\alpha$. When implementing the coproduct, we will often use Sweedler notation and write
$\Delta(\U)=\U_{(1)}\otimes \U_{(2)}$. In both cases the notation serves to keep track of the position in each of the spaces in the tensor product. For example, $\mathcal F_{op}=\tau\mathcal F=f_\alpha\otimes f^\alpha$ and $\Delta^{op}(\U)=\tau\circ\Delta(\U)=\U_{(2)}\otimes \U_{(1)}$, where we omit possible signs that may appear in the supersymmetric case to avoid burdening the notation.
Sometimes we will also use a notation that indicates explicitly the spaces in the tensor products where the objects are acting. For example, for the twist we may write $\mathcal F_{12}$ and therefore for $\mathcal{F}_{op}$ we have $\mathcal F_{21}$.

\subsection{Hopf algebra representations on fields}
As already mentioned, for us the Hopf algebra of interest will be the universal enveloping algebra of symmetries acting on the fields of the field theory. These will therefore transform under a certain representation of the Hopf algebra. 

\subsubsection{The  $\mu$-product for the ordinary multiplication of fields}
The fields themselves can be multiplied with the ordinary product of functions in the undeformed case, which we denote by $\mu$, so that $\mu(\Phi_1(x)\otimes \Phi_2(x))=\Phi_1(x)\Phi_2(x)$.
An important property that is used already in the undeformed setting is that whenever an element of the universal enveloping algebra acts on the product of two fields, then this is implemented as
\begin{equation}\label{eq:Umu}
    \hat\U(\Phi_1\Phi_2)=\hat\U\mu(\Phi_1\otimes \Phi_2)=\mu(\Delta(\hat\U)(\Phi_1\otimes \Phi_2)).
\end{equation}
This may be understood as the Leibniz rule. It is easy to check the above identity for an element of $\mathfrak g$, and therefore generalise it to a generic one of $U(\mathfrak g)$.
The generalisation to more fields is
\begin{equation}\label{eq:UmuN}
    \hat\U(\Phi_1\Phi_2\cdots \Phi_n)=\mu^{(n)}(\Delta^{(n)}(\hat\U)(\Phi_1\otimes \Phi_2\otimes \cdots \otimes \Phi_n),
\end{equation}
where $\mu^{(n)}=\mu(\mu^{(n-1)}\otimes 1)$ and $\mu^{(2)}=\mu$.

We remark that extra care should be taken when applying~\eqref{eq:Umu} when another element of the universal enveloping algebra is already inside the product $\mu$. This comment is particularly important when acting with $\hat\U$ on the star product of fields, because that already comes with $\hat{\mathcal F}$ inside $\mu$. Taking $\U\in U(\mathfrak g)$ and $\W\in U(\mathfrak g)\otimes U(\mathfrak g)$, it is easy to see that, while for the standard action via Lie derivatives the rule is
\begin{equation}\label{eq:UmuF-Lieder}
    \mathcal L_U\mu(\mathcal L_{W}(\Phi_1\otimes \Phi_2))=\mu(\mathcal L_{\Delta(U)}\mathcal L_{W}(\Phi_1\otimes \Phi_2)),
\end{equation}
for active transformations the rule must be
\begin{equation}\label{eq:UmuF-active}
    \hat\U\mu(\hat{\W}(\Phi_1\otimes \Phi_2))= \mu\left(\hat{\W}\Delta(\hat \U)(\Phi_1\otimes \Phi_2)\right).
\end{equation}
This is in fact just a simple consequence of the extra antipode that appears in the composition rule of active transformations
\begin{equation}
    \begin{aligned}
            \hat\U\mu(\hat{\W}(\Phi_1\otimes \Phi_2))
            &=\mathcal L_{S(U)}\mu\left(\mathcal L_{(S\otimes S)W}(\Phi_1\otimes \Phi_2)\right)\\
            &=\mu\left(\mathcal L_{\Delta(S(U))}\mathcal L_{(S\otimes S)W}(\Phi_1\otimes \Phi_2)\right)\\
            &=\mu\left(\mathcal L_{(S\otimes S)\Delta(U)}\mathcal L_{(S\otimes S)W}(\Phi_1\otimes \Phi_2)\right)\\
            &=\mu\left(\hat{\W}\Delta(\hat \U)(\Phi_1\otimes \Phi_2)\right),
    \end{aligned}
\end{equation}
where we used first~\eqref{eq:UmuF-Lieder}, then the composition of the coproduct with the antipode under the assumption (valid in our case) that $\Delta=\tau\circ\Delta$, and finally the composition rule of active transformations. For consistency, we extend~\eqref{eq:UmuF-active} to all active transformations, including those that are not represented by Lie derivatives.

Another important remark is that when having the composition of two elements $U_1,U_2\in U(\mathfrak g)$ it is the \emph{outermost} one on which we should implement the Leibniz rule
\begin{equation}
\hat\U_1\hat \U_2 \mu\left(\Phi_1\otimes \Phi_2\right) = \hat \U_2 \mu\left(\Delta(\hat\U_1)(\Phi_1\otimes \Phi_2) \right). 
\end{equation}
In fact, this is consistent with the rules of Lie derivatives
\begin{equation}
    \begin{aligned}
        \hat\U_1\hat \U_2 \mu\left(\Phi_1\otimes \Phi_2\right) 
        &=\mathcal L_{S(U_2)}\mathcal L_{S(U_1)} \mu\left(\Phi_1\otimes \Phi_2\right) 
        =\mathcal L_{S(U_2)} \mu\left(\mathcal L_{\Delta (S(U_1))}(\Phi_1\otimes \Phi_2)\right) \\
        &= \hat \U_2 \mu\left(\Delta(\hat\U_1)(\Phi_1\otimes \Phi_2) \right),
    \end{aligned}
\end{equation}
and it makes~\eqref{eq:Umu} consistent with the fact that the map from the abstract to the active picture is a Hopf algebra homomorphism. In fact, taking for example $\U=\U_1\U_2$ one may check
\begin{equation}
\begin{aligned}
    \hat\U\mu(\Phi_1\otimes \Phi_2)
    &= \hat\U_1\hat \U_2 \mu\left(\Phi_1\otimes \Phi_2\right)
     = \hat \U_2 \mu\left(\Delta(\hat\U_1)(\Phi_1\otimes \Phi_2) \right)
     =  \mu\left(\Delta(\hat\U_1)\Delta(\hat\U_2)(\Phi_1\otimes \Phi_2) \right)\\
    &= \mu\left(\Delta(\hat \U)(\Phi_1\otimes \Phi_2)\right).
\end{aligned}
\end{equation}

\subsubsection{Integration by parts and generalised Leibniz rule}\label{app:IBP}
Let us prove that given two functions/fields $\Phi_i(x)$ then
\begin{equation}\label{eq:IBP}
    \hat\U(\Phi_1)\Phi_2 = \hat\U_{(1)}\left(\Phi_1\ S(\hat\U_{(2)})(\Phi_2)\right),
\end{equation}
where we are using the Sweedler notation $\Delta(\hat\U)=\hat\U_{(1)}\otimes \hat\U_{(2)}$ and in the above equation $\hat\U_{(1)}$ acts on all the expression inside parenthesis while $S(\hat\U_{(2)})$ only acts on $\Phi_2(x)$. In fact, remembering that the action on a product of functions has to be implemented with the coproduct as in~\eqref{eq:UmuF-active}, the right-hand-side becomes 
\begin{equation}
    \begin{aligned}
        \hat\U_{(1)}\left(\Phi_1\ S(\hat\U_{(2)})(\Phi_2)\right)
        &=\mu\left(\left(id\otimes S(\hat\U_{(2)})\right)\Delta(\hat\U_{(1)})(\Phi_1\otimes\Phi_2)\right)\\
        &=\mu\left((id\otimes m)\left(id\otimes id\otimes S\right)(\Delta\otimes id)\Delta(\hat\U)(\Phi_1\otimes\Phi_2)\right)\\
        &=\mu\left((id\otimes m)\left(id\otimes id\otimes S\right)(id\otimes \Delta)\Delta(\hat\U)(\Phi_1\otimes\Phi_2)\right)\\
        &=\mu\left(\left(id\otimes m(id\otimes S) \Delta\right)\Delta(\hat\U)(\Phi_1\otimes\Phi_2)\right)\\
        &=\mu\left(\left(id\otimes id\, \epsilon\right)\Delta(\hat\U)(\Phi_1\otimes\Phi_2)\right)\\
        &=\mu\left(\hat\U(\Phi_1)\otimes\Phi_2)\right)=\hat\U(\Phi_1)\Phi_2,
    \end{aligned}
\end{equation}
where to get the last line we used the counit axiom, before that we used the antipode axiom, and before that coassociativity. 
The above relation can be seen as a generalisation of the rule of integration by parts when using the Leibniz rule. In fact, if we take $\hat\U=\hat\X$ with $\Delta(\hat\X)=\hat\X_{(1)}\otimes \hat\X_{(2)}=\hat\X\otimes 1+1\otimes \hat\X$ and $S(\hat\X)=-\hat\X$, then one has $\hat\X(\Phi_1)\Phi_2=\hat\X(\Phi_1\Phi_2)-\Phi_1\hat\X(\Phi_2)$.  The  identity in~\eqref{eq:IBP}, however, is more general and it is valid also for  a generic element $\U$ of the universal enveloping algebra. 
The above identity can also be generalised to the case when another element $\hat \U'$ of the universal enveloping algebra is already acting on $\Phi_2$. We write it explicitly to avoid confusion with the composition rules of active transformations. In particular, using~\eqref{eq:UmuF-active} one has
\begin{equation}\label{eq:IBP-withUp}
     \hat\U(\Phi_1) \hat\U'(\Phi_2) = \hat\U_{(1)}\left(\Phi_1\ (\hat\U'S(\hat\U_{(2)}))(\Phi_2)\right).
\end{equation}

Sometimes we will use the above relation to identify terms that we will call ``total derivatives''. The terminology will be motivated by the fact that in our construction they will indeed be just derivatives. We will say that a term is a total derivative if it is of the form $\hat\V(\ldots)$ and $\hat\V$ is a quite generic element of the universal enveloping algebra that, however, cannot contain terms proportional to the unit element $1$ in the summand. In other words, $\hat\V(\ldots)$ is a total derivative if $\hat\V$ is written as the sum of products of elements $\hat\X_a$ of the conformal symmetry algebra. It is not a total derivative if, for example, $\hat \V\propto 1$ or $\hat\V=1+\hat\X_a$, etc.

Essentially, the $\hat\V$'s corresponding to total derivatives are the non-trivial symmetry transformations in $U(\mathfrak g)$, that however leave the undeformed action invariant.
As argued in section~\ref{sec:twisted-symm}, see~\eqref{eq:tot-der-tw}, they leave invariant also the star-deformed action. In those cases the Lagrangian density will be invariant possibly up to total derivatives.

Taking this into account,  a more useful  identity than~\eqref{eq:IBP-withUp} is 
\begin{equation}\label{eq:IBP-new}
    \hat\U(\Phi_1)\hat\U'(\Phi_2)=\Phi_1 (\hat\U'S(\hat\U))(\Phi_2)+\text{total derivative},\qquad 
     \quad \forall \hat\U\in U(\mathfrak g) .
\end{equation}
To prove the above identity, consider first that 
\begin{equation}\label{eq:Delta-def-1}
    \hat \U_{(1)}\otimes \hat \U_{(2)} = \Delta(\hat\U)=1\otimes \hat\U+\hat{ \V}_{(1)}'\otimes \hat{ \V}_{(2)}',
\end{equation}
where in $\hat{ \V}_{(1)}'$ we only have elements of the universal enveloping algebra different from 1. In fact, one can easily be convinced that any element of the universal enveloping algebra is of that form because
\begin{equation}
    \begin{aligned}
        &\Delta(1)=1\otimes 1,\\
        &\Delta(\hat\X)=1\otimes \hat\X+\hat\X\otimes 1,\qquad \hat\X\in\mathfrak g,\\
        &\Delta(\hat\X\hat \Y)=1\otimes \hat\X\hat\Y+(\hat\X\hat\Y\otimes 1+\hat\X\otimes \hat\Y+\hat\Y\otimes \hat\X),\qquad \hat\X,\hat\Y\in\mathfrak g,\\
    \end{aligned}
\end{equation}
and to show~\eqref{eq:Delta-def-1} for a generic element of the universal enveloping algebra it is an easy proof by induction.
Then
\begin{equation}
    \hat\U(\Phi_1)\hat\U'(\Phi_2)= \hat\U_{(1)}\left(\Phi_1\ (\hat\U'S(\hat\U_{(2)}))(\Phi_2)\right)=\Phi_1\ \hat\U'S(\hat\U)(\Phi_2)+\hat{ \V}_{(1)}'\left(\Phi_1\ \hat\U'S(\hat{ \V}_{(2)}')(\Phi_2)\right),
\end{equation}
where we first used~\eqref{eq:IBP-withUp} and then~\eqref{eq:Delta-def-1}. Indeed we see that the last term is a total derivative.

\subsubsection{Rewriting of Ward identities}\label{app:Ward}
Here we rewrite the familiar Ward identities in the language of Hopf algebras.
Consider the correlation function
\begin{equation}
   \braket{0|Tj^\mu(x)\Phi_1(x_1)\Phi_2(x_2)\ldots\Phi_N(x_N)|0},
\end{equation}
where $j^\mu(x)$ is the Noether current of a global symmetry that we assume is not anomalous at the quantum level. We will denote by $\hat\X$ the corresponding symmetry generator. Ward identities may be obtained by taking the derivative of this expression and obtaining
\begin{equation}\label{eq:calc-ward}
\begin{aligned}
  \partial_\mu^x &\braket{0|Tj^\mu(x)\Phi_1(x_1)\Phi_2(x_2)\ldots\Phi_N(x_N)|0}=\braket{0|T\partial_\mu^xj^\mu(x)\Phi_1(x_1)\Phi_2(x_2)\ldots\Phi_N(x_N)|0}\\
  &\qquad+\sum_{k=1}^N\delta(x^0-x_k^0)\braket{0|T\Phi_1(x_1)\Phi_2(x_2)\ldots[j^0(x),\Phi_k(x_k)]\ldots\Phi_N(x_N)|0},
\end{aligned}
\end{equation}
where the commutator becomes an anticommutator if both $j^0$ and $\Phi_k$ are fermionic. We allow also for this possibility because in general we are interested also in supersymmetry transformations, but for simplicity we will maintain the notation with the commutator. Other signs are taken care of by the time ordering.

The first line on the right-hand-side is zero if the symmetry is not anomalous. If now we integrate over the whole spacetime ($\int d^4x\ldots$), the left-hand-side gives zero because it is a total derivative. On the right-hand-side, the integration over $dx^0$ allows us to use the delta function, while the integration over the spatial part ($\int d^3x\ldots$) allows us to identify the corresponding charge $Q=\int d^3x j^0(x)$, so that we obtain
\begin{equation}
    \sum_{k=1}^N\braket{0|T\Phi_1(x_1)\Phi_2(x_2)\ldots[Q,\Phi_k(x_k)]\ldots\Phi_N(x_N)|0}=0.
\end{equation}
Now we identify the commutator between the charge and the field with the action of the symmetry generator $\hat\X$ on the field itself, and we conclude that 
\begin{equation}
    \begin{aligned}
\Delta^{(N)}(\hat\X)\braket{0|T\Phi_1(x_1)\Phi_2(x_2)\ldots\Phi_N(x_N)|0}&=\sum_{k=1}^N\braket{0|T\Phi_1(x_1)\Phi_2(x_2)\ldots\hat\X(\Phi_k(x_k))\ldots\Phi_N(x_N)|0}\\&=0,
    \end{aligned}
\end{equation}
where we used the $N$-fold coproduct defined in appendix~\ref{app:Hopf}.

As an example, consider $\hat\X\in\mathfrak g$ in the conformal algebra and let us take the scalar propagator. One has
\begin{equation}
  \Delta(\hat\X) \Delta_F(x-y)=-\Bigg[\left(X(x)^\rho\partial^x_\rho +\frac{\Delta_\phi}{d}\partial^x_\rho X(x)^\rho \right)+\left(X(y)^\rho\partial^y_\rho   +\frac{\Delta_\phi}{d}\partial^y_\rho X(y)^\rho \right)\Bigg]\frac{1}{|x-y|^2}=0,
\end{equation}
where we used $d=4,\Delta_\phi=1$ and we set the mass to zero to respect conformal invariance, so that the above equation is valid for translations, Lorentz, dilatations and special conformal transformations. 

In fact, the reason why the above identity works also for dilatations and special conformal transformations is that $\hat\X$ is implemented by the Weyl-Lie derivative: without the extra contribution with the weight (the scaling dimension of the field) and using the standard Lie derivatives only, one would not get invariance of the propagator.

When dealing with a set of identical fields with bosonic or fermionic statistics, the correlation functions naturally satisfy symmetry/antisymmetry properties as a consequence of the time ordering of those fields. The Ward identities naturally inherit such (anti)symmetry properties. For example, when checking the conformal invariance of the propagator of the vector field one should calculate
\begin{equation}
\begin{aligned}
    \Delta(\hat\X)\Delta^F_{\mu\nu}(x-y)&=-\Bigg[\left(X(x)^\rho\partial^x_\rho \delta_{(\mu}^{\mu'} +\partial^x_{(\mu} X(x)^{\mu'} +\frac{\Delta_A-1}{d}\partial^x_\rho X(x)^\rho \delta_{(\mu}^{\mu'}\right)\delta_{\nu)}^{\nu'}\\
    &\qquad+\delta_{(\mu}^{\mu'}\left(X(y)^\rho\partial^y_\rho \delta_{\nu)}^{\nu'} +\partial^y_{\nu)} X(y)^{\nu'}  +\frac{\Delta_A-1}{d}\partial^y_\rho X(y)^\rho \delta_{\nu)}^{\nu'}\right)\Bigg]\frac{\eta_{\mu'\nu'}}{|x-y|^2}\\
    &=-\Bigg[\left(X(x)^\rho\partial^x_\rho +\frac{\Delta_A}{d}\partial^x_\rho X(x)^\rho \right)+\left(X(y)^\rho\partial^y_\rho   +\frac{\Delta_A}{d}\partial^y_\rho X(y)^\rho \right)\Bigg]\frac{\eta_{\mu\nu}}{|x-y|^2}\\
    &=0,
\end{aligned}
\end{equation}
so that there is an explicit symmetrisation of the indices $\mu,\nu$ on the right-hand-side, and where we used the conformal Killing equation, the fact that $d=4,\Delta_A=1$ and the invariance of the scalar propagator.

\section{Drinfel'd twists}\label{app:Drinf}

\subsection{Definitions and relations to Yang-Baxter}
We refer to~\cite{drinfeld1983constant} for the original reference and to~\cite{Giaquinto:1994jx,Kulish2009} for useful presentations. A Drinfel'd twist is an invertible object $\mathcal F\in \mathcal H\otimes \mathcal H$ satisfying the cocycle condition
\begin{equation}
    (\mathcal F\otimes 1)(\Delta\otimes id)\mathcal F=(1\otimes \mathcal F)(id\otimes \Delta)\mathcal F,
\end{equation}
and the normalisation condition $(\epsilon\otimes id)\mathcal F=1\otimes 1=(id\otimes\epsilon)\mathcal F$. Taking $\mathcal F=f^\alpha\otimes f_\alpha$ and using Sweedler's notation, the cocycle condition may be written as
\begin{equation}
    f^\beta f_{(1)}^\alpha\otimes f_\beta f_{(2)}^\alpha\otimes f_\alpha
    =f^\alpha\otimes f^\beta f_{\alpha(1)}\otimes f_\beta f_{\alpha(2)}.
\end{equation}
We will often denote the inverse of the twist but putting a bar on it
\begin{equation}
    \mathcal F^{-1}=\bar{\mathcal F}=\bar f^\alpha\otimes \bar f_\alpha.
\end{equation}
We notice that the above properties for $\mathcal F$ imply the cocycle condition written as
\begin{equation}
    (\Delta\otimes id)(\bar{\mathcal F})(\bar{\mathcal F}\otimes 1)=(id\otimes \Delta)(\bar{\mathcal F})(1\otimes \bar{\mathcal F}),
\end{equation}
and the normalisation conditions $(\epsilon\otimes id)\bar{\mathcal F}=1\otimes 1=(id\otimes\epsilon)\bar{\mathcal F}$.\footnote{
For example, one has $
    1=(\epsilon\otimes id)(\mathcal F\mathcal F^{-1})=(\epsilon\otimes id)(\mathcal F)(\epsilon\otimes id)(\mathcal F^{-1})=(\epsilon\otimes id)(\mathcal F^{-1})$,
where we used the fact that the counit $\epsilon$ is an algebra homomorphism, and finally the normalisation for $\mathcal F$.}

We will always consider Drinfel'd twists continuously connected to the identity, i.e.~they will depend on a deformation parameter $\xi$ such that $\lim_{\xi\to 0}\mathcal F=1\otimes 1$. If we define
\begin{equation}
    \Delta_{\mathcal F}(\U)=\mathcal F\Delta(\U)\mathcal F^{-1},
\end{equation}
and
\begin{equation}
    S_\mathcal{F}(\U)=V^{-1}S(\U)V,\qquad 
    V^{-1}=f^\alpha S(f_\alpha),\qquad
    V=S(\bar f^\alpha)\bar f_\alpha,
\end{equation}
then we give rise to a new Hopf algebra $\mathcal H_{\mathcal F}$, that we may call a ``twisted'' version of the original one. In particular, the cocycle condition for $\mathcal F$ ensures that $\Delta_{\mathcal F}$ is coassociative.

To show that indeed $V^{-1}=f^\alpha S(f_\alpha)$ is the inverse of $V=S(\bar f^\alpha)\bar f_\alpha$ we can reproduce the proof of~\cite{Aschieri:2005zs}
\begin{equation}\label{eq:proof-uum1}
\begin{aligned}
    V^{-1}V&=f^\alpha S(f_\alpha)S(\bar f^\beta)\bar f_\beta\\
    &=\bar f^\gamma \epsilon(\bar f_\gamma)f^\alpha S(f_\alpha)S(\bar f^\beta)\bar f_\beta\\
    &=\bar f^\gamma f^\alpha S(f_\alpha)S(\bar f^\beta)\epsilon(\bar f_\gamma)\bar f_\beta\\
    &=\bar f^\gamma f^\alpha S(f_\alpha)S(\bar f^\beta)(m(S\otimes 1)\Delta(\bar f_\gamma))\bar f_\beta\\
    &=\bar f^\gamma f^\alpha S(f_\alpha)S(\bar f^\beta)S(\bar f_{\gamma{(1)}})\bar f_{\gamma{(2)}}\bar f_\beta\\
    &=\bar f^\gamma f^\alpha S(\bar f_{\gamma{(1)}}\bar f^\beta f_\alpha)\bar f_{\gamma{(2)}}\bar f_\beta\\
    &=m(m\otimes id)(id\otimes S\otimes id)((id\otimes \Delta)\bar{\mathcal F})(1\otimes\bar{\mathcal F})(\mathcal F\otimes 1)\\
    &=m(m\otimes id)(id\otimes S\otimes id)((\Delta\otimes id)\bar{\mathcal F})(\bar{\mathcal F}\otimes1)(\mathcal F\otimes 1)\\
    &=m(m\otimes id)(id\otimes S\otimes id)((\Delta\otimes id)\bar{\mathcal F})\\
    &=m((m(id\otimes S)\Delta\otimes id)\bar{\mathcal F})\\
     &=m(\epsilon\otimes id)\bar{\mathcal F})=\epsilon(\bar f^\alpha)\bar f_\alpha=1,
\end{aligned}
\end{equation}
where in the first and in the last step we used the normalisation conditions for the inverse twist.

Two twists $\mathcal F,\mathcal F'$ are said to be equivalent if there exists an invertible element $w\in \mathcal H$ such that
\begin{equation}
    \mathcal F'=(w^{-1}\otimes w^{-1})\mathcal F\ \Delta(w).
\end{equation}

Consider now the twist written as
\begin{equation}
    \mathcal F=1\otimes 1+\sum_{i=1}^\infty \xi^i\mathcal F_i,
\end{equation}
and define 
\begin{equation}
    r=\frac{1}{2}\left(\mathcal F_1-\tau(\mathcal F_1)\right),
\end{equation}
the (graded) anti-symmetrisation of  $\mathcal F_1$ (i.e.~the first non-trivial term in the expansion), so that $r^{op}=\tau(r)=-r$. Drinfel'd proved the following statements:
\begin{enumerate}
    \item $r$ is a unitary (i.e.~$r^{op}=-r$) solution of the classical Yang-Baxter equation (CYBE)
    \begin{equation}\label{eq:CYBE}
        [r_{12},r_{13}]+[r_{12},r_{23}]+[r_{13},r_{23}]=0,
    \end{equation}
    where the subindices indicate the position of the spaces in the three-fold tensor product $\mathcal H\otimes \mathcal H\otimes \mathcal H$ where the $r$-matrix acts.
    \item $r$ characterises equivalent classes of twists. In fact, if $\mathcal F'$ is a twist equivalent to $\mathcal F$ then $ r'=r$.
    \item It is always possible to put the twist in an ``$r$-symmetric form''. To be more precise, it means that given $\mathcal F$ it is always possible to find an equivalent twist $\mathcal F'$  such that the first non-trivial term in the expansion is precisely the $r$-matrix, $\mathcal F'=1\otimes 1+\xi \, r+\sum_{i=2}^\infty \xi^i\mathcal F'_i$. Moreover one has alternating symmetry properties for the higher terms, meaning that $\mathcal F'_i=(-1)^i\tau(\mathcal F'_i)$.
    \item $\mathcal R=\mathcal{F}_{21}^{-1}\mathcal F$ is a unitary (i.e. $\mathcal R_{21}\mathcal R_{12}=1$) solution of the quantum Yang-Baxter equation
    \begin{equation}\label{eq:YBE}
        \mathcal R_{12}\, \mathcal R_{13}\, \mathcal R_{23}\, =\mathcal R_{23}\, \mathcal R_{13}\, \mathcal R_{12}. 
    \end{equation}
    \item Given a solution $r$ of the classical Yang-Baxter equation, it is always possible to ``integrate'' it to a full twist of the form $1\otimes 1+\xi\, r+\sum_{i=2}^\infty \xi^i\mathcal F'_i$. 
\end{enumerate}
To summarise, Drinfel'd twists continuously connected to the identity are in one-to-one correspondence with solutions of the classical Yang-Baxter equation. Given a Drinfel'd twist $\mathcal F$ it is straightforward to construct the corresponding $r$-matrix. However, the vice versa
is in general more difficult: given an $r$ there is no constructive procedure to obtain the twist $\mathcal F$.

We will say that the twist is $r$-symmetric if $\mathcal F_1=r$, i.e.
\begin{equation}
    \mathcal F = 1\otimes 1+\xi\, r+\mathcal O(\xi^2)=1\otimes 1+\xi\, r^{ab}\, \X_a\wedge \X_b+\mathcal O(\xi^2),
\end{equation}
where we use the graded wedge product $\X_a\wedge \X_b=\X_a\otimes \X_b-(-1)^{F(a)F(b)}\X_b\otimes \X_a$. This means that $r^{ba}=-r^{ab}$ when $\X_a,\X_b$ have even grading (i.e.~they are bosonic, $F(a)=F(b)=0$) and $r^{ba}=+r^{ab}$ when $\X_a,\X_b$ have odd grading (i.e.~they are fermionic, $F(a)=F(b)=1$). 

\subsection{Identities for the unimodularity conditions}\label{app:ident-un}
When considering an equivalent twist $\mathcal F'$ as in~\eqref{eq:Fp-equiv}, we find
\begin{equation}\label{eq:mu1SFp-app}
    \begin{aligned}
        V'\equiv m(S\otimes 1)\mathcal F'^{-1}&=S( w)S(\bar f^\alpha)S(\bar w_{(1)}^\beta )\bar w_{{(2)\beta}}\bar f_\alpha  w\\
        &=S( w)S(\bar f^\alpha)(m(S\otimes 1)\Delta(\bar w))\bar f_\alpha  w\\
        &=\epsilon(\bar w)S( w)S(\bar f^\alpha)\bar f_\alpha  w\\
        &=S( w)Vw,
    \end{aligned}
\end{equation}
where we used the antilinearity of the antipode, the antipode identity $m(S\otimes 1)\Delta=i\epsilon$, and the fact that $\epsilon(\bar w)=1$.

Given the definition of $V$ in~\eqref{eq:SF} and applying $m(S\otimes 1)$  or $m(1\otimes S)$ to $\mathcal F\mathcal F^{-1}=\mathcal F^{-1}\mathcal F=1$, one may find the identities
\begin{equation}\label{eq:identities-V-gen}
    \begin{aligned}
        &V=S(\bar f^\alpha)\bar f_\alpha,\qquad
        &&V^{-1}=f^{\alpha}S(f_\alpha),\\
        &S(f^\alpha)Vf_\alpha=1,\qquad
        &&\bar f^\alpha V^{-1}S(\bar f_\alpha)=1,\\
        &f^\alpha\bar f^\beta S(\bar f_\beta)S(f_\alpha)=f_\alpha\bar f_\beta S(\bar f^\beta)S(f^\alpha)=1,
        &&S(\bar f^\alpha)S(f^\beta)f_\beta\bar f_\alpha=S(\bar f_\alpha)S(f_\beta)f^\beta\bar f^\alpha=1.
    \end{aligned}
\end{equation}
If we also demand the $\mathcal R$-unimodularity condition $S(V)=V$, then we also have
\begin{equation}\label{eq:identities-SVV}
    \begin{aligned}
        &V=S(\bar f_\alpha)\bar f^\alpha,\qquad
        &&V^{-1}=f_{\alpha}S(f^\alpha),\\
        &S(f_\alpha)Vf^\alpha=1,\qquad
        &&\bar f_\alpha V^{-1}S(\bar f^\alpha)=1.
    \end{aligned}
\end{equation}
If in addition we demand the stronger $\mathcal F$-unimodularity condition $V=1$, then we have the eight  conditions
\begin{equation}
\begin{aligned}
    &f^\alpha S(f_\alpha)=f_\alpha S(f^\alpha)=\bar f^\alpha S(\bar f_\alpha)=\bar f_\alpha S(\bar f^\alpha)\\
    =&S(f_\alpha)f^\alpha = S(f^\alpha)f_\alpha= S(\bar f_\alpha)\bar f^\alpha= S(\bar f^\alpha)\bar f_\alpha=1.
\end{aligned}
\end{equation}

\subsection{The $n$-fold twist}
As reviewed in appendix~\ref{app:ass-star}, it is useful to introduce the concept of the ``$n$-fold twist'' 
\begin{equation}\label{eq:n-F}
\begin{aligned}
     {{\mathcal F}}^{(n)}&= ({{\mathcal F}}^{(n-1)}\otimes 1)(\Delta^{(n-1)}\otimes id)({{\mathcal F}})\\
    &=\overrightarrow{\prod_{i=1}^{n-1}}(\Delta^{(i)}\otimes id^{\otimes n-i})({{\mathcal F}}\otimes1^{\otimes n-i-1}),
\end{aligned}
\end{equation}
where the arrow indicates that the factors from smaller to larger $i$ should be ordered from left to right, and it is assumed that $\Delta^{(1)}=id$.
For example, for the cases $n=3,4$ one has
\begin{equation}\label{eq:3-F}
    {{\mathcal F}}^{(3)}\equiv({{\mathcal F}}\otimes 1) (\Delta\otimes id)({{\mathcal F}}),
\end{equation}
and
\begin{equation}
    {\mathcal F}^{(4)}=({\mathcal F}\otimes 1\otimes 1)\cdot (\Delta\otimes 1\otimes 1)({\mathcal F}\otimes 1)
    \cdot (\Delta^{(3)}\otimes 1)({\mathcal F}).
\end{equation}
Obviously, ${\mathcal F}^{(2)}={\mathcal F}$.
Another useful way to think of ${\mathcal F}^{(n)}$ is to write it as
\begin{equation}\label{eq:Fn-FkFl}
    {{\mathcal F}}^{(n)}=({{\mathcal F}}^{(k)}\otimes {{\mathcal F}}^{(l)})(\Delta^{(k)}\otimes \Delta^{(l)})({{\mathcal F}}).
\end{equation}
This is compatible with the previous expression if we declare ${\mathcal F}^{(1)}=id$.

\subsection{The opposite twist}
For later convenience, we will define also the $n$-fold generalisation of the opposite twist $\mathcal F_{op}=\tau\circ\mathcal F$. We will take it to be 
\begin{equation}\label{eq:n-Fop}
\begin{aligned}
    \mathcal F_{op}^{(n)}&= ({\mathcal F}_{op}^{(n-1)}\otimes 1)(\Delta^{(n-1)}\otimes id)({\mathcal F}_{op})\\
    &=\overrightarrow{\prod_{i=1}^{n-1}}(\Delta^{(i)}\otimes id^{\otimes n-i})(\mathcal{ F}_{op}\otimes1^{\otimes n-i-1}).
\end{aligned}
\end{equation}
In our case, the opposite twist satisfies the identity
\begin{equation}\label{eq:VSF=FopV}
    (\bar V\otimes \bar V)(S\otimes S)(\bar{\mathcal F})=\mathcal F_{op}\Delta(\bar V).
\end{equation}
This can be proved by noticing that in the presence of the twist one has the Hopf algebra property $(S_{\mathcal F}\otimes S_{\mathcal F})\circ \Delta_{\mathcal F}=\tau\circ\Delta_{\mathcal{F}}\circ S_{\mathcal F}$. Then, on a generic element $h$ of the Hopf algebra one has
\begin{equation}
    \begin{aligned}
        (S_{\mathcal F}\otimes S_{\mathcal F})\circ \Delta_{\mathcal F}(h)
        &=(\bar V\otimes \bar V)(S\otimes S)(\mathcal F\Delta(h)\bar{\mathcal F})(V\otimes V)\\
        &=(\bar V\otimes \bar V)(S\otimes S)(\bar{\mathcal F})(S\otimes S)(\Delta(h))(S\otimes S)(\mathcal F)(V\otimes V)\\
        &=(\bar V\otimes \bar V)(S\otimes S)(\bar{\mathcal F})\Delta^{op}(S(h))(S\otimes S)(\mathcal F)(V\otimes V),
    \end{aligned}
\end{equation}
and 
\begin{equation}
    \begin{aligned}
   \tau\circ\Delta_{\mathcal{F}}\circ S_{\mathcal F}(h)
   &= \tau\left(\mathcal F\Delta(\bar V S(h)V)\bar{\mathcal F}\right)\\
   &=\tau \left(\mathcal F\Delta(\bar V)\Delta(S(h))\Delta(V)\bar{\mathcal F}\right)\\
   &=\mathcal F_{op}\Delta^{op}(\bar V)\Delta^{op}(S(h))\Delta^{op}(V)\bar{\mathcal F}_{op}.
    \end{aligned}
\end{equation}
The two expressions agree for generic $h$ if~\eqref{eq:VSF=FopV} holds, where we also used the fact that the coproduct $\Delta$ on the universal enveloping algebra is cocommutative, $\Delta^{op}=\Delta$.

Equation~\eqref{eq:VSF=FopV} may be generalised to the case of $n$-fold twists. The identity is
\begin{equation}\label{eq:VSF=FopV-n}
    \bar V^{\otimes n}S^{\otimes n}(\bar{\mathcal F}^{(n)})=\mathcal F_{op}^{(n)}\Delta^{(n)}(\bar V).
\end{equation}
Given that we have already proved the case $n=2$, we may prove it for generic $n$ by induction by assuming that it holds at $n-1$. We have
\begin{equation}
    \begin{aligned}
    \bar V^{\otimes n}S^{\otimes n}(\bar{\mathcal F}^{(n)})
    &=\bar V^{\otimes n}S^{\otimes n}( (\Delta^{(n-1)}\otimes id)(\bar{\mathcal F})(\bar{\mathcal F}^{(n-1)}\otimes 1))\\
    &=\bar V^{\otimes n}S^{\otimes (n-1)}(\bar{\mathcal F}^{(n-1)}\otimes 1)S^{\otimes n}( (\Delta^{(n-1)}\otimes id)(\bar{\mathcal F}))\\
    &=(\mathcal F_{op}^{(n-1)}\Delta^{(n-1)}(\bar V)\otimes \bar V)S^{\otimes n}( (\Delta^{(n-1)}\otimes id)(\bar{\mathcal F}))\\
    &=(\mathcal F_{op}^{(n-1)}\Delta^{(n-1)}(\bar V)\otimes \bar V)( (\Delta^{(n-1)}\otimes id)((S\otimes S)\bar{\mathcal F}))\\
    &=(\mathcal F_{op}^{(n-1)}\Delta^{(n-1)}(\bar V)\otimes \bar V)( (\Delta^{(n-1)}\otimes id)((V\otimes V)\mathcal F_{op}\Delta(\bar V)))\\
    &=(\mathcal F_{op}^{(n-1)}\otimes 1)
    ( (\Delta^{(n-1)}\otimes id)(\mathcal F_{op}))
    ( (\Delta^{(n-1)}\otimes id)(\Delta(\bar V)))\\
    &=\mathcal F_{op}^{(n)}\Delta^{(n)}(\bar V).
     \end{aligned}
\end{equation}
We first used the expression~\eqref{eq:n-F} for the $n$-fold inverse twist, then the anti-homomorphism property of the antipode, then~\eqref{eq:VSF=FopV-n} at $n-1$, then the $n$-fold generalisation of $(S\otimes S)\Delta=\Delta S$, then~\eqref{eq:VSF=FopV}, the homomorphism property of the coproduct, and finally the definitions of the $n$-fold opposite twist and coproduct.

By inversion,~\eqref{eq:VSF=FopV-n} implies
\begin{equation}\label{eq:VSF=FopV-n-inverse}
    S^{\otimes n}({\mathcal F}^{(n)}) V^{\otimes n}=\Delta^{(n)}( V)\bar{\mathcal F}_{op}^{(n)},
\end{equation}
which is an identity that we will use in the main text.

\section{More details on the star product}\label{app:star}

In this appendix we will work exclusively with the star product in the active picture.

\subsection{Associativity of the star product}\label{app:ass-star}
It is possible to prove that the cocycle condition implies that 
\begin{equation}
    \Phi_1\hstar(\Phi_2\hstar\Phi_3)=(\Phi_1\hstar\Phi_2)\hstar\Phi_3,
\end{equation}
which then justifies the writing simply as $\Phi_1\hstar\Phi_2\hstar\Phi_3$ without parenthesis. In fact
\begin{equation}\label{eq:ass-star-pr}
\begin{aligned}
    \Phi_1\hstar(\Phi_2\hstar\Phi_3)
    &=\mu(\hat {{\mathcal F}}\Phi_1\otimes \mu(\hat {{\mathcal F}}\Phi_2\otimes \Phi_3))\\
    &=\mu(\hat { f}^\alpha\Phi_1\otimes\hat { f}_\alpha\mu(\hat { f}^\beta\Phi_2\otimes\hat { f}_\beta\Phi_3))\\
    &=\mu(\hat { f}^\alpha\Phi_1\otimes\mu((\hat { f}^\beta\otimes\hat { f}_\beta)\Delta(\hat { f}_\alpha)(\Phi_2\otimes\Phi_3)))\\
    &=\mu(1\otimes \mu)(1\otimes \hat {{\mathcal F}})(id\otimes \Delta)(\hat {{\mathcal F}})(\Phi_1\otimes \Phi_2\otimes \Phi_3)\\
    &=\mu(1\otimes \mu)(\hat {{\mathcal F}}\otimes 1)(\Delta\otimes id)(\hat {{\mathcal F}})(\Phi_1\otimes \Phi_2\otimes \Phi_3)\\
    &=\mu(\mu\otimes 1)(\hat {{\mathcal F}}\otimes 1)(\Delta\otimes id)(\hat {{\mathcal F}})(\Phi_1\otimes \Phi_2\otimes \Phi_3)\\
    &=(\Phi_1\hstar\Phi_2)\hstar\Phi_3,
\end{aligned}
\end{equation}
where we used the identity~\eqref{eq:UmuF-active}, the cocycle condition for $\hat {{\mathcal F}}$~\eqref{eq:cocycle}, and the associativity of $\mu$.
The above property justifies introducing the notation 
\begin{equation}
    \hat{{\mathcal F}}^{(3)}\equiv(\hat{{\mathcal F}}\otimes 1) (\Delta\otimes id)(\hat{{\mathcal F}}),
\end{equation}
as in~\eqref{eq:3-F} so that 
\begin{equation}
    \Phi_1\hstar\Phi_2\hstar\Phi_3=\mu^{(3)}(\hat {{\mathcal F}}^{(3)}(\Phi_1\otimes \Phi_2\otimes \Phi_3)).
\end{equation}
More generally one may write the star product of $n$ fields as
\begin{equation}
    \Phi_1\hstar\Phi_2\hstar\cdots \hstar\Phi_n=\mu^{(n)}(\hat {{\mathcal F}}^{(n)}(\Phi_1\otimes \Phi_2\otimes\cdots\otimes  \Phi_n)),
\end{equation}
where the ``$n$-fold twist'' is given in~\eqref{eq:n-F}. Considering a generic bracketing of $k+l=n$ fields $(\Phi_1\hstar\Phi_2\hstar\cdots \hstar\Phi_k)\hstar(\Phi_1\hstar\Phi_2\hstar\cdots \hstar\Phi_l)$, one gets the relation
\begin{equation}\label{eq:Fn-FkFl-hat}
    \hat{{\mathcal F}}^{(n)}=(\hat{{\mathcal F}}^{(k)}\otimes \hat{{\mathcal F}}^{(l)})(\Delta^{(k)}\otimes \Delta^{(l)})(\hat{{\mathcal F}}),
\end{equation}
as in~\eqref{eq:Fn-FkFl},
where the coproducts arise by pulling the twist of the ``last'' star product inside the $\mu^{(k)}$ and $\mu^{(l)}$ products.

\subsection{Tensorial properties of star products under spacetime symmetries}\label{app:tens-spt}

In this section we give the details on the tensorial nature of the star products under the spacetime symmetry transformations. This discussion is in fact similar to the one in section 7 of~\cite{Aschieri:2005yw}, discussing the tensor calculus for the deformed algebra of diffeomorphisms.
To be general, considering tensors $\Phi_{\mu_1\cdots \mu_m}^{\nu_1\cdots \nu_n}$ and $\tilde\Phi_{\mu_1\cdots \mu_{\tilde m}}^{\nu_1\cdots \nu_{\tilde n}}$ transforming as in~\eqref{eq:WeylLieTensor}, we first notice that
\begin{equation}
    \begin{aligned}
        \mathcal L^W_{(S\otimes S)\Delta(\X)}(\Phi_{\mu_1\cdots \mu_m}^{\nu_1\cdots \nu_n}\otimes \tilde\Phi_{\mu_1\cdots \mu_{\tilde m}}^{\nu_1\cdots \nu_{\tilde n}})&=
        \mathcal L^W_{-\X}\Phi_{\mu_1\cdots \mu_m}^{\nu_1\cdots \nu_n}\otimes \tilde\Phi_{\mu_1\cdots \mu_{\tilde m}}^{\nu_1\cdots \nu_{\tilde n}}+ \Phi_{\mu_1\cdots \mu_m}^{\nu_1\cdots \nu_n}\otimes \mathcal L^W_{-\X}\tilde\Phi_{\mu_1\cdots \mu_{\tilde m}}^{\nu_1\cdots \nu_{\tilde n}}.
    \end{aligned}
\end{equation}
Now, to simplify the notation, let us write
\begin{equation}
    (\Phi\hat\star\tilde\Phi)_{\mu_1\cdots \mu_m\mu_{m+1}\cdots \mu_{m+\tilde m}}^{\nu_1\cdots \nu_n\nu_{n+1}\cdots \nu_{n+\tilde n}}
    \equiv(\Phi_{\mu_1\cdots \mu_m}^{\nu_1\cdots \nu_n}\hat\star\tilde\Phi_{\mu_{m+1}\cdots \mu_{m+\tilde m}}^{\nu_{n+1}\cdots \nu_{n+\tilde n}})
\end{equation}
and then following~\eqref{eq:Xofstar} we have
\begin{equation}\label{eq:sptXofstar}
    \begin{aligned}
        \hat\X((\Phi\hat\star\tilde\Phi)_{\mu_1\cdots  \mu_{m+\tilde m}}^{\nu_1\cdots  \nu_{n+\tilde n}})&=-X^\rho\partial_\rho  (\Phi\hat\star\tilde\Phi)_{\mu_1\cdots  \mu_{m+\tilde m}}^{\nu_1\cdots  \nu_{n+\tilde n}}
        -\sum_{i=1}^{m+\tilde m}\partial_{\mu_i} X^\rho (\Phi\hat\star\tilde\Phi)_{\mu_1\cdots \mu_{i-1}\rho\mu_{i+1}\cdots \mu_{m+\tilde m}}^{\nu_1\cdots \nu_{n+\tilde n}} \\
&
+\sum_{j=1}^{n+\tilde n}\partial_\rho X^{\nu_j} (\Phi\hat\star\tilde\Phi)_{\mu_1\cdots \mu_{m+\tilde m}}^{\nu_1\cdots \nu_{j-1}\rho\nu_{j+1}\cdots \nu_{n+\tilde n}} \\
&-\frac{\Delta_\Phi+\Delta_{\tilde\Phi}+n+\tilde n-m-\tilde m}{d}\partial_\rho X^\rho\ (\Phi\hat\star\tilde\Phi)_{\mu_1\cdots \mu_{m+\tilde m}}^{\nu_1\cdots \nu_{n+\tilde n}}.
    \end{aligned}
\end{equation}
We conclude that the star product of tensors also transforms as a tensor.

Notice that the discussion above includes also the case of fermionic fields. In fact, considering the fermion bilinear $\bar\Psi\gamma^{\mu_1\cdots\mu_n}\Psi$ where $\gamma^{\mu_1\cdots\mu_n}$ denotes the antisymmetrised product of $n$ gamma matrices, and using the identity $[\gamma^{\alpha\beta},\gamma^{\mu_1\cdots\mu_n}]=-4\sum_{i=1}^n \eta^{\mu_i[\alpha}\gamma^{\mu_1\cdots\beta]\cdots\mu_n}$ (where it is understood that the index $\beta$ replaces $\mu_i$ and only $\alpha,\beta$ are antisymmetrised)
\begin{equation}
 \mathcal L^W_{\X}\bar\Psi\gamma^{\mu_1\cdots\mu_n}\Psi=X^\mu\partial_\mu \bar\Psi\gamma^{\mu_1\cdots\mu_n}\Psi-\sum_{i=1}^n\partial_\rho X^{\mu_i}\bar\Psi\gamma^{\mu_1\cdots\rho\cdots\mu_n}\Psi+\frac{2\Delta_{\Psi}+n}{4}\, \partial_\mu X^\mu\, \bar\Psi\gamma^{\mu_1\cdots\mu_n}\Psi,
\end{equation}
where we also used the conformal Killing equation~\eqref{eq:conf-Kill-eq}. The above one is indeed the Weyl-Lie derivative of a contravariant tensor of rank $n$.

\subsection{Star gauge transformations}\label{app:star-gauge-tr}
Let us start with matter fields in the fundamental representation of the gauge group. They may be scalars or fermionic spinors. We may write the gauge transformations as
\begin{equation}
    \delta_\epsilon\Phi=i\epsilon\hat\star\Phi.
\end{equation}
As remarked in the main text, we have
\begin{equation}
    \partial_\mu\delta_\epsilon\Phi=i\partial_\mu\epsilon\hat\star\Phi+i\epsilon\hat\star\partial_\mu\Phi,
\end{equation}
and then one may introduce the star-covariant derivative
\begin{equation}
    D_\mu\Phi=\partial_\mu\Phi-iA\hat\star\Phi,
\end{equation}
with the gauge field transforming as
\begin{equation}
    \delta_\epsilon A_\mu=i[\epsilon\stackrel{\hat\star}{,}A_\mu]+\partial_\mu\epsilon,
\end{equation}
where we use the star-commutator
\begin{equation}
    [\Phi_1\stackrel{\hat\star}{,}\Phi_2]=\Phi_1\hat\star\Phi_2-\Phi_2\hat\star\Phi_1,
\end{equation}
because we have
\begin{equation}
    \delta_\epsilon (D_\mu\Phi )= i\epsilon\hat\star D_\mu\Phi.
\end{equation}
Similarly, when dealing with scalars in the adjoint representations of the gauge group
\begin{equation}
    \delta_\epsilon\Phi = i[\epsilon\stackrel{\hat\star}{,}\Phi]
\end{equation}
we have the star-covariant derivative
\begin{equation}
    D_\mu\Phi=\partial_\mu\Phi-i[A\stackrel{\hat\star}{,}\Phi],
\end{equation}
because
\begin{equation}
    \delta_\epsilon (D_\mu\Phi )= i[\epsilon\stackrel{\hat\star}{,} D_\mu\Phi].
\end{equation}
We may also define the field strength of the gauge field as 
\begin{equation}
    F_{\mu\nu} = \partial_\mu A_\nu-\partial_\nu A_\mu-i[A_\mu\stackrel{\hat\star}{,}A_\nu],
\end{equation}
because then
\begin{equation}
    \delta_\epsilon F_{\mu\nu}=i[\epsilon\stackrel{\hat\star}{,}  F_{\mu\nu}].
\end{equation}
Importantly, when dealing with the $U(1)$ gauge group, there are some non-trivial $\hstar$-commutators that become trivial in the undeformed limit.



\begin{thebibliography}{10}
\ifx\href\asklfhas\newcommand{\href}[2]{#2}\fi
\ifx\arxivref\asklfhas\newcommand{\arxivref}[2]{\href{http://arxiv.org/abs/#1}{#2}}\fi
\ifx\doiref\asklfhas\newcommand{\doiref}[2]{\href{http://dx.doi.org/#1}{#2}}\fi
\raggedright
\small
\parskip 0pt

\bibitem{Szabo:2001kg}
R.~J.~Szabo,
\textit{``{Quantum field theory on noncommutative spaces}''},
\textsf{\doiref{10.1016/S0370-1573(03)00059-0}{Phys.~Rept.~378,~207~(2003)}},
\texttt{\arxivref{hep-th/0109162}{hep-th/0109162}}.

\bibitem{Douglas:2001ba}
M.~R.~Douglas and N.~A.~Nekrasov,
\textit{``{Noncommutative field theory}''},
\textsf{\doiref{10.1103/RevModPhys.73.977}{Rev.~Mod.~Phys.~73,~977~(2001)}},
\texttt{\arxivref{hep-th/0106048}{hep-th/0106048}}.

\bibitem{Szabo:2025mxr}
R.~J.~Szabo,
\textit{``{Noncommutative Geometry of Gravity, Strings and Fields: A Panoramic
  Overview}''},
\texttt{\arxivref{2511.22672}{arxiv:2511.22672}}.

\bibitem{Hersent:2022gry}
K.~Hersent, P.~Mathieu and J.-C.~Wallet,
\textit{``{Gauge theories on quantum spaces}''},
\textsf{\doiref{10.1016/j.physrep.2023.03.002}{Phys.~Rept.~1014,~1~(2023)}},
\texttt{\arxivref{2210.11890}{arxiv:2210.11890}}.

\bibitem{Vitale:2023znb}
P.~Vitale, M.~Adamo, R.~Dekhil and D.~Fern{\'a}ndez-Silvestre,
\textit{``{Introduction to noncommutative field and gauge theory}''},
\textsf{\doiref{10.22323/1.440.0007}{PoS~QG-MMSchools,~007~(2024)}},
\texttt{\arxivref{2309.17369}{arxiv:2309.17369}}.

\bibitem{Wallet:2025xbp}
J.-C.~Wallet,
\textit{``{Noncommutative Gauge Theories: Yang-Mills extensions and beyond - An
  overview}''},
\texttt{\arxivref{2510.19112}{arxiv:2510.19112}}.

\bibitem{Snyder:1946qz}
H.~S.~Snyder,
\textit{``{Quantized space-time}''},
\textsf{\doiref{10.1103/PhysRev.71.38}{Phys.~Rev.~71,~38~(1947)}}.

\bibitem{Minwalla:1999px}
S.~Minwalla, M.~Van~Raamsdonk and N.~Seiberg,
\textit{``{Noncommutative perturbative dynamics}''},
\textsf{\doiref{10.1088/1126-6708/2000/02/020}{JHEP~0002,~020~(2000)}},
\texttt{\arxivref{hep-th/9912072}{hep-th/9912072}}.

\bibitem{Giotopoulos:2021ieg}
G.~Giotopoulos and R.~J.~Szabo,
\textit{``{Braided symmetries in noncommutative field theory}''},
\textsf{\doiref{10.1088/1751-8121/ac5dad}{J.~Phys.~A~55,~353001~(2022)}},
\texttt{\arxivref{2112.00541}{arxiv:2112.00541}}.

\bibitem{drinfeld_YBESolutions_1983}
V.~Drinfel'd,
\textit{``On constant quasi-classical solutions of the Yang-Baxter quantum
  equation''},
\textsf{Sov.~Math.~Dokl.~28,~667~(1983)}.

\bibitem{Aschieri:2005zs}
P.~Aschieri, M.~Dimitrijevic, F.~Meyer and J.~Wess,
\textit{``{Noncommutative geometry and gravity}''},
\textsf{\doiref{10.1088/0264-9381/23/6/005}{Class.~Quant.~Grav.~23,~1883~(2006)}},
\texttt{\arxivref{hep-th/0510059}{hep-th/0510059}}.

\bibitem{Dimitrijevic:2011jg}
M.~Dimitrijevic and L.~Jonke,
\textit{``{A Twisted look on kappa-Minkowski: U(1) gauge theory}''},
\textsf{\doiref{10.1007/JHEP12(2011)080}{JHEP~1112,~080~(2011)}},
\texttt{\arxivref{1107.3475}{arxiv:1107.3475}}.

\bibitem{Moyal:1949sk}
J.~E.~Moyal,
\textit{``{Quantum mechanics as a statistical theory}''},
\textsf{\doiref{10.1017/S0305004100000487}{Proc.~Cambridge~Phil.~Soc.~45,~99~(1949)}}.

\bibitem{Groenewold:1946kp}
H.~J.~Groenewold,
\textit{``{On the Principles of elementary quantum mechanics}''},
\textsf{\doiref{10.1016/S0031-8914(46)80059-4}{Physica~12,~405~(1946)}}.

\bibitem{Guica:2017mtd}
M.~Guica, F.~Levkovich-Maslyuk and K.~Zarembo,
\textit{``{Integrability in dipole-deformed ${\mathcal{N}=4}$ super
  Yang–Mills}''},
\textsf{\doiref{10.1088/1751-8121/aa8491}{J.~Phys.~A~A50,~394001~(2017)}},
\texttt{\arxivref{1706.07957}{arxiv:1706.07957}}.

\bibitem{Meier:2023kzt}
T.~Meier and S.~J.~van~Tongeren,
\textit{``{Quadratic Twist-Noncommutative Gauge Theory}''},
\textsf{\doiref{10.1103/PhysRevLett.131.121603}{Phys.~Rev.~Lett.~131,~121603~(2023)}},
\texttt{\arxivref{2301.08757}{arxiv:2301.08757}}.

\bibitem{Aschieri:2009ky}
P.~Aschieri and L.~Castellani,
\textit{``{Noncommutative D=4 gravity coupled to fermions}''},
\textsf{\doiref{10.1088/1126-6708/2009/06/086}{JHEP~0906,~086~(2009)}},
\texttt{\arxivref{0902.3817}{arxiv:0902.3817}}.

\bibitem{Aschieri:2005yw}
P.~Aschieri, C.~Blohmann, M.~Dimitrijevic, F.~Meyer, P.~Schupp and J.~Wess,
\textit{``{A Gravity theory on noncommutative spaces}''},
\textsf{\doiref{10.1088/0264-9381/22/17/011}{Class.~Quant.~Grav.~22,~3511~(2005)}},
\texttt{\arxivref{hep-th/0504183}{hep-th/0504183}}.

\bibitem{Wess:2006cm}
J.~Wess,
\textit{``{Differential calculus and gauge transformations on a deformed
  space}''},
\textsf{\doiref{10.1007/s10714-007-0459-6}{Gen.~Rel.~Grav.~39,~1121~(2007)}},
\texttt{\arxivref{hep-th/0607251}{hep-th/0607251}}.

\bibitem{Dimitrijevic:2014dxa}
M.~Dimitrijevic, L.~Jonke and A.~Pachol,
\textit{``{Gauge Theory on Twisted $\kappa$-Minkowski: Old Problems and
  Possible Solutions}''},
\textsf{\doiref{10.3842/SIGMA.2014.063}{SIGMA~10,~063~(2014)}},
\texttt{\arxivref{1403.1857}{arxiv:1403.1857}}.

\bibitem{Beisert:2010jr}
N.~Beisert et~al.,
\textit{``{Review of AdS/CFT Integrability: An Overview}''},
\textsf{\doiref{10.1007/s11005-011-0529-2}{Lett.~Math.~Phys.~99,~3~(2012)}},
\texttt{\arxivref{1012.3982}{arxiv:1012.3982}}.

\bibitem{Bena:2003wd}
I.~Bena, J.~Polchinski and R.~Roiban,
\textit{``{Hidden symmetries of the AdS$_5 \times$S$^5$ superstring}''},
\textsf{\doiref{10.1103/PhysRevD.69.046002}{Phys.~Rev.~D~D69,~046002~(2004)}},
\texttt{\arxivref{hep-th/0305116}{hep-th/0305116}}.

\bibitem{Minahan:2002ve}
J.~A.~Minahan and K.~Zarembo,
\textit{``{The Bethe ansatz for N=4 superYang-Mills}''},
\textsf{\doiref{10.1088/1126-6708/2003/03/013}{JHEP~0303,~013~(2003)}},
\texttt{\arxivref{hep-th/0212208}{hep-th/0212208}}.

\bibitem{Beisert:2005tm}
N.~Beisert,
\textit{``{The su$(2|2)$ dynamic S-matrix}''},
\textsf{\doiref{10.4310/ATMP.2008.v12.n5.a1}{Adv.Theor.Math.Phys.~12,~945~(2008)}},
\texttt{\arxivref{hep-th/0511082}{hep-th/0511082}}.

\bibitem{Bombardelli:2009ns}
D.~Bombardelli, D.~Fioravanti and R.~Tateo,
\textit{``{Thermodynamic Bethe Ansatz for planar AdS/CFT: A Proposal}''},
\textsf{\doiref{10.1088/1751-8113/42/37/375401}{J.~Phys.~A~A42,~375401~(2009)}},
\texttt{\arxivref{0902.3930}{arxiv:0902.3930}}.

\bibitem{Arutyunov:2009ur}
G.~Arutyunov and S.~Frolov,
\textit{``{Thermodynamic Bethe Ansatz for the AdS(5) x S(5) Mirror Model}''},
\textsf{\doiref{10.1088/1126-6708/2009/05/068}{JHEP~0905,~068~(2009)}},
\texttt{\arxivref{0903.0141}{arxiv:0903.0141}}.

\bibitem{Gromov:2009tv}
N.~Gromov, V.~Kazakov and P.~Vieira,
\textit{``{Exact Spectrum of Anomalous Dimensions of Planar N=4 Supersymmetric
  Yang-Mills Theory}''},
\textsf{\doiref{10.1103/PhysRevLett.103.131601}{Phys.~Rev.~Lett.~103,~131601~(2009)}},
\texttt{\arxivref{0901.3753}{arxiv:0901.3753}}.

\bibitem{Gromov:2013pga}
N.~Gromov, V.~Kazakov, S.~Leurent and D.~Volin,
\textit{``{Quantum Spectral Curve for Planar $\mathcal{N} =4$ Super-Yang-Mills
  Theory}''},
\textsf{\doiref{10.1103/PhysRevLett.112.011602}{Phys.~Rev.~Lett.~112,~011602~(2014)}},
\texttt{\arxivref{1305.1939}{arxiv:1305.1939}}.

\bibitem{Klimcik:2002zj}
C.~Klimcik,
\textit{``{Yang-Baxter sigma models and dS/AdS T duality}''},
\textsf{\doiref{10.1088/1126-6708/2002/12/051}{JHEP~0212,~051~(2002)}},
\texttt{\arxivref{hep-th/0210095}{hep-th/0210095}}.

\bibitem{Klimcik:2008eq}
C.~Klimcik,
\textit{``{On integrability of the Yang-Baxter sigma-model}''},
\textsf{\doiref{10.1063/1.3116242}{J.~Math.~Phys.~50,~043508~(2009)}},
\texttt{\arxivref{0802.3518}{arxiv:0802.3518}}.

\bibitem{Delduc:2013qra}
F.~Delduc, M.~Magro and B.~Vicedo,
\textit{``{An integrable deformation of the AdS$_5 \times$S$^5$ superstring
  action}''},
\textsf{\doiref{10.1103/PhysRevLett.112.051601}{Phys.Rev.Lett.~112,~051601~(2014)}},
\texttt{\arxivref{1309.5850}{arxiv:1309.5850}}.

\bibitem{Kawaguchi:2014qwa}
I.~Kawaguchi, T.~Matsumoto and K.~Yoshida,
\textit{``{Jordanian deformations of the $AdS_5 x S^5$ superstring}''},
\textsf{\doiref{10.1007/JHEP04(2014)153}{JHEP~1404,~153~(2014)}},
\texttt{\arxivref{1401.4855}{arxiv:1401.4855}}.

\bibitem{Matsumoto:2014gwa}
T.~Matsumoto and K.~Yoshida,
\textit{``{Integrability of classical strings dual for noncommutative gauge
  theories}''},
\textsf{\doiref{10.1007/JHEP06(2014)163}{JHEP~1406,~163~(2014)}},
\texttt{\arxivref{1404.3657}{arxiv:1404.3657}}.

\bibitem{Matsumoto:2015jja}
T.~Matsumoto and K.~Yoshida,
\textit{``{Yang{\textendash}Baxter sigma models based on the CYBE}''},
\textsf{\doiref{10.1016/j.nuclphysb.2015.02.009}{Nucl.~Phys.~B~B893,~287~(2015)}},
\texttt{\arxivref{1501.03665}{arxiv:1501.03665}}.

\bibitem{vanTongeren:2015soa}
S.~J.~van~Tongeren,
\textit{``{On classical Yang-Baxter based deformations of the AdS$_{5}$ ×
  S$^{5}$ superstring}''},
\textsf{\doiref{10.1007/JHEP06(2015)048}{JHEP~1506,~048~(2015)}},
\texttt{\arxivref{1504.05516}{arxiv:1504.05516}}.

\bibitem{Beisert:2005if}
N.~Beisert and R.~Roiban,
\textit{``{Beauty and the twist: The Bethe ansatz for twisted N=4 SYM}''},
\textsf{\doiref{10.1088/1126-6708/2005/08/039}{JHEP~0508,~039~(2005)}},
\texttt{\arxivref{hep-th/0505187}{hep-th/0505187}}.

\bibitem{vanTongeren:2013gva}
S.~J.~van~Tongeren,
\textit{``{Integrability of the ${\rm Ad}{{{\rm S}}_{5}}\times {{{\rm S}}^{5}}$
  superstring and its deformations}''},
\textsf{\doiref{10.1088/1751-8113/47/43/433001}{J.Phys.~A47,~433001~(2014)}},
\texttt{\arxivref{1310.4854}{arxiv:1310.4854}}.

\bibitem{Borsato:2025smn}
R.~Borsato and M.~G.~Fern{\'a}ndez,
\textit{``{Jordanian deformation of the non-compact and $
  \mathfrak{s}{\mathfrak{l}}_2 $-invariant XXX$_{-1/2}$ spin-chain}''},
\textsf{\doiref{10.1007/JHEP08(2025)074}{JHEP~2508,~074~(2025)}},
\texttt{\arxivref{2503.24223}{arxiv:2503.24223}}.

\bibitem{Driezen:2025dww}
S.~Driezen and A.~Molines,
\textit{``{Jordanian spin chains for twisted strings in AdS5{\texttimes}S5}''},
\textsf{\doiref{10.1103/b1cg-6s5n}{Phys.~Rev.~D~112,~106001~(2025)}},
\texttt{\arxivref{2507.13911}{arxiv:2507.13911}}.

\bibitem{Driezen:2025izd}
S.~Driezen, F.~Levkovich-Maslyuk and A.~Molines,
\textit{``{Integrability for the spectrum of Jordanian AdS/CFT}''},
\textsf{\doiref{10.1007/JHEP04(2026)052}{JHEP~2604,~052~(2026)}},
\texttt{\arxivref{2511.11521}{arxiv:2511.11521}}.

\bibitem{Borsato:2026ypo}
R.~Borsato and M.~Garc{\'\i}a~Fern{\'a}ndez,
\textit{``{Groenewold-Moyal twists, integrable spin-chains and AdS/CFT}''},
\texttt{\arxivref{2604.07291}{arxiv:2604.07291}}.

\bibitem{vanTongeren:2018vpb}
S.~J.~Van~Tongeren,
\textit{``{On Yang--Baxter models, twist operators, and boundary
  conditions}''},
\textsf{\doiref{10.1088/1751-8121/aac8eb}{J.~Phys.~A~51,~305401~(2018)}},
\texttt{\arxivref{1804.05680}{arxiv:1804.05680}}.

\bibitem{Borsato:2021fuy}
R.~Borsato, S.~Driezen and J.~L.~Miramontes,
\textit{``{Homogeneous Yang-Baxter deformations as undeformed yet twisted
  models}''},
\textsf{\doiref{10.1007/JHEP04(2022)053}{JHEP~2204,~053~(2022)}},
\texttt{\arxivref{2112.12025}{arxiv:2112.12025}}.

\bibitem{Lunin:2005jy}
O.~Lunin and J.~M.~Maldacena,
\textit{``{Deforming field theories with $U(1) \times U(1)$ global symmetry and
  their gravity duals}''},
\textsf{\doiref{10.1088/1126-6708/2005/05/033}{JHEP~0505,~033~(2005)}},
\texttt{\arxivref{hep-th/0502086}{hep-th/0502086}}.

\bibitem{Leigh:1995ep}
R.~G.~Leigh and M.~J.~Strassler,
\textit{``{Exactly marginal operators and duality in four-dimensional N=1
  supersymmetric gauge theory}''},
\textsf{\doiref{10.1016/0550-3213(95)00261-P}{Nucl.~Phys.~B~447,~95~(1995)}},
\texttt{\arxivref{hep-th/9503121}{hep-th/9503121}}.

\bibitem{Maldacena:1999mh}
J.~M.~Maldacena and J.~G.~Russo,
\textit{``{Large N limit of noncommutative gauge theories}''},
\textsf{\doiref{10.1088/1126-6708/1999/09/025}{JHEP~9909,~025~(1999)}},
\texttt{\arxivref{hep-th/9908134}{hep-th/9908134}}.

\bibitem{Hashimoto:1999ut}
A.~Hashimoto and N.~Itzhaki,
\textit{``{Noncommutative Yang-Mills and the AdS / CFT correspondence}''},
\textsf{\doiref{10.1016/S0370-2693(99)01037-0}{Phys.~Lett.~B~B465,~142~(1999)}},
\texttt{\arxivref{hep-th/9907166}{hep-th/9907166}}.

\bibitem{vanTongeren:2015uha}
S.~J.~van~Tongeren,
\textit{``{Yang–Baxter deformations, AdS/CFT, and twist-noncommutative gauge
  theory}''},
\textsf{\doiref{10.1016/j.nuclphysb.2016.01.012}{Nucl.~Phys.~B~B904,~148~(2016)}},
\texttt{\arxivref{1506.01023}{arxiv:1506.01023}}.

\bibitem{vanTongeren:2016eeb}
S.~J.~van~Tongeren,
\textit{``{Almost abelian twists and AdS/CFT}''},
\textsf{\doiref{10.1016/j.physletb.2016.12.002}{Phys.~Lett.~B~B765,~344~(2017)}},
\texttt{\arxivref{1610.05677}{arxiv:1610.05677}}.

\bibitem{Araujo:2017jkb}
T.~Araujo, I.~Bakhmatov, E.~O.~Colg\'{a}in, J.~Sakamoto, M.~M.~Sheikh-Jabbari
  and K.~Yoshida,
\textit{``{Yang-Baxter $\sigma$-models, conformal twists, and noncommutative
  Yang-Mills theory}''},
\textsf{\doiref{10.1103/PhysRevD.95.105006}{Phys.~Rev.~D~D95,~105006~(2017)}},
\texttt{\arxivref{1702.02861}{arxiv:1702.02861}}.

\bibitem{Araujo:2017jap}
T.~Araujo, I.~Bakhmatov, E.~{\'O}.~Colg{\'a}in, J.-i.~Sakamoto,
  M.~M.~Sheikh-Jabbari and K.~Yoshida,
\textit{``{Conformal twists, Yang{\textendash}Baxter
  {\ensuremath{\sigma}}-models {\&} holographic noncommutativity}''},
\textsf{\doiref{10.1088/1751-8121/aac195}{J.~Phys.~A~51,~235401~(2018)}},
\texttt{\arxivref{1705.02063}{arxiv:1705.02063}}.

\bibitem{Meier:2023lku}
T.~Meier and S.~J.~van~Tongeren,
\textit{``{Gauge theory on twist-noncommutative spaces}''},
\textsf{\doiref{10.1007/JHEP12(2023)045}{JHEP~2312,~045~(2023)}},
\texttt{\arxivref{2305.15470}{arxiv:2305.15470}}.

\bibitem{FILK199653}
T.~Filk,
\textit{``Divergencies in a field theory on quantum space''},
\textsf{\doiref{https://doi.org/10.1016/0370-2693(96)00024-X}{Physics~Letters~B~376,~53~(1996)}},
\href{https://www.sciencedirect.com/science/article/pii/037026939600024X}{\texttt{https://www.sciencedirect.com/science/article/pii/037026939600024X}}.

\bibitem{StijnJulio}
J.~Cabello~Gil and S.~van~Tongeren,
to appear.

\bibitem{Galperin:2001seg}
A.~S.~Galperin, E.~A.~Ivanov, V.~I.~Ogievetsky and E.~S.~Sokatchev,
\textit{``{Harmonic superspace}''},
Cambridge University Press (2007).

\bibitem{drinfeld1983constant}
V.~G.~Drinfeld,
\textit{``Constant quasiclassical solutions of the Yang--Baxter quantum
  equation''},
in: \textit{``Doklady Akademii Nauk''},
531--535p.

\bibitem{Giaquinto:1994jx}
A.~Giaquinto and J.~J.~Zhang,
\textit{``{Bialgebra actions, twists, and universal deformation formulas}''},
\textsf{\doiref{10.1016/S0022-4049(97)00041-8}{J.~Pure~Appl.~Algebra~128,~133~(1998)}},
\texttt{\arxivref{hep-th/9411140}{hep-th/9411140}}.

\bibitem{Kulish2009}
P.~Kulish,
\textit{``Twist Deformations of Quantum Integrable Spin Chains''},
in: \textit{``Noncommutative Spacetimes: Symmetries in Noncommutative Geometry
  and Field Theory''},
Springer Berlin Heidelberg (2009),
Berlin, Heidelberg,
167--190p,
\href{https://doi.org/10.1007/978-3-540-89793-4\_9}{\texttt{https://doi.org/10.1007/978-3-540-89793-4\_9}}.

\bibitem{tolstoy2004chainsextendedjordaniantwists}
V.~N.~Tolstoy,
\textit{``Chains of extended Jordanian twists for Lie superalgebras''},
\texttt{\arxivref{math/0402433}{math/0402433}},
\href{https://arxiv.org/abs/math/0402433}{\texttt{https://arxiv.org/abs/math/0402433}}.

\bibitem{Drinfeld:1985rx}
V.~G.~Drinfeld,
\textit{``{Hopf algebras and the quantum Yang-Baxter equation}''},
\textsf{Sov.~Math.~Dokl.~32,~254~(1985)}.

\bibitem{Aschieri:2006ye}
P.~Aschieri, M.~Dimitrijevic, F.~Meyer, S.~Schraml and J.~Wess,
\textit{``{Twisted gauge theories}''},
\textsf{\doiref{10.1007/s11005-006-0108-0}{Lett.~Math.~Phys.~78,~61~(2006)}},
\texttt{\arxivref{hep-th/0603024}{hep-th/0603024}}.

\bibitem{Vassilevich:2006tc}
D.~V.~Vassilevich,
\textit{``{Twist to close}''},
\textsf{\doiref{10.1142/S0217732306020755}{Mod.~Phys.~Lett.~A~21,~1279~(2006)}},
\texttt{\arxivref{hep-th/0602185}{hep-th/0602185}}.

\bibitem{Kontsevich:1997vb}
M.~Kontsevich,
\textit{``{Deformation quantization of Poisson manifolds. 1.}''},
\textsf{\doiref{10.1023/B:MATH.0000027508.00421.bf}{Lett.~Math.~Phys.~66,~157~(2003)}},
\texttt{\arxivref{q-alg/9709040}{q-alg/9709040}}.

\bibitem{Borsato:2025jre}
R.~Borsato and T.~Meier,
\textit{``{Non-commutative deformations of gauge theories via Drinfel'd twists
  of the scale symmetry}''},
\texttt{\arxivref{2512.04162}{arxiv:2512.04162}}.

\bibitem{Aschieri:2009zz}
P.~Aschieri, M.~Dimitrijevic, P.~Kulish, F.~Lizzi and J.~Wess,
\textit{``{Noncommutative spacetimes: Symmetries in noncommutative geometry and
  field theory}''}.

\bibitem{Osten:2016dvf}
D.~Osten and S.~J.~van~Tongeren,
\textit{``{Abelian Yang–Baxter deformations and TsT transformations}''},
\textsf{\doiref{10.1016/j.nuclphysb.2016.12.007}{Nucl.~Phys.~B~B915,~184~(2017)}},
\texttt{\arxivref{1608.08504}{arxiv:1608.08504}}.

\bibitem{Gomis:2000zz}
J.~Gomis and T.~Mehen,
\textit{``{Space-time noncommutative field theories and unitarity}''},
\textsf{\doiref{10.1016/S0550-3213(00)00525-3}{Nucl.~Phys.~B~591,~265~(2000)}},
\texttt{\arxivref{hep-th/0005129}{hep-th/0005129}}.

\bibitem{Seiberg:2000gc}
N.~Seiberg, L.~Susskind and N.~Toumbas,
\textit{``{Space-time noncommutativity and causality}''},
\textsf{\doiref{10.1088/1126-6708/2000/06/044}{JHEP~0006,~044~(2000)}},
\texttt{\arxivref{hep-th/0005015}{hep-th/0005015}}.

\bibitem{Meier:2025tjq}
T.~Meier and S.~J.~van~Tongeren,
\textit{``{Integrable Spin Chains in Twisted Maximally Supersymmetric
  Yang-Mills Theory}''},
\textsf{\doiref{10.1103/lhfj-cw9d}{Phys.~Rev.~Lett.~136,~051601~(2026)}},
\texttt{\arxivref{2507.18626}{arxiv:2507.18626}}.

\bibitem{Seiberg:2003yz}
N.~Seiberg,
\textit{``{Noncommutative superspace, N = 1/2 supersymmetry, field theory and
  string theory}''},
\textsf{\doiref{10.1088/1126-6708/2003/06/010}{JHEP~0306,~010~(2003)}},
\texttt{\arxivref{hep-th/0305248}{hep-th/0305248}}.

\bibitem{Borowiec:2008se}
A.~Borowiec, J.~Lukierski and V.~N.~Tolstoy,
\textit{``{New twisted quantum deformations of D=4 super-Poincare algebra}''},
\texttt{\arxivref{0803.4167}{arxiv:0803.4167}},
in: \textit{``{7th International Workshop on Supersymmetries and Quantum
  Symmetries}''},
205--216p.

\bibitem{Borsato:2016ose}
R.~Borsato and L.~Wulff,
\textit{``{Target space supergeometry of $\eta$ and $\lambda$-deformed
  strings}''},
\textsf{\doiref{10.1007/JHEP10(2016)045}{JHEP~1610,~045~(2016)}},
\texttt{\arxivref{1608.03570}{arxiv:1608.03570}}.

\bibitem{vanTongeren:2019dlq}
S.~J.~van~Tongeren,
\textit{``{Unimodular jordanian deformations of integrable superstrings}''},
\textsf{\doiref{10.21468/SciPostPhys.7.1.011}{SciPost~Phys.~7,~011~(2019)}},
\texttt{\arxivref{1904.08892}{arxiv:1904.08892}}.

\bibitem{Borsato:2022ubq}
R.~Borsato and S.~Driezen,
\textit{``{All Jordanian deformations of the $AdS_5 \times S^5$
  superstring}''},
\textsf{\doiref{10.21468/SciPostPhys.14.6.160}{SciPost~Phys.~14,~160~(2023)}},
\texttt{\arxivref{2212.11269}{arxiv:2212.11269}}.

\bibitem{Wulff:2018aku}
L.~Wulff,
\textit{``{Trivial solutions of generalized supergravity vs non-abelian
  T-duality anomaly}''},
\textsf{\doiref{10.1016/j.physletb.2018.04.025}{Phys.~Lett.~B781,~417~(2018)}},
\texttt{\arxivref{1803.07391}{arxiv:1803.07391}}.

\bibitem{Hronek:2020skb}
S.~Hronek and L.~Wulff,
\textit{``{Relaxing unimodularity for Yang-Baxter deformed strings}''},
\textsf{\doiref{10.1007/JHEP10(2020)065}{JHEP~2010,~065~(2020)}},
\texttt{\arxivref{2007.15663}{arxiv:2007.15663}}.

\bibitem{Gross:2000ba}
D.~J.~Gross, A.~Hashimoto and N.~Itzhaki,
\textit{``{Observables of noncommutative gauge theories}''},
\textsf{\doiref{10.4310/ATMP.2000.v4.n4.a4}{Adv.~Theor.~Math.~Phys.~4,~893~(2000)}},
\texttt{\arxivref{hep-th/0008075}{hep-th/0008075}}.

\bibitem{Kowalski-Glikman:2004fsz}
J.~Kowalski-Glikman,
\textit{``{Introduction to doubly special relativity}''},
\textsf{\doiref{10.1007/11377306_5}{Lect.~Notes~Phys.~669,~131~(2005)}},
\texttt{\arxivref{hep-th/0405273}{hep-th/0405273}}.

\bibitem{Hoare:2018ngg}
B.~Hoare and F.~K.~Seibold,
\textit{``{Supergravity backgrounds of the $\eta$-deformed AdS$_2 \times S^2
  \times T^6 $ and AdS$_5 \times S^5$ superstrings}''},
\textsf{\doiref{10.1007/JHEP01(2019)125}{JHEP~1901,~125~(2019)}},
\texttt{\arxivref{1811.07841}{arxiv:1811.07841}}.

\bibitem{Freedman:2012zz}
D.~Z.~Freedman and A.~Van~Proeyen,
\textit{``{Supergravity}''},
Cambridge Univ. Press (2012),
Cambridge, UK.

\end{thebibliography}
\end{document}